%% file: main.tex
\documentclass{article}

\PassOptionsToPackage{numbers, compress}{natbib}

\usepackage[final]{neurips_2025}

\usepackage[utf8]{inputenc} 
\usepackage[T1]{fontenc}    
\usepackage{microtype}
\usepackage{graphicx}
\usepackage{booktabs} 
\usepackage{multirow}
\usepackage{caption}
\usepackage{subcaption}
\usepackage{collcell}
\usepackage{rotating}
\usepackage{makecell}
\usepackage{soul}
\usepackage{enumitem}
\usepackage{pifont}
\usepackage{makecell}
\usepackage{xcolor}
\usepackage{xspace}
\usepackage{amsmath}
\usepackage{amssymb}
\usepackage{amsfonts}       
\usepackage{mathtools}
\usepackage{amsthm}
\usepackage{nicefrac}       
\usepackage{thm-restate}
\usepackage{wrapfig}
\usepackage{float}

\usepackage{nomencl}
\makenomenclature

\usepackage{hyperref}
\usepackage{url}

\usepackage[capitalize,noabbrev]{cleveref}

\newtheorem{assumption}{Assumption}
\newcommand{\val}[2]{#1{\scriptsize $\pm$#2}}

\newcommand{\circled}[1]{\textcircled{\scriptsize #1}}

\definecolor{pinkball}{HTML}{FF93CD}
\definecolor{greenball}{HTML}{47FF9D}

\title{\algo: Temporal Chained-Hashing Watermark for Time Series Data}

\author{
Zhi Wen Soi$^{1,}$\thanks{Equal contribution.}\;\, \quad Chaoyi Zhu$^{2,*}$ \quad Fouad Abiad$^{2}$ \quad Aditya Shankar$^{2}$\vspace{0.1cm}\\
\textbf{Jeroen M. Galjaard}$^{2}$ \quad \textbf{Huijuan Wang}$^{2}$ \quad \textbf{Lydia Y. Chen}$^{1,2,}$\thanks{Corresponding author.}\vspace{0.3cm}\\
$^{1}$University of Neuchâtel \quad $^{2}$Delft University of Technology\vspace{0.15cm}\\
\texttt{\{zhi.soi, yiyu.chen\}@unine.ch}\\
\texttt{\{c.zhu-2, f.abiad, a.shankar, j.m.galjaard, h.wang\}@tudelft.nl}\\
}

\begin{document}
\include{group}

\maketitle

\colourcross{red}
\colourcheck{teal}

\begin{abstract}
Synthetic time series generated by diffusion models enable sharing privacy-sensitive datasets, such as patients' functional MRI records. Key criteria for synthetic data include high data utility and traceability to verify the data source. Recent watermarking methods embed in homogeneous latent spaces, but state-of-the-art time series generators operate in data space, making latent-based watermarking incompatible. This creates the challenge of watermarking directly in data space while handling feature heterogeneity and temporal dependencies. We propose \algo, the first watermarking algorithm for multivariate time series diffusion models. To handle temporal dependence and spatial heterogeneity, \algo embeds a temporal chained-hashing watermark directly within the temporal-feature data space. 
The other unique feature is the $\epsilon$-exact inversion, which addresses the non-uniform reconstruction error distribution across features from inverting the diffusion process to detect watermarks. We derive the error bound of inverting multivariate time series while preserving robust watermark detectability. 
We extensively evaluate \algo on its impact on synthetic data quality, watermark detectability, and robustness under various post-editing attacks, against five datasets and baselines of different temporal lengths. Our results show that \algo achieves improvements of 61.96\% in context-FID score, and 8.44\% in correlational scores against the strongest state-of-the-art baseline, while remaining consistently detectable.
Our code is available at \url{https://github.com/soizhiwen/TimeWak}.
\end{abstract}

\input{src/introduction}

\input{src/relatedwork}
\input{src/method}

\input{src/experiments}
\input{src/conclusion}

\bibliography{refs}
\bibliographystyle{plainnat}

\input{src/appendix}

\end{document}

%% file: group.tex
\newlist{todolist}{itemize}{2}
\setlist[todolist]{label=$\square$}

\newcommand{\algo}{\texttt{TimeWak}\xspace}

\newif\ifinternal
\internaltrue
 %
\ifinternal
\newcommand{\mynote}[2]{
    \fbox{
        \bfseries\footnotesize\textcolor{red}{#1}}{
            {\textcolor{cyan}{{#2}}}}{\bfseries\footnotesize}
            \typeout{#2}}

\newcommand\td[1]{\mynote{TODO}{\textcolor{cyan}{#1}}
 
}
\else
\newcommand{\mynote}[2]{}
\fi
 
\newcommand{\lc}[1]{\mynote{Lydia}{\textcolor{magenta}{~#1}}}
 
\newcommand{\je}[1]{\mynote{Jeroen}{\textcolor{blue}{~#1}}}

\newcommand{\cz}[1]{\mynote{Chaoyi}{\textcolor{orange}{~#1}}}

\newcommand{\br}[1]{\mynote{Robert}{\textcolor{green}{~#1}}}

\newcommand{\fa}[1]{\mynote{Fouad}{\textcolor{olive}{~#1}}}

\newcommand{\zw}[1]{\mynote{Zhi Wen}{\textcolor{purple}{~#1}}}

\newcommand{\as}[1]{\mynote{Aditya}{\textcolor{teal}{~#1}}}

\newcommand*\colourcheck[1]{%
  \expandafter\newcommand\csname #1check\endcsname{\textcolor{#1}{\ding{52}}}%
}
\newcommand*\colourcross[1]{%
  \expandafter\newcommand\csname #1cross\endcsname{\textcolor{#1}{\ding{54}}}%
}

%% file: src/introduction.tex
\section{Introduction}

Multivariate time series data drive key applications in healthcare~\cite{morid2020learning}, finance~\cite{li2020financial}, and science~\cite{suh_2024_timeautodiff}.
However, access to real-world datasets is often restricted by privacy regulations, limited availability, and high acquisition costs. To address these issues, synthetic time series generated by models are increasingly adopted as practical alternatives~\cite{guo_genAI, suh_2024_timeautodiff}.
Among generative techniques, \textit{diffusion} models~\cite{DDPM} have gained prominence for producing high-quality samples, often outperforming the mainstream Generative Adversarial Networks and Variational Autoencoders~\cite{diffusion_vs_GAN, kollovieh_2023}.

Beyond generation quality, \textit{traceability} is equally critical, as it ensures verifiability and safeguards against misuse~\cite{liu2023watermarking, zhu2025tabwak}. In this context, \textit{watermarking} has become the de-facto approach for tracking and auditing synthetic data~\cite{gaussianshading, liu2023watermarking}. The challenge lies in striking a delicate balance: embedding imperceptible signals that preserve the quality of generated content while remaining detectable, even under post-processing~\cite{zhu_duwak_2024}. 
Recent works embed watermarks \textit{during} generation by adding them into the \textit{latent} space, offering advantages such as generality and lightweight computation~\cite{gaussianshading, zhu2025tabwak}. However, latent-space watermarks are not always viable, especially since many state-of-the-art (SOTA) time series generators operate within the data space~\cite{tashiro_csdi_2021, alcaraz2022, yuan2024diffusion}. Additionally, latent generators introduce a trade-off in detectability, as diffusion inversion and the encode-decode cycle are inherently lossy and can degrade the watermarks~\cite{shankar2024silofuse}.

While generation-time watermarks have proven effective for images~\cite{gaussianshading} and tables~\cite{zhu2025tabwak}, their applicability to time series data remains unexplored. Multivariate time series data possess temporal dependencies and heterogeneous features, like gender versus income. The ensuing challenges are twofold: (i) embedding watermarks \textit{directly} in the data space while preserving inter- \textit{and} intra-variate temporal dependencies of the generated time series, and (ii) ensuring accurate watermark detection despite the lossy nature of diffusion inversion, whilst handling mixed feature types.

We propose \algo, the first \textit{generation-time} watermark for multivariate time series diffusion models, featuring \textbf{temporal chained-hashing} with \textbf{$\epsilon$-exact inversion}. \algo first embeds cyclic watermark patterns, i.e., the positional seeds of Gaussian noises, along the temporal direction.
First, we \textit{chained-hash} the seeds along the temporal axis, then shuffle the seeds across features to maintain temporal correlations while preserving the unique characteristics of each feature. 
To ensure reliable watermark detection, we introduce an $\epsilon$-exact inversion strategy that makes a practical concession in the otherwise exactly invertible diffusion process: the \textit{Bi-Directional Integration Approximation} (BDIA)~\cite{zhang_BDIA_2024}.
We further provide theoretical guarantees on the resulting inversion error in Appendix~\ref{app:proof}. 



We evaluate \algo against five SOTA watermarking methods on five datasets under varying temporal lengths. \algo achieves the best detectability under six post-editing attack configurations with the minimum data quality degradation. Additionally, results show that \algo preserves temporal and cross-variate dependencies with high quality with near-exact watermark bit reconstruction. To summarize, we list our contributions as follows:
\setlist[itemize]{leftmargin=1em}
\begin{itemize}
    \item We propose \algo, the first generation-time watermarking scheme for multivariate time series diffusion models,  preserving realistic spatio-temporal dependencies while remaining detectable.
    \item To preserve temporal characteristics and boost robustness against post-processing operations, we design a \textit{temporal chained-hashing} scheme that embeds watermark seeds along the temporal direction, followed by a shuffle across features.
    \item For robust detectability, we propose \textit{$\epsilon$-exact} inversion by extending BDIA sampling into a data space diffusion generator, and provide a theoretical error bound analysis.
    \item Our extensive evaluation 
    shows that \algo achieves up to \textbf{61.96\%} better context-FID scores and \textbf{8.44\%} better correlation scores compared to the strongest SOTA watermarking method.
\end{itemize}

%% file: src/relatedwork.tex
\section{Related work}
We summarize related works on watermarking diffusion models according to their generating method (post-processing or generation-time generation), and data modality. To the best of our knowledge, our work is the first generation-time watermarking scheme for time series diffusion models.

\textbf{Watermarking diffusion models.} Watermarking has become a critical solution for tracing and authenticating machine-generated content. \textit{Post-generation} techniques embed watermarks after synthesis, often degrading the generated data's quality due to direct modifications~\cite{cox2007digital, zhang2019robust}. Alternatively, recent advancements embed watermarks within the training process. Studies such as \textit{Stable Signature}~\cite{fernandez2023stable} and \textit{FixedWM}~\cite{liu2023watermarking} \textit{fine-tune} diffusion models to embed and extract watermarks. However, these methods modify model parameters, risking overfitting which hurts generalizability. 

\textbf{Watermarking images.} \textit{Tree-Ring} (TR)~\cite{treering} embeds the watermark during sampling by modifying the initial noise latent vector in the Fourier space. However, it disrupts the Gaussian noise distribution, reducing the diversity and quality of generated samples. \textit{Gaussian Shading} (GS)~\cite{gaussianshading} improves robustness by embedding watermarks directly in the latent space using invertible transformations. However, GS is tailored towards images and requires reversing synthetic samples into the latent space for detection, which is a noisy and error-prone process that limits the detection accuracy.

\textbf{Watermarking tables.} \textit{TabWak}~\cite{zhu2025tabwak} watermarks tabular data in latent space using seeded \textit{self-cloning}, \textit{shuffling} with a secret key, and a \textit{valid-bit} mechanism. However, it relies on latent models and does not account for temporal dependencies in a time series. Furthermore, its detectability is limited by the invertibility of the diffusion process and the lossy conversion to and from latent representations.

%% file: src/method.tex
\section{\algo}

\begin{figure}
\setlength\fboxsep{1.5pt}
\centering
\includegraphics[width=\linewidth]{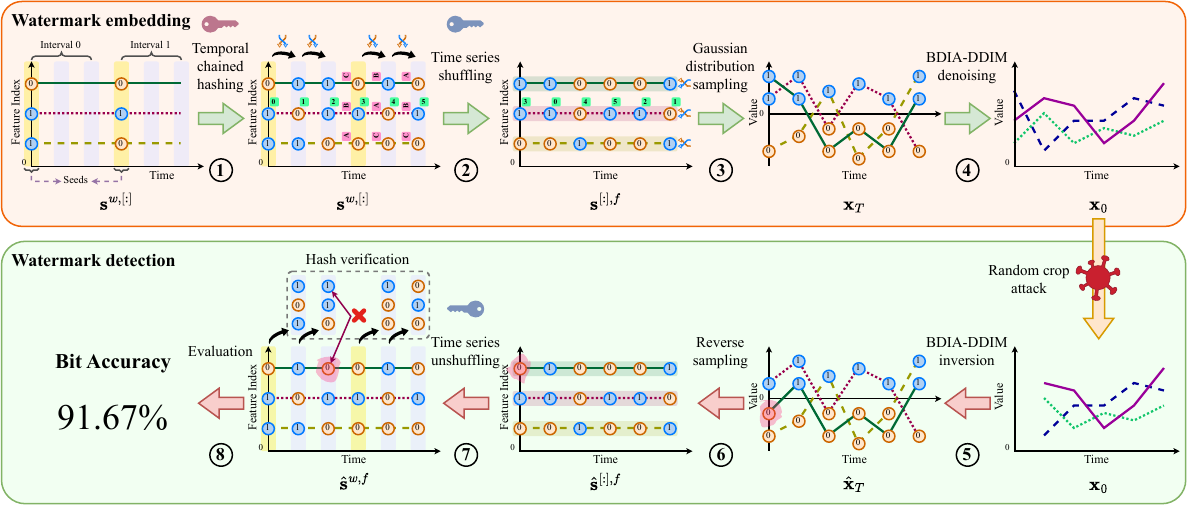}
\caption{Overview of \algo. First, we assign random seeds at the beginning of each interval. \circled{1} Temporally chained-hashing. A, B, and C (\colorbox{pinkball}{pink}) show seeds being copied from the previous step and the feature order shuffled. \circled{2} Shuffling the seeds for each series. Positional indices are highlighted in \colorbox{greenball}{green}. \circled{3} Constructing an initial Gaussian noise. \circled{4} Generating multivariate time series. \circled{5} Reversing the diffusion process. \circled{6} Recovering the watermark seed. \circled{7} Unshuffling the seeds in the opposite way they were shuffled. \circled{8} Bit accuracy between the hash and recovered seed.}
\label{fig:timewak}
\vspace{-10pt}
\end{figure}

We first highlight the unique challenges of watermarking multivariate time series, motivating the design of \algo, shown in \cref{fig:timewak}. Then we introduce \algo's key novelties: (i) temporal chained-hashing the watermark seeds cyclically along the temporal axis; (ii) shuffling the seeds across features, accounting for feature heterogeneity in the time series; and (iii) an adapted BDIA-DDIM sampling method with a theoretically bounded $\epsilon$-exact inversion, enhancing robust detectability. Key notations are summarized in \cref{appx:notation}.

\subsection{Time series diffusion and observations}

\begin{figure}
    \centering
    \begin{subfigure}{0.245\linewidth}
        \centering
        \includegraphics[width=\linewidth]{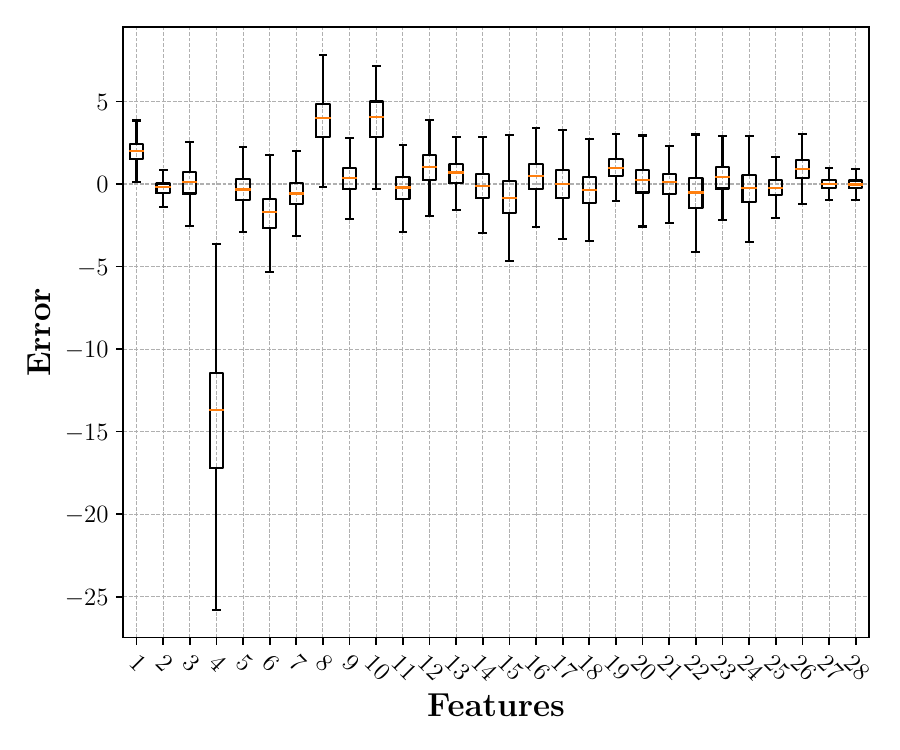}
        \caption{Feature (Energy)}\label{subfig:feature_energy}
    \end{subfigure}
    \hfill
    \begin{subfigure}{0.245\linewidth}
        \centering
        \includegraphics[width=\linewidth]{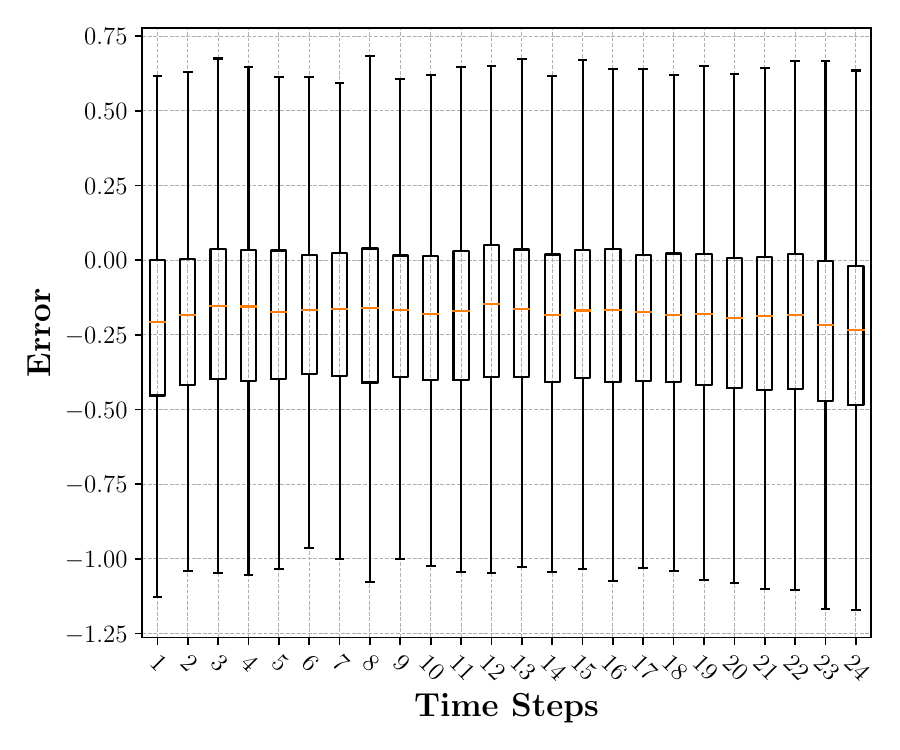}
        \caption{Temporal (Energy)}\label{subfig:temporal_energy}
    \end{subfigure}
    \hfill
    \begin{subfigure}{0.245\linewidth}
        \centering
        \includegraphics[width=\linewidth]{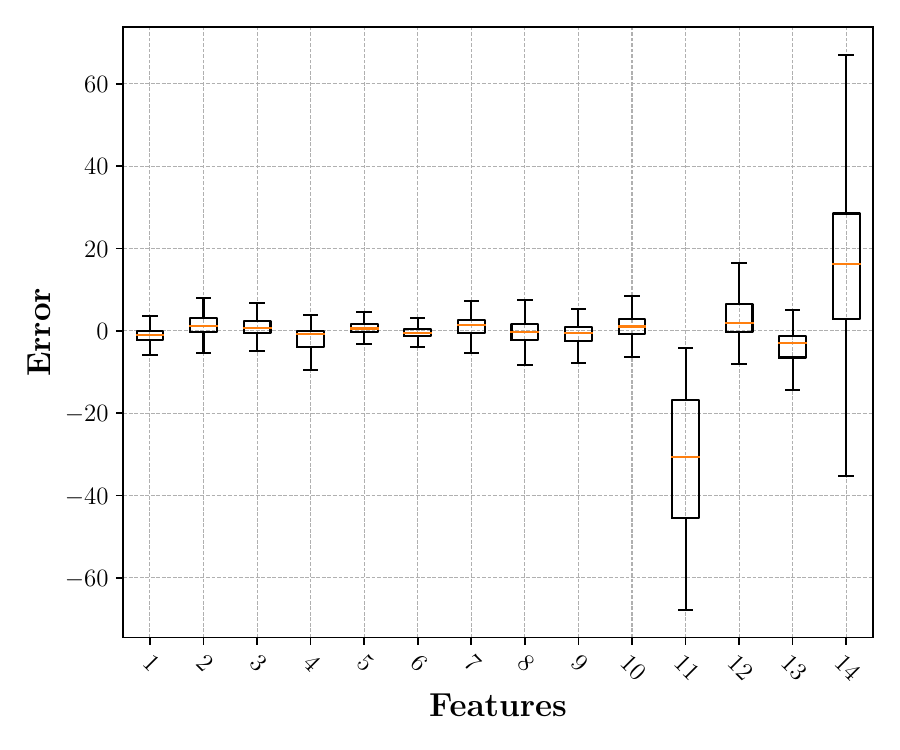}
        \caption{Feature (MuJoCo)}\label{subfig:feature_mujoco}
    \end{subfigure}
    \hfill
    \begin{subfigure}{0.245\linewidth}
        \centering
        \includegraphics[width=\linewidth]{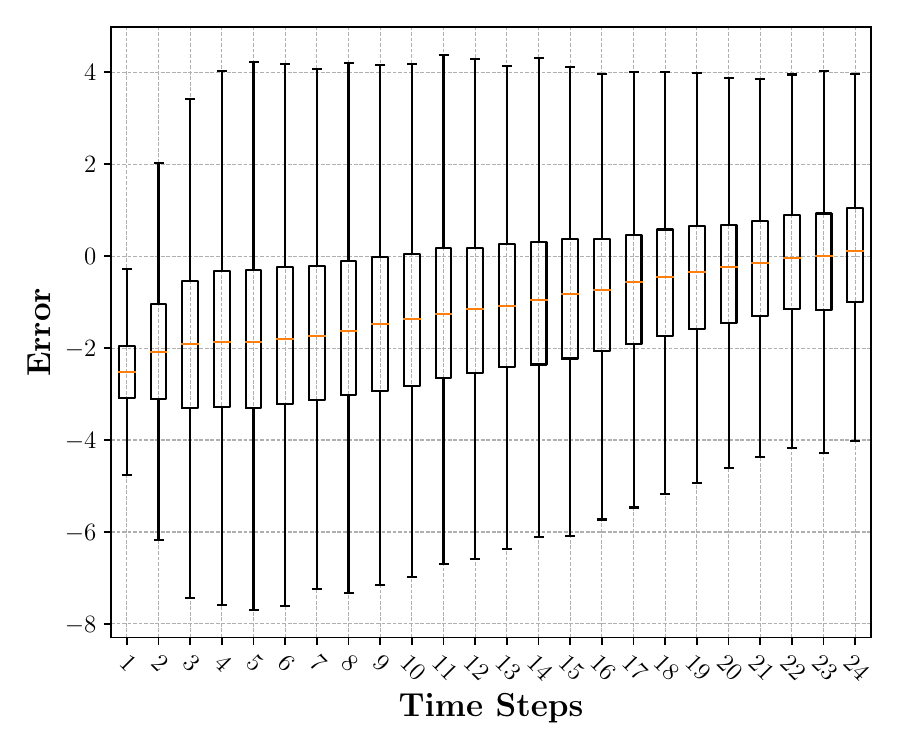}
        \caption{Temporal (MuJoCo)}\label{subfig:temporal_mujoco}
    \end{subfigure}

\caption{Average reconstruction error distribution across feature indices and timesteps on Diffusion-TS with DDIM and DDIM inversion. Reconstruction error is the signed absolute difference between reconstructed and original values.}\label{fig:error}
\vskip -0.2in
\end{figure}

\textbf{Time series diffusion.} 
 We define a time series sample of \( F \) features (variates) and \( W \) timesteps as \( \mathbf{x}_0 \in \mathbb{R}^{W \times F} \), with \( \mathbf{x}_0^{w,f} \) denoting the value of feature $f$ at  timestep $w$. Unlike image and tabular diffusion, SOTA time series diffusion models operate on the data space, which have heterogeneous features, e.g., income vs. gender, and temporal dependence~\cite{tashiro_csdi_2021, alcaraz2022, yuan2024diffusion}.  
These time series generators use \textit{Denoising Diffusion Implicit Models} (DDIM)~\cite{DDIM_ICLR21} to synthesize time series starting from  Gaussian noise, 
\(\mathbf{x}_T\), by iteratively denoising over $T$ steps, i.e., \( \mathbf{x}_T, \mathbf{x}_{T-1}, \dots, \mathbf{x}_0 \). Specifically, DDIM sets the state $\mathbf{x}_{t-1}$ at diffusion step $t-1$ as follows:
\begin{align}
    \mathbf{x}_{t-1} &= \alpha_{t-1}\left(\frac{\mathbf{x}_t - \sigma_t \hat{\boldsymbol{\epsilon}}_{\boldsymbol{\theta}}(\mathbf{x}_t, t)}{\alpha_t}\right) + \sigma_{t-1} \hat{\boldsymbol{\epsilon}}_{\boldsymbol{\theta}}(\mathbf{x}_t, t), \label{eq:DDIM}
\end{align}
where $\alpha_t$ and $\sigma_t$ are time-dependent diffusion coefficients, and $\hat{\boldsymbol{\epsilon}}_{\boldsymbol{\theta}}$ represents the model's noise estimate. DDIM approximates $\mathbf{x}_t$ as follows:
\begin{align}
    \mathbf{x}_t = \alpha_t\left(\frac{\mathbf{x}_{t-1} - \sigma_{t-1} \hat{\boldsymbol{\epsilon}}_{\boldsymbol{\theta}}(\mathbf{x}_t, t)}{\alpha_{t-1}}\right) + \sigma_t \hat{\boldsymbol{\epsilon}}_{\boldsymbol{\theta}}(\mathbf{x}_t, t) 
  \approx \alpha_t\left(\frac{\mathbf{x}_{t-1} - \sigma_{t-1} \hat{\boldsymbol{\epsilon}}_{\boldsymbol{\theta}}(\mathbf{x}_{t-1}, t)}{\alpha_{t-1}}\right) + \sigma_t \hat{\boldsymbol{\epsilon}}_{\boldsymbol{\theta}}(\mathbf{x}_{t-1}, t).
\end{align}
However, this approximation introduces errors, producing inconsistencies between the forward and backward processes. It also introduces the following time series-specific watermarking challenges:

\textbf{Spatial heterogeneity.} Watermarks must be embedded directly within the temporal and feature spaces of the data. Features can be very diverse, e.g., gender vs. income distribution, which increases the difficulty of detecting watermarks. Specifically, the key detection step inverts the time series back to Gaussian noise. Unfortunately, this inversion process is inexact, yielding reconstruction errors during noise-estimation. Figure~\ref{fig:error} shows the impact of heterogeneity on the reconstruction errors for the Energy and MuJoCo datasets. Due to spatial heterogeneity, the reconstruction errors across the features vary significantly more than they do along the temporal axis.
Existing tabular watermarks~\cite{zhu2025tabwak} implicitly assume a uniform distribution across features and compare watermark seeds across features, which prevents reliable watermark verification in multivariate time series.

\textbf{Temporal dependence.} Time series consist of values that are inherently correlated across timesteps. It is critical to preserve such temporal consistencies when generating time series. Consequently, reconstruction errors are not fully independent across timesteps within each sample, as errors at neighbouring timesteps often exhibit stronger correlations than more distant ones.
To ensure robustness, solutions must embed the watermark in a way that respects these temporal dependencies while remaining detectable. This requires designing watermarking strategies that align with the sequential nature of time series diffusion models, invalidating the applicability of existing watermark approaches that neglect the temporal dependence of time series data.


\subsection{\algo algorithm} 
To address the challenges of spatial heterogeneity and temporal dependence, we propose \algo, a method that enables per-sample watermark detection while mitigating non-uniform reconstruction errors across features and preserving temporal structure.
Through a structured propagation mechanism, \algo enhances the watermark's robustness, even in the presence of inversion errors.

\textbf{Overview.}
We begin with a high-level overview of \algo's watermarking pipeline, which consists of four main stages: watermark embedding, time series generation, inversion, and detection. Each stage involves multiple steps that we describe briefly here and explain in detail in the subsequent sections. Following Figure~\ref{fig:timewak}, the complete process works as follows:
\setlist[enumerate]{leftmargin=1.5em}
\begin{enumerate}
    \item \textbf{Embedding I:} generating watermark seeds (\(\mathbf{s}\)). We first split a multivariate time series into intervals along the time axis, then we randomly sample seeds $\mathbf{s}^{w,[:]}$ (with values in \{0, 1\}) at the start of each interval. \circled{1} We temporally chain-hash the seeds in a cyclic manner across timesteps until the end of the current interval, by applying a unique permutation key at each timestep. Then, \circled{2} we independently shuffle the seeds for each feature using distinct permutation keys.
    \item \textbf{Embedding II:} generating time series from the watermarked seeds (\(\mathbf{s} + \mathbf{x}_T \rightarrow  \mathbf{x}_0\)). \circled{3} Sampling from a feature-wise pseudo-random Gaussian distribution based on all the aforementioned seeds, where the seeds determine the sign of the sampled values ($\mathbf{s}^{w,f}=1$ becomes positive, $\mathbf{s}^{w,f}=0$ becomes negative).
    These noise signals are used as input to a BDIA variant of a DDIM diffusion model, which is then \circled{4} used to generate a multivariate time series. An attack, such as a random crop attack, may occur at this stage. 
    \item \textbf{Detection I}: inversion of the time series (\( \mathbf{x}_0 \rightarrow \hat{\mathbf{x}}_T \)). \circled{5} The inverse BDIA-DDIM process is applied to the time series, inverting it to Gaussian noise \( \hat{\mathbf{x}}_T \). Then \circled{6} we reverse-sample each time series to get the seeds (positive values become 1, negative values become 0); we now have the shuffled seed features.
    \item \textbf{Detection II:} watermark detection (\( \hat{\mathbf{x}}_T \rightarrow \hat{\mathbf{s}} \)). Given shuffled seeded features, we \circled{7} unshuffle the seeds in the opposite way they were shuffled (using the inverse of the permutation keys of step \circled{2}), to obtain the retrieved seed features $\hat{\mathbf{s}}$. We then \circled{8} verify the hash of these retrieved seed features by comparing them with the original temporal chained-hash seeds $\mathbf{s}$, for which we compute the bit accuracy between the hashed and recovered versions of each seed.
\end{enumerate}

\subsubsection{Chained-hashing watermark seeds (\texorpdfstring{\(\mathbf{s} + \mathbf{x}_T \rightarrow  \mathbf{x}_0\)}))}
Existing watermarking methods assign a watermark seed to each feature dimension with \( L \) bits, forming a seed matrix of dimensions \( W \times F \), denoted as \(\mathbf{s} \in \mathbb{R}^{W \times F }\), where \( W\) and \(F\) are the respective total timesteps and total features of the time series~\cite{gaussianshading, zhu2025tabwak}.
However, such approaches 
lack the ability to leverage temporal dependencies and may introduce inconsistencies across timesteps. 
To improve the watermark's temporal coherence, we partition the time series data into \( n = \lfloor W / H  \rfloor \) non-overlapping intervals, each of length $H$. 
At the start of each interval, the watermark seed across all features, \(\mathbf{s}^{kH+1, [:]} \in \mathbb{R}^{F}\), is sampled from a discrete uniform distribution $\mathcal{U}\left(\left\{0, L-1\right\}\right)^{F}$, with $[:]$ denoting all indices along a dimension.

Within each interval, the watermark seed evolves over timesteps using a \textbf{temporal chained-hashing} mechanism. Specifically, for all features, the seed at timestep \( w \) is recursively derived from the seed at the previous timestep \( w-1 \), ensuring temporal consistency. Formally, for \( k = 0, \dots, n-1 \), we initialize the watermark seed as:
\begin{equation}
\mathbf{s}^{w,[:]} =
\begin{cases} 
\mathcal{U}\left(\left\{0, L-1\right\}\right)^{F} & \textbf{if } w = kH+1, \\
\mathcal{H}\left(\kappa, w, \mathbf{s}^{w-1, [:]}\right) & \textbf{otherwise},
\end{cases}
\end{equation}
where \( \kappa \) is a cryptographic key controlling the hashing process, \( \mathcal{H} \) is a deterministic permutation hash function ensuring temporal consistency, and \( \mathcal{U}\left(\left\{0, L-1\right\}\right)^{F}\) is a vector of $F$ i.i.d discrete uniform samples over $\{0,1,\ldots,L-1\}$. The parameter \( n = \lfloor W  / H  \rfloor\) denotes the total integer count of intervals, while \( k \) indexes the intervals, ranging from \( 0 \) to \( n-1 \). While temporal chaining preserves coherence across timesteps by linking each seed to its past, it may lead to repetitive patterns across intervals. To increase diversity, we further permute the seeds along the temporal axis for each feature:
\begin{equation}
\mathbf{s}^{[:],f} \leftarrow \pi_\kappa(\mathbf{s}^{[:], f}),
\end{equation}
where $\pi_\kappa$ is a permutation function parameterized by the cryptographic key $\kappa$.
This step preserves inter-feature seed correlations while adding generation diversity.

After obtaining the watermark seed, we 
construct an initial Gaussian noise sample as follows. 
First, we draw a variable from the continuous uniform distribution \( u \sim \mathcal{U}(0,1) \) and use it to generate the noise variable \( \mathbf{x}^{w,f}_{T} \) at diffusion step $T$ as:
\begin{equation}
\mathbf{x}_{T}^{w,f} = \Phi^{-1} \left(\frac{u + \mathbf{s}^{w,f}}{L} \right),
\end{equation}
where \( \Phi^{-1}(\cdot) \) is the percent point function (PPF) of the standard Gaussian distribution \( \Phi(\cdot) \), and \( \mathbf{s}^{w, f} \) is the watermark seed for feature \( f \) at timestep \( w \). 
Finally, the final time series sample $\mathbf{x}_{0}$ is obtained by denoising the initial noise $\mathbf{x}_{T}$ with the learned diffusion model. 

\subsubsection{\texorpdfstring{$\epsilon$}--Exact inversion (\texorpdfstring{\(\mathbf{x}_0 \rightarrow \hat{\mathbf{x}}_T\)}))}
Here, we propose a near-lossless inversion procedure by adopting 
the Bi-directional Integration Approximation (BDIA) technique~\cite{zhang_BDIA_2024}, a novel approach to address inconsistencies in DDIM inversion~\cite{song_2021}. We introduce a practical approximation in BDIA by removing the assumption of known \(\mathbf{x_1}\), and derive the bound of the inversion error. BDIA improves upon DDIM by jointly leveraging the forward and backward diffusion updates.
Specifically, obtaining each $\mathbf{x}_{t-1}$ as a linear combination of $(\mathbf{x}_{t+1}, \mathbf{x}_t, \hat{\boldsymbol{\epsilon}}_{\boldsymbol{\theta}}(\mathbf{x}_t, t))$, where $\hat{\boldsymbol{\epsilon}}_{\boldsymbol{\theta}}$ represents the noise estimator of the diffusion model:
\begin{align}
    \mathbf{x}_{t-1} &= \gamma\left(\mathbf{x}_{t+1} - \mathbf{x}_t\right) - \gamma\left(\frac{\mathbf{x}_t}{a_{t+1}} - \frac{b_{t+1}}{a_{t+1}} \hat{\boldsymbol{\epsilon}}_{\boldsymbol{\theta}}(\mathbf{x}_t, t) - \mathbf{x}_t\right) + \left(a_t \mathbf{x}_t + b_t \hat{\boldsymbol{\epsilon}}_{\boldsymbol{\theta}}(\mathbf{x}_t, t)\right),\label{eq:BDIA}
\end{align}
where $\gamma \in [0,1]$, $a_t$ and $b_t$ are differentiable functions of $t$ with bounded derivatives.
Consequently, the inversion process can be directly calculated without approximation as follows:
\begin{equation}
    \mathbf{x}_{t+1} = \frac{\mathbf{x}_{t-1}}{\gamma} - \frac{1}{\gamma} \left(a_t \mathbf{x}_t + b_t \hat{\boldsymbol{\epsilon}}_{\boldsymbol{\theta}}(\mathbf{x}_t, t)\right) + \left(\frac{\mathbf{x}_t}{a_{t+1}} - \frac{b_{t+1}}{a_{t+1}} \hat{\boldsymbol{\epsilon}}_{\boldsymbol{\theta}}(\mathbf{x}_t, t)\right).\label{eq:BDIA-Inversion}
\end{equation}
By design, the introduced symmetry ensures time-reversible updates, meaning that if $\mathbf{x}_t$ and $\mathbf{x}_{t-1}$ are known, $\mathbf{x}_{t+1}$ can be computed without error.
However, obtaining an exact inversion requires knowing both $\mathbf{x}_1$ and $\mathbf{x}_0$, the latter of which is available only as the model's denoised sample in practice. 

To address this limitation, we introduce an adaptation to BDIA: we directly approximate $\mathbf{x}_1$ by equating it to $\mathbf{x}_0$.
This seemingly simple estimation effectively enables practical application of BDIA while maintaining reasonable accuracy.
To quantify the estimation error, we establish \cref{thm:original}, which demonstrates that the final error remains bounded in terms of the initial estimation $\epsilon=\mathbf{x}_1-\mathbf{x}_2$. We refer to this property as `$\epsilon$-exact' inversion. We defer our proof of \cref{thm:original} to \cref{app:proof}.

\begin{restatable}{theorem}{mytheorem}\label{thm:original}
Let \(\{\mathbf{x}_t\}_{t=0}^{T}\) be the sequence of diffusion states governed by the BDIA-DDIM recurrence for a given dataset, following Equation~\eqref{eq:BDIA-Inversion}. Given the noise estimator $\hat{\boldsymbol{\epsilon}}_{\boldsymbol{\theta}}$ follows Assumption~\ref{assump}. Suppose that instead of the exact terminal state \(\mathbf{x}_1\), an approximation of \(\mathbf{x}_1\), termed as \(\mathbf{x}_1^{\text{approx}}\), is used with a small perturbation \(\epsilon\), given by:
\begin{equation}
    \mathbf{x}_1^{\text{approx}} = \mathbf{x}_{2} = \mathbf{x}_1^{\text{orig}} + \epsilon.
\end{equation}

Let the propagated error at time \(t\) be defined as,
\begin{equation}
    \delta_t = \|\mathbf{x}_t^{\text{approx}} - \mathbf{x}_t^{\text{orig}}\|.
\end{equation}
Then, for \(t \geq 1\), the error is bounded by,
\begin{equation}
    \|\delta_T\| \leq |\epsilon| \prod_{t=1}^{T-1} \left( \left| \frac{1}{\gamma} - \frac{a_t}{\gamma} + \frac{1}{a_{t+1}} \right| + \frac{b_t}{\gamma} \Delta_{t} + \frac{b_{t+1}}{a_{t+1}} \Delta_{t} \right),
\end{equation}
where \(\Delta_t\) quantifies the sensitivity of the noise estimator at timestep $t$.
\end{restatable}

\begin{assumption}[Lipschitz continuity of the noise estimator]\label{assump}
There exists a time-dependent constant \(\Delta_{t} > 0\) and \(\delta_t > 0\) such that, for any diffusion state \(\mathbf{x}_t^{\text{orig}}\) encountered during sampling from a given dataset and any \(\mathbf{x}_t^{\text{approx}}\) satisfying \(\|\mathbf{x}_t^{\text{approx}} - \mathbf{x}_t^{\text{orig}}\| \leq \delta_t\), the noise estimator $\hat{\boldsymbol{\epsilon}}_{\boldsymbol{\theta}}$ satisfies the Lipschitz condition:
\begin{equation}
    \|\hat{\boldsymbol{\epsilon}}_{\boldsymbol{\theta}}(\mathbf{x}_t^{\text{approx}}, t) - \hat{\boldsymbol{\epsilon}}_{\boldsymbol{\theta}}(\mathbf{x}_t^{\text{orig}}, t)\| \leq \Delta_{t} \left\|\mathbf{x}_t^{\text{approx}} - \mathbf{x}_t^{\text{orig}}\right\|.
\end{equation}
\end{assumption}

\subsubsection{Watermark detection (\texorpdfstring{\( \mathbf{x}_0 \rightarrow \hat{\mathbf{x}}_T \rightarrow \hat{\mathbf{s}} \)}))}  
To verify the presence of \algo's watermark in a generated time series, we first recover the initial noise used in the generative process via \textbf{diffusion inversion}. Given a time series instance $\mathbf{x}_{0}$, we estimate the initial noise $\hat{\mathbf{x}}_{T}$ through inversion. The watermark seed at each timestep and feature is then recovered by inverting the Gaussian mapping as follows:
\begin{equation}
\hat{\mathbf{s}}^{w,f} = \lfloor L \cdot \Phi(\hat{\mathbf{x}}_{T}^{w,f}) \rfloor,
\end{equation}
where \( \Phi(\cdot) \) is the cumulative distribution function (CDF) of the standard Gaussian distribution. This operation reconstructs the discrete watermark values embedded during sample generation.

Since the watermarking mechanism applies a feature-wise permutation controlled by a cryptographic key, the extracted watermark values are initially shuffled across feature dimensions. We restore the original seed assignments along the timesteps using the inverse permutation: 
\begin{equation}
\hat{\mathbf{s}}^{[:],f} \leftarrow \pi_{\kappa}^{-1}(\hat{\mathbf{s}}^{[:],f}).
\end{equation}
Beyond feature-space consistency, the extracted watermark sequence should exhibit structured temporal dependencies. Specifically, the watermark seed at each step must follow a predefined hash function, given by:
\begin{equation}
\forall w \neq kH + 1, \quad \hat{\mathbf{s}}^{w,[:]} = \mathcal{H}(\kappa, w, \hat{\mathbf{s}}^{w-1,[:]}).
\end{equation}
To quantify detection confidence, we compute the \textbf{bit accuracy} of the extracted watermark sequence, measuring the proportion of correctly recovered bits:
\begin{equation}
\text{Acc} = \frac{1}{|\mathcal{W}^*| F} \sum_{w \in \mathcal{W}^*} \sum_{f=1}^{F} \mathbb{I} \Big[ \hat{\mathbf{s}}^{w,f} = \mathbf{s}^{w,f} \Big],
\end{equation}
where \( \mathbb{I}[\,\cdot\,] \) is an indicator function that evaluates to 1 if the extracted bit matches the ground-truth watermark, and \(\mathcal{W}^* = \{ w \mid w \neq kH+1 \}\) represents the valid timesteps for comparison.
By combining diffusion inversion, feature unshuffling, and temporal consistency verification, this detection framework ensures robust identification of watermarked time series samples. 

To assess the statistical significance, we compute the Z-score to measure how strongly the observed bit accuracy deviates from the expected accuracy under a null hypothesis, detailed in Appendix~\ref{appx:eval_metrics}.

%% file: src/experiments.tex
\section{Evaluation}\label{sec:evaluation}

\subsection{Experiments setup}\label{subsec:experiments}

\textbf{Datasets.} We use five time series datasets to evaluate \algo's impact on generation quality, watermark detection accuracy, and robustness towards post-editing operations. These are: \textit{Stocks}~\cite{yoon2019time}, \textit{ETTh}~\cite{zhou2021informer}, \textit{MuJoCo}~\cite{tunyasuvunakool2020dm_control}, \textit{Energy}~\cite{appliances_energy_prediction_374}, and \textit{fMRI}~\cite{smith2011network}. Additional dataset details are in \cref{appx:datasets}.

\textbf{Metrics.} \textit{Synthetic data quality:} \underline{Context-FID} score~\cite{jeha2022psa} measures the closeness between the real and synthetic time series distributions using the Fréchet distance~\cite{frechet1957}. \underline{Correlational} score~\cite{liao2020conditional} measures the cross-correlation error between the real and synthetic multivariates. \underline{Discriminative} score~\cite{yoon2019time} trains a classifier to distinguish between synthetic and real data, with low scores implying they are indistinguishable. \underline{Predictive} score~\cite{yoon2019time} measures the downstream task performance by training a sequence model on the synthetic data and evaluating on real data. \textit{Watermark detectability:} \underline{Z-score} quantifies the difference in mean values between synthetic data with and without the watermark, with larger positive values indicating better detectability. \underline{TPR@X\%FPR} measures the True Positive Rate (TPR) at a fixed False Positive Rate (FPR) of X\% in detecting watermarked time series. We provide additional details on these metrics in \cref{appx:eval_metrics}.


\textbf{Baselines.} We compare against three sampling-based diffusion watermarks: TR~\cite{treering}, GS~\cite{gaussianshading}, and TabWak~\cite{zhu2025tabwak}. We also compare with $\text{TabWak}^\top$, an adaptation of TabWak that transposes the representation and watermarks along the temporal axis instead of feature-wise. We also compare against a post-generation watermarking method for time series, \textit{Heads Tails Watermark} (HTW), which embeds a watermark by slightly adjusting the time series values by assigning a `heads' or `tails' based on a predefined ratio on the proportion of `heads' values~\cite{van_schaik_2024}. Detailed implementations of these methods are provided in Appendix~\ref{appx:baselines}. Hardware specifications are detailed in Appendix~\ref{appx:training_sampling}. 


\subsection{Synthetic data quality and watermark detectability}

\input{src/tables/main_exp}
When evaluating synthetic time series quality, \cref{tab:main_exp} shows that \algo consistently delivers top-tier performance across all metrics, outperforming or comparable to other baselines like HTW and TabWak$^\top$. While HTW sometimes surpasses un-watermarked data, possibly due to subtle perturbations introduced during watermarking that unintentionally bring synthetic samples closer to the ground truth, it fails to offer strong detectability, as reflected in its low Z-scores. In contrast, \algo and TabWak$^\top$ offer a far more favorable trade-off between quality and detectability. When benchmarked against TabWak$^\top$, \algo shows substantial gains, achieving up to 61.96\% better Context-FID score on MuJoCo and 8.44\% better correlation score on fMRI. Moreover, low discriminative and predictive scores further emphasize that \algo's watermarking remains imperceptible and does not degrade downstream utility. This is made possible by its temporal chained-hashing mechanism, which precisely embeds the watermark while preserving both temporal structure and inter-variate relationships. Meanwhile, traditional image-based watermarking methods such as TR and GS perform poorly across all quality metrics. These methods struggle with time series data because they are optimized for spatial domains. Time series, however, are governed by temporal continuity and feature heterogeneity, thus requiring fundamentally different treatment.

\algo achieves significant improvements in detection performance. Using $\epsilon$-exact inversion via BDIA-DDIM, it reconstructs high-fidelity noise estimates $\hat{\mathbf{x}}_T$ that closely resemble the ground truth $\mathbf{x}_T$. This results in consistently higher Z-scores across all datasets, outperforming all baselines, except on the fMRI dataset, where TabWak$^\top$ slightly edges out. Unlike GS, which maintains moderate detection at the cost of quality, or TR, which fails on both fronts, \algo delivers strong detectability without compromising fidelity.

\begin{figure}[ht!]
\centering

\begin{subfigure}{0.195\linewidth}
    \centering
    \includegraphics[width=\linewidth]{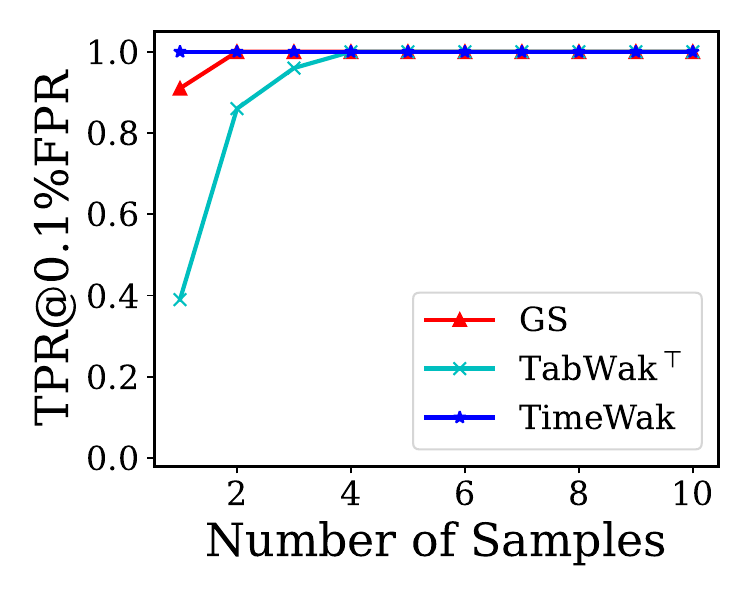}
    \caption{Stocks}\label{subfig:main_tpr_stocks_64}
\end{subfigure}
\begin{subfigure}{0.195\linewidth}
    \centering
    \includegraphics[width=\linewidth]{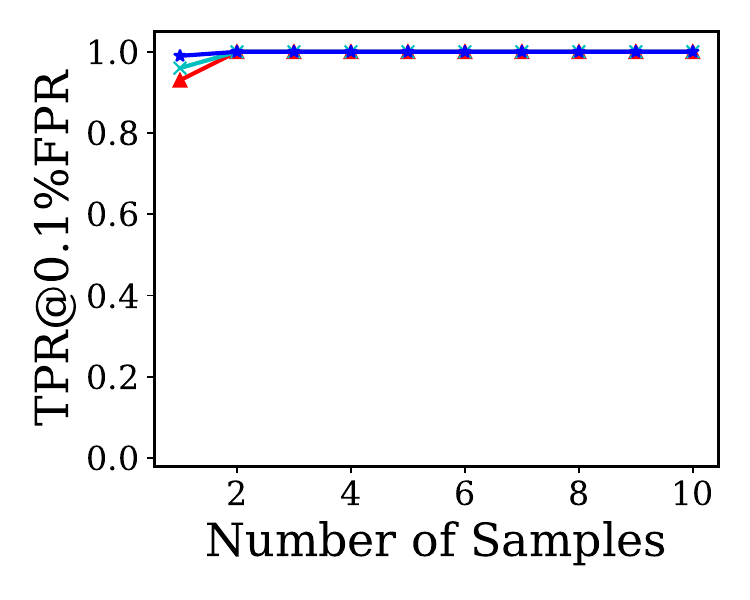}
    \caption{ETTh}\label{subfig:main_tpr_etth_64}
\end{subfigure}
\begin{subfigure}{0.195\linewidth}
    \centering
    \includegraphics[width=\linewidth]{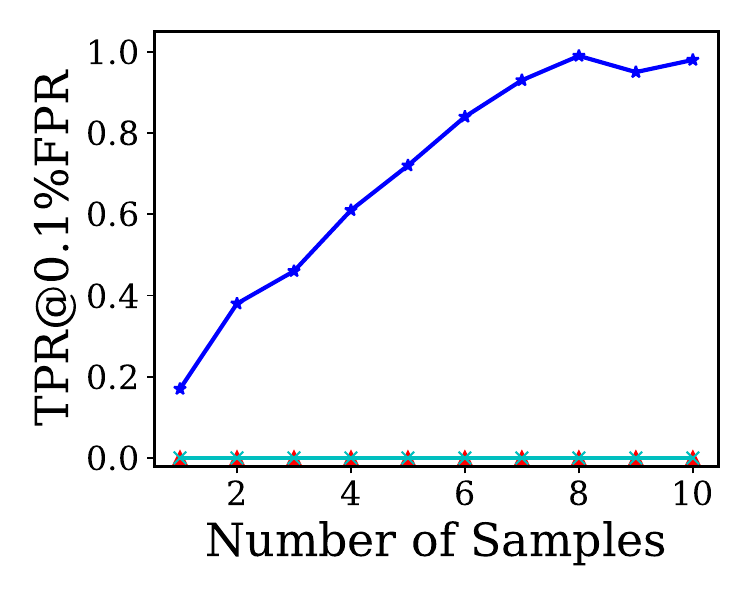}
    \caption{MuJoCo}\label{subfig:main_tpr_mujoco_64}
\end{subfigure}
\begin{subfigure}{0.195\linewidth}
    \centering
    \includegraphics[width=\linewidth]{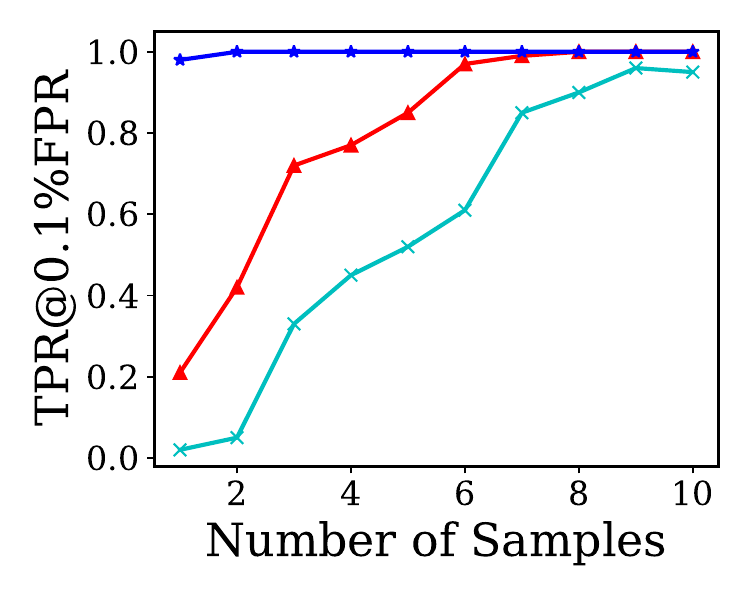}
    \caption{Energy}\label{subfig:main_tpr_energy_64}
\end{subfigure}
\begin{subfigure}{0.195\linewidth}
    \centering
    \includegraphics[width=\linewidth]{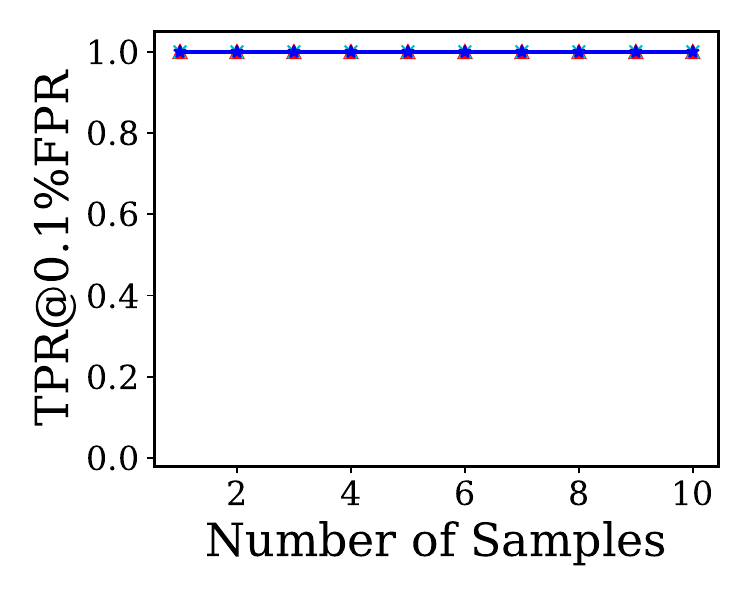}
    \caption{fMRI}\label{subfig:main_tpr_fmri_64}
\end{subfigure}

\caption{TPR@0.1\%FPR against number of samples across five datasets under 64-length sequences.}
\label{fig:main_tprs}
\vskip -0.15in
\end{figure}

This strength is further shown in \cref{fig:main_tprs}, which plots TPR@0.1\%FPR on 64-length sequences as a function of the number of samples. Across all settings, \algo consistently outperforms GS and TabWak$^\top$, achieving significantly higher TPR values. Notably, \algo reaches a perfect TPR of 1.0 in all cases, requiring only one sample in three settings and two in the remaining one. This high sensitivity makes \algo well-suited for real-world use, where only limited samples may be available. Additional results are provided in \cref{appx:tprs}.

\subsection{Robustness against post-editing attacks}

\input{src/tables/main_attack_64}

To evaluate the robustness, we first design a set of post-editing attacks. \textit{Offsetting} perturbs the time series by adding a constant offset to each feature based on 5\% or 30\% of its magnitude, and applied uniformly across all timesteps. \textit{Random cropping} masks out a subregion of the time series by a fixed proportion (5\% or 30\%), along the rows and columns, similar to the image domain~\cite{gaussianshading}. The \textit{min-max insertion} attack perturbs the series by randomly replacing a proportion of points (5\% or 30\%) in each feature with random values drawn uniformly between the feature's minimum and maximum values.



\cref{tab:main_attack_64} presents the Z-scores of 64-length watermarked synthetic time series data under these attacks, and averaged over 100 trials. Random cropping at 30\% proves especially challenging, with several methods showing negative Z-scores. Nevertheless, \algo demonstrates the best overall robustness, consistently outperforming all baselines across most attack scenarios while maintaining high generation quality and accurate watermark detection. In contrast, although HTW has better quality, it does poorly under attacks, indicating a struggle in balancing trade-offs between quality and robustness.

Interestingly, TR’s detection scores further improve under certain post-processing attacks. In particular, significant gains are observed under random cropping and min-max insertion, likely due to the inherent robustness of watermarking in the Fourier domain. However, its overall performance lags behind \algo.
Both TabWak and TabWak$^\top$ show significant degradation under attacks, particularly on MuJoCo and Energy datasets, where detection frequently fails. GS overall demonstrates strong robustness, maintaining detectability under all attacks except for 30\% cropping on the ETTh dataset. However, it produces low-quality synthetic samples, highlighting the need for a time series specific watermark that can navigate the trade-offs between generation quality and watermark robustness.





%% file: src/tables/main_exp.tex
\begin{table}

\centering
\caption{Results of synthetic time series quality and watermark detectability. No watermarking (`W/O') is included. Quality metrics are for 24-length sequences. Best results are in \textbf{bold}, and second-best are \underline{underlined}.}
\label{tab:main_exp}
\vskip 0.15in

\resizebox{\linewidth}{!}{
\begin{tabular}{l l c c c c c c c}

\toprule

& & \multicolumn{4}{c}{Quality Metric $\downarrow$} & \multicolumn{3}{c}{Z-score $\uparrow$} \\

\cmidrule(r){3-6}
\cmidrule(l){7-9}

Dataset & Method & Context-FID & Correlational & Discriminative & Predictive & 24-length  & 64-length & 128-length \\

\midrule

\multirow{7}{*}{Stocks} & W/O & \val{0.258}{0.047} & \val{0.027}{0.015} & \val{0.120}{0.049} & \val{0.038}{0.000} & - & - & - \\
& TR & \val{1.069}{0.231} & \val{0.091}{0.007} & \val{0.209}{0.056} & \val{0.039}{0.000} & \val{0.43}{0.04} & \val{0.40}{0.08} & \val{0.08}{0.10} \\
& GS & \val{8.802}{2.415} & \val{0.052}{0.026} & \val{0.403}{0.031} & \val{0.041}{0.003} & \underline{\val{86.07}{0.74}} & \underline{\val{148.92}{1.08}} & \underline{\val{172.23}{1.08}} \\
& HTW & \underline{\val{0.279}{0.052}} & \underline{\val{0.017}{0.003}} & \underline{\val{0.122}{0.034}} & \textbf{\val{0.037}{0.000}} & \val{4.45}{0.62} & \val{7.34}{0.94} & \val{10.39}{1.33} \\
& TabWak & \val{0.292}{0.064} & \val{0.017}{0.012} & \val{0.124}{0.024} & \underline{\val{0.038}{0.000}} & \val{-67.22}{1.17} & \val{16.14}{0.89} & \val{-10.49}{1.28} \\
& TabWak$^\top$ & \val{0.314}{0.071} & \textbf{\val{0.016}{0.017}} & \val{0.132}{0.022} & \textbf{\val{0.037}{0.000}} & \val{55.39}{0.86} & \val{88.81}{0.82} & \val{129.39}{0.90} \\
& \algo & \textbf{\val{0.277}{0.019}} & \val{0.020}{0.018} & \textbf{\val{0.120}{0.039}} & \underline{\val{0.038}{0.000}} & \textbf{\val{182.10}{0.73}} & \textbf{\val{395.34}{1.24}} & \textbf{\val{550.05}{1.18}} \\

\cmidrule{1-9}

\multirow{7}{*}{ETTh} & W/O & \val{0.232}{0.018} & \val{0.086}{0.023} & \val{0.093}{0.016} & \val{0.120}{0.009} & - & - & - \\
& TR & \val{1.570}{0.102} & \val{0.187}{0.017} & \val{0.283}{0.020} & \val{0.134}{0.004} & \val{7.84}{0.12} & \val{7.73}{0.13} & \val{6.18}{0.16} \\
& GS & \val{4.530}{0.393} & \val{0.433}{0.017} & \val{0.390}{0.015} & \val{0.169}{0.007} & \val{101.07}{1.17} & \underline{\val{197.41}{2.10}} & \underline{\val{327.47}{4.44}} \\
& HTW & \underline{\val{0.243}{0.024}} & \textbf{\val{0.077}{0.025}} & \val{0.103}{0.003} & {\val{0.123}{0.002}} & \val{3.43}{0.83} & \val{5.08}{1.48} & \val{6.84}{2.22} \\
& TabWak & \val{0.251}{0.027} & \val{0.335}{0.029} & \textbf{\val{0.085}{0.013}} & \val{0.125}{0.002} & \val{-14.95}{1.08} & \val{-6.16}{1.18} & \val{-20.57}{0.96} \\
& TabWak$^\top$ & \val{0.450}{0.057} & \underline{\val{0.116}{0.020}} & \underline{\val{0.096}{0.014}} & \textbf{\val{0.120}{0.005}} & \underline{\val{109.35}{0.91}} & \val{162.44}{1.11} & \val{235.03}{1.48} \\
& \algo & \textbf{\val{0.237}{0.017}} & \val{0.212}{0.043} & \val{0.102}{0.014} & \underline{\val{0.122}{0.002}} & \textbf{\val{134.83}{0.95}} & \textbf{\val{236.08}{1.63}} & \textbf{\val{340.36}{2.06}} \\

\cmidrule{1-9}

\multirow{7}{*}{MuJoCo} & W/O & \val{0.065}{0.011} & \val{0.419}{0.084} & \val{0.032}{0.026} & \val{0.008}{0.001} & - & - & - \\
& TR & \val{1.512}{0.179} & \val{1.153}{0.065} & \val{0.261}{0.069} & \val{0.015}{0.003} & \val{1.38}{0.03} & \val{1.49}{0.04} & \val{1.31}{0.04} \\
& GS & \val{6.548}{1.267} & \val{1.327}{0.061} & \val{0.474}{0.006} & \val{0.014}{0.002} & \val{21.13}{0.77} & \underline{\val{10.13}{0.62}} & \underline{\val{39.63}{0.71}} \\
& HTW & \val{0.261}{0.067} & \underline{\val{0.493}{0.056}} & \val{0.413}{0.024} & \val{0.010}{0.002} & \val{2.89}{0.54} & \val{3.41}{0.96} & \val{4.20}{1.37} \\
& TabWak & \val{0.545}{0.122} & \val{0.975}{0.061} & \val{0.207}{0.046} & \val{0.009}{0.001} & \underline{\val{31.30}{1.07}} & \val{1.26}{0.96} & \val{-3.00}{1.04} \\
& TabWak$^\top$ & \underline{\val{0.234}{0.032}} & \textbf{\val{0.463}{0.059}} & \underline{\val{0.123}{0.011}} & \textbf{\val{0.007}{0.001}} & \val{-4.85}{0.87} & \val{-4.51}{0.85} & \val{3.91}{0.88} \\
& \algo & \textbf{\val{0.089}{0.017}} & \val{0.532}{0.137} & \textbf{\val{0.044}{0.021}} & \underline{\val{0.008}{0.001}} & \textbf{\val{85.69}{1.08}} & \textbf{\val{56.45}{1.26}} &\textbf{ \val{123.36}{1.43}} \\

\cmidrule{1-9}

\multirow{7}{*}{Energy} & W/O & \val{0.118}{0.021} & \val{1.245}{0.236} & \val{0.137}{0.014} & {\val{0.253}{0.000}} & - & - & - \\
& TR & \val{0.649}{0.128} & \val{3.870}{0.537} & \val{0.455}{0.017} & \val{0.337}{0.007} & \val{9.51}{0.09} & \val{17.29}{0.11} & \val{22.58}{0.19} \\
& GS & \val{1.480}{0.273} & \val{3.831}{0.272} & \val{0.494}{0.004} & \val{0.330}{0.004} & \underline{\val{51.22}{0.88}} & \underline{\val{68.67}{1.20}} & \underline{\val{45.42}{1.05}} \\
& HTW & \textbf{\val{0.099}{0.009}} & \textbf{\val{1.312}{0.280}} & \underline{\val{0.138}{0.019}} & \textbf{\val{0.253}{0.000}} & \val{3.06}{0.36} & \val{4.30}{0.68} & \val{5.42}{1.00} \\
& TabWak & \val{0.179}{0.027} & \val{2.724}{0.203} & \val{0.162}{0.011} & \val{0.255}{0.001} & \val{3.26}{0.89} & \val{3.86}{1.02} & \val{0.57}{0.87} \\
& TabWak$^\top$ & \val{0.213}{0.024} & \underline{\val{1.740}{0.290}} & \textbf{\val{0.129}{0.013}} & \val{0.265}{0.004} & \val{40.82}{0.81} & \val{46.68}{0.86} & \val{26.00}{1.12} \\
& \algo & \underline{\val{0.121}{0.016}} & \val{1.977}{0.750} & \val{0.142}{0.008} & \underline{\val{0.254}{0.000}} & \textbf{\val{231.28}{1.45}} & \textbf{\val{267.53}{2.60}} & \textbf{\val{245.37}{2.88}} \\

\cmidrule{1-9}

\multirow{7}{*}{fMRI} & W/O & \val{0.190}{0.006} & \val{1.952}{0.087} & \val{0.132}{0.027} & \val{0.100}{0.000} & - & - & - \\
& TR & \val{2.474}{0.341} & \val{13.312}{0.254} & \val{0.496}{0.003} & \val{0.146}{0.004} & \val{6.49}{0.05} & \val{8.25}{0.05} & \val{9.94}{0.04} \\
& GS & \val{0.714}{0.051} & \val{14.628}{0.052} & \val{0.499}{0.001} & \val{0.108}{0.001} & \underline{\val{420.02}{1.44}} & \val{321.52}{0.59} & \underline{\val{701.90}{0.72}} \\
& HTW & \textbf{\val{0.180}{0.011}} & \textbf{\val{1.900}{0.047}} & \underline{\val{0.140}{0.019}} & \textbf{\val{0.100}{0.000}} & \val{4.32}{0.22} & \val{6.80}{0.41} & \val{9.43}{0.61} \\
& TabWak & \val{0.326}{0.042} & \val{6.825}{0.395} & \val{0.452}{0.092} & \val{0.112}{0.000} & \val{84.02}{1.04} & \val{204.16}{0.82} & \val{47.29}{0.83} \\
& TabWak$^\top$ & \val{0.350}{0.014} & \val{2.191}{0.095} & \val{0.208}{0.049} & \underline{\val{0.101}{0.000}} & \textbf{\val{464.67}{0.50}} & \textbf{\val{743.33}{0.55}} & \textbf{\val{1031.96}{0.79}} \\
& \algo & \underline{\val{0.199}{0.010}} & \underline{\val{2.006}{0.053}} & \textbf{\val{0.122}{0.033}} & \textbf{\val{0.100}{0.000}} & \val{379.51}{0.82} & \underline{\val{595.68}{1.03}} & \val{526.81}{13.12} \\

\bottomrule

\end{tabular}}
\vskip -0.1in
\end{table}

%% file: src/tables/main_attack_64.tex
\begin{table}
\centering
\caption{Results of robustness against post-editing attacks. Average Z-score on 64-length sequences, including un-attacked scores from \cref{tab:main_exp}. Best results are in \textbf{bold}, and second-best are \underline{underlined}.}
\label{tab:main_attack_64}
\vskip 0.08in

\resizebox{\linewidth}{!}{
\begin{tabular}{l l c c c c c c c}
\toprule

& & \multirow[b]{2}{*}{\makecell{Without \\ Attack}}  & \multicolumn{2}{c}{Offset $\uparrow$} & \multicolumn{2}{c}{Random Crop $\uparrow$} & \multicolumn{2}{c}{Min-Max Insertion $\uparrow$} \\

\cmidrule(r){4-5}
\cmidrule(rl){6-7}
\cmidrule(l){8-9}

Dataset & Method & & 5\% & 30\% & 5\% & 30\% & 5\% & 30\% \\

\midrule

\multirow{6}{*}{Stocks} & TR & \val{0.40}{0.08} & \val{0.35}{0.07} & \val{0.17}{0.09} & \underline{\val{57.09}{1.09}} & \textbf{\val{76.87}{1.29}} & \val{0.99}{0.10} & \val{5.43}{0.18} \\
& GS & \underline{\val{148.92}{1.08}} & \underline{\val{152.81}{1.33}} & \underline{\val{164.23}{1.54}} & \val{42.53}{1.11} & \underline{\val{32.80}{0.78}} & \underline{\val{136.73}{0.90}} & \underline{\val{77.20}{0.86}} \\
& HTW & \val{7.34}{0.94} & \val{7.34}{0.94} & \val{7.34}{0.94} & \val{-0.81}{0.42} & \val{-0.92}{0.32} & \val{6.41}{1.16} & \val{2.72}{1.15} \\
& TabWak & \val{16.14}{0.89} & \val{25.74}{1.04} & \val{45.02}{1.40} & \val{40.59}{1.16} & \val{19.71}{1.06} & \val{19.11}{0.75} & \val{30.87}{0.45} \\
& TabWak$^\top$ & \val{88.81}{0.82} & \val{88.84}{0.85} & \val{85.87}{0.86} & \val{30.55}{0.92} & \val{1.71}{0.87} & \val{65.25}{0.79} & \val{15.10}{0.56} \\
& \algo & \textbf{\val{395.34}{1.24}} & \textbf{\val{375.96}{0.96}} & \textbf{\val{371.04}{1.02}} & \textbf{\val{78.20}{2.23}} & \val{10.40}{1.05} & \textbf{\val{296.60}{1.62}} & \textbf{\val{83.74}{1.55}} \\

\cmidrule{1-9}

\multirow{6}{*}{ETTh} & TR & \val{7.73}{0.13} & \val{7.96}{0.11} & \val{8.83}{0.12} & \val{27.22}{0.39} & \underline{\val{38.53}{0.32}} & \val{21.86}{0.26} & \underline{\val{49.85}{0.64}} \\
& GS & \underline{\val{197.41}{2.10}} & \underline{\val{186.62}{2.21}} & \underline{\val{159.04}{2.20}} & \textbf{\val{105.11}{2.09}} & \val{-39.87}{1.24} & \textbf{\val{182.03}{2.09}} & \textbf{\val{105.23}{1.21}} \\
& HTW & \val{5.08}{1.48} & \val{5.08}{1.48} & \val{5.08}{1.48} & \val{0.54}{1.33} & \val{-1.99}{0.80} & \val{4.47}{1.45} & \val{2.03}{1.08} \\
& TabWak & \val{-6.16}{1.18} & \val{-2.75}{1.29} & \val{4.92}{1.52} & \underline{\val{84.37}{2.67}} & \textbf{\val{88.07}{2.59}} & \val{4.29}{0.94} & \val{36.83}{0.71} \\
& TabWak$^\top$ & \val{162.44}{1.11} & \val{157.75}{1.09} & \val{135.66}{1.23} & \val{70.38}{1.16} & \val{10.79}{1.14} & \val{99.23}{0.96} & \val{9.01}{0.80} \\
& \algo & \textbf{\val{236.08}{1.63}} & \textbf{\val{243.86}{1.61}} & \textbf{\val{207.90}{1.70}} & \val{75.35}{1.18} & \val{2.47}{1.08} & \underline{\val{171.51}{1.46}} & \val{27.78}{1.22} \\

\cmidrule{1-9}

\multirow{6}{*}{MuJoCo} & TR & \val{1.49}{0.04} & \val{1.46}{0.03} & \val{1.54}{0.03} & \val{7.14}{0.08} & \underline{\val{7.17}{0.07}} & \val{6.09}{0.11} & \textbf{\val{14.83}{0.22}} \\
& GS & \underline{\val{10.13}{0.62}} & \underline{\val{10.35}{0.69}} & \underline{\val{8.76}{0.75}} & \underline{\val{28.99}{0.80}} & \textbf{\val{10.09}{0.91}} & \underline{\val{9.35}{0.76}} & \underline{\val{10.40}{1.07}} \\
& HTW & \val{3.41}{0.96} & \val{3.41}{0.96} & \val{3.41}{0.96} & \val{1.13}{0.78} & \val{-1.32}{0.67} & \val{3.09}{0.91} & \val{1.67}{0.65} \\
& TabWak & \val{1.26}{0.96} & \val{0.48}{0.89} & \val{5.48}{1.22} & \val{-1.75}{1.21} & \val{-6.65}{0.71} & \val{-0.57}{0.91} & \val{-0.31}{0.59} \\
& TabWak$^\top$ & \val{-4.51}{0.85} & \val{-3.46}{0.99} & \val{-0.24}{0.94} & \val{-32.20}{1.04} & \val{-42.74}{1.25} & \val{-34.16}{0.90} & \val{-80.63}{1.05} \\
& \algo & \textbf{\val{56.45}{1.26}} & \textbf{\val{58.90}{1.36}} & \textbf{\val{51.53}{1.31}} & \textbf{\val{49.30}{1.20}} & \val{1.85}{1.12} & \textbf{\val{36.59}{1.19}} & \val{10.35}{1.18} \\

\cmidrule{1-9}

\multirow{6}{*}{Energy} & TR & \val{17.29}{0.11} & \val{16.74}{0.11} & \val{15.09}{0.09} & \textbf{\val{128.26}{0.39}} & \textbf{\val{148.31}{0.29}} & \val{37.72}{0.20} & \textbf{\val{75.09}{0.29}} \\
& GS & \underline{\val{68.67}{1.20}} & \underline{\val{63.93}{1.09}} & \underline{\val{54.84}{1.09}} & \underline{\val{66.30}{0.99}} & \underline{\val{96.29}{0.80}} & \underline{\val{66.40}{1.27}} & \underline{\val{53.06}{1.65}} \\
& HTW & \val{4.30}{0.68} & \val{4.30}{0.68} & \val{4.30}{0.68} & \val{0.40}{0.48} & \val{-0.60}{0.34} & \val{3.87}{0.66} & \val{2.03}{0.48} \\
& TabWak & \val{3.86}{1.02} & \val{-5.32}{1.01} & \val{-8.99}{1.03} & \val{-8.31}{0.65} & \val{9.94}{0.60} & \val{4.21}{0.78} & \val{2.67}{0.36} \\
& TabWak$^\top$ & \val{46.68}{0.86} & \val{43.10}{0.84} & \val{42.38}{0.89} & \val{12.26}{1.01} & \val{-49.34}{1.74} & \val{11.87}{0.85} & \val{-43.15}{0.71} \\
& \algo & \textbf{\val{267.53}{2.60}} & \textbf{\val{296.74}{2.49}} & \textbf{\val{191.63}{2.13}} & \val{4.91}{0.96} & \val{15.73}{0.96} & \textbf{\val{195.37}{1.97}} & \val{34.26}{0.99} \\

\cmidrule{1-9}

\multirow{6}{*}{fMRI} & TR & \val{8.25}{0.05} & \val{8.22}{0.05} & \val{8.10}{0.05} & \val{8.24}{0.05} & \val{7.84}{0.05} & \val{11.45}{0.06} & \val{23.36}{0.10} \\
& GS & \val{321.52}{0.59} & \val{319.93}{0.60} & \val{312.57}{0.62} & \val{286.05}{1.13} & \val{116.24}{0.87} & \val{320.52}{1.66} & \underline{\val{275.34}{2.20}} \\
& HTW & \val{6.80}{0.41} & \val{6.80}{0.41} & \val{6.80}{0.41} & \val{4.82}{0.46} & \val{-0.70}{0.46} & \val{5.95}{0.45} & \val{2.56}{0.41} \\
& TabWak & \val{204.16}{0.82} & \val{205.01}{0.91} & \val{215.64}{1.07} & \val{154.27}{2.15} & \textbf{\val{452.61}{3.14}} & \val{248.19}{3.95} & \textbf{\val{297.10}{8.67}} \\
& TabWak$^\top$ & \textbf{\val{743.33}{0.55}} & \textbf{\val{743.16}{0.59}} & \textbf{\val{742.43}{0.53}} & \textbf{\val{636.28}{0.67}} & \underline{\val{317.24}{1.18}} & \textbf{\val{614.25}{0.77}} & \val{224.27}{0.85} \\
& \algo & \underline{\val{595.68}{1.03}} & \underline{\val{601.68}{0.81}} & \underline{\val{601.53}{0.96}} & \underline{\val{459.66}{1.00}} & \val{112.68}{0.85} & \underline{\val{498.39}{0.84}} & \val{189.43}{1.00} \\

\bottomrule

\end{tabular}
}
\vskip -0.15in
\end{table}

%% file: src/conclusion.tex
\section{Conclusion}
\label{sec:conclusion}

Motivated by the need to ensure the traceability of synthetic time series, we propose \algo, the first watermarking algorithm for multivariate time series diffusion models. \algo embeds seeds through a temporally chained-hash and feature-wise shuffling in data space, preserving the temporal and feature dependencies and enhancing the watermark detectability. To address non-uniform error distribution in the time series diffusion process,  we optimize \algo for $\epsilon$-exact inversion and provide the bounded error analysis. Compared to multiple SOTA watermarking algorithms, \algo balances synthetic data quality, watermark detectability, and robustness against post-editing attacks. Extensive evaluations on five datasets show that \algo improves context-FID score by 61.96\% and correlational scores by 8.44\%, against the strongest SOTA baseline, while maintaining strong detectability.


\textbf{Limitations.} While \algo does not natively support streaming data, it can be applied to streaming scenarios by processing data in small, fixed-length windows and watermarking each window independently. However, finer-grained cases, such as watermarking at the per-timestep level, remain unexplored. This represents a key limitation of \algo and an important direction for future work.





 \section*{Acknowledgments}

This research is partly funded by Priv-GSyn project, 200021E\_229204 of Swiss National Science Foundation, the DEPMAT project, P20-22 / N21022, of the research programme Perspectief of the Dutch Research Council, and ASM International NV.

%% file: src/appendix.tex
\clearpage
\appendix

\input{src/appx_nomenclature}

\input{src/appx_diffusion}

\input{src/appx_proof}

\section{Experiment details}
\label{appx:exp_details}


\subsection{Datasets}
\label{appx:datasets}

\cref{tab:appx_dataset} shows the details of all datasets used in the experiments.

\input{src/tables/appx_datasets}


\subsection{Evaluation metrics}
\label{appx:eval_metrics}

\paragraph{Context-FID} \citet{jeha2022psa} introduced the Context-FID score, which is a refined adaptation of the Fréchet Inception Distance (FID) used for evaluating the similarity between real and synthetic time series distributions. Unlike traditional FID, which relies on the Inception model as a feature extractor for images, Context-FID uses TS2Vec~\cite{yue2022ts2vec}, which is a specialized time series embedding model. \citet{yue2022ts2vec} demonstrated that models with lower Context-FID scores tend to perform well in downstream tasks, revealing a strong correlation between Context-FID and the forecasting performance of generative models. Ultimately, a lower Context-FID score signifies a closer resemblance between real and synthetic distributions.

\paragraph{Correlational} We calculates the covariance between the $i^{th}$ and $j^{th}$ features of a time series using the following equation~\cite{liao2020conditional}:
\begin{equation}
    \mathrm{Cov}_{i,j} = \frac{1}{W}\sum\limits_{t = 1}^W {\mathrm{K}_i^t\mathrm{K}_i^t}  - \left( {\frac{1}{W}\sum\limits_{t = 1}^W {\mathrm{K}_i^t} } \right)\left( {\frac{1}{W}\sum\limits_{t = 1}^W {\mathrm{K}_{j}^t}} \right).
\end{equation}
where \(W\) is the total number of time steps, \(\mathrm{K}_i^t\) and \(\mathrm{K}_j^t\) are the values of the \(i^{th}\) and \(j^{th}\) features at time step \(t\), and the summations compute the average product of these values, subtracting the product of their individual means. To assess the correlation between real and synthetic time series, we use the following metric~\cite{yuan2024diffusion}:
\begin{equation}
    \frac{1}{{10}}\sum\limits_{i,j}^d {\left| {\frac{{{\mathrm{Cov}} _{i,j}^{R}}}{{\sqrt {{\mathrm{Cov}} _{i,i}^{R}} {\mathrm{Cov}} _{j,j}^{R}}} - \frac{{{\mathrm{Cov}} _{i,j}^{S}}}{{\sqrt {{\mathrm{Cov}} _{i,i}^{S}} {\mathrm{Cov}} _{j,j}^{S}}}} \right|},
\end{equation}
where \(\mathrm{Cov}\) is the covariance between its subscripts (\(\langle i,i \rangle\), \(\langle j,j \rangle\), \(\langle i,j \rangle\)), where $i$ and $j$ are the features in the real (denoted by $R$) and synthetic (denoted by $S$) time series data, and \(d\) is the total number of features, with the summation taken over all feature pairs.

\paragraph{Discriminative} The discriminative score is computed as $\mid$accuracy $-$ 0.5$\mid$, quantifying the model’s ability to distinguish between real and synthetic time series. A lower score indicates better performance, as it indicates greater difficulty in differentiation, implying a higher degree of similarity between the two distributions. To ensure consistency, we follow the experimental setup of TimeGAN~\citep{yoon2019time} by using a two-layer GRU-based neural network as the classifier.

\paragraph{Predictive} The predictive score is evaluated using the mean absolute error (MAE) between the predicted and actual values on the test data. Again, we use the experimental setup of TimeGAN~\citep{yoon2019time} by using a two-layer GRU-based neural network for sequence prediction.

\paragraph{Z-score} The Z-score is a statistical measure used to assess watermark detectability by quantifying the deviation between watermarked and non-watermarked samples. It facilitates hypothesis testing, where the null hypothesis \( H_0 \) states that a given sample is not watermarked by the corresponding watermarking method. A sufficiently high positive Z-score provides evidence against \( H_0 \), suggesting the presence of a watermark.

For sample-wise bit accuracy in time series, the Z-score is computed as 
$
Z = \frac{\mu_{\text{Acc, W}} - \mu_{\text{Acc, NW}}}{\sigma_{\text{Acc, NW}} / \sqrt{n}}
$, where \( \mu_{\text{Acc, W}} \) and \( \mu_{\text{Acc, NW}} \) are the mean bit accuracy of watermarked and non-watermarked samples, respectively, \( \sigma_{\text{Acc, NW}} \) is the standard deviation of bit accuracy in the non-watermarked samples, and \( n \) is the number of watermarked samples. Under \( H_0 \), the expected difference in means is negligible, resulting in a Z-score close to zero. A large positive Z-score provides statistical evidence for the presence of a watermark.

For the Tree-Ring watermarking method, the Z-score is computed in the Fourier domain instead of the sample-wise bit accuracy domain. It measures the deviation in the amplitude spectrum between watermarked and non-watermarked samples, given by
\(
Z = \frac{\mu_{\mathcal{F}_{\text{NW}}} - \mu_{\mathcal{F}_{\text{W}}}}{\sigma_{\mathcal{F}_{\text{NW}}}}
\)
where \( \mathcal{F}_{\text{W}} \) and \( \mathcal{F}_{\text{NW}} \) denote the Fourier amplitude spectrum of watermarked and non-watermarked samples, respectively, with \( \mu \) and \( \sigma \) representing their mean and standard deviation. Since this method does not rely on per-sample bit accuracy, the test statistic is independent of \( n \), and the computation utilizes the opposite tail of the distribution.

\paragraph{TPR@X\%FPR} This metric measures the True Positive Rate (TPR) at a specified False Positive Rate (FPR) of X\%, where X\% denotes a fixed false positive threshold. It reflects the effectiveness of watermark detection by quantifying how reliably watermarked samples are identified under controlled false positive conditions. A higher TPR@X\%FPR indicates stronger detection performance and greater robustness of the watermarking method.


\subsection{Baselines}
\label{appx:baselines}
\input{src/appx_baselines}


\subsection{Training and sampling}
\label{appx:training_sampling}

All code implementations are done in PyTorch (version 2.3.1) using a single NVIDIA GeForce RTX 2080 Graphics Card coupled with an Intel(R) Xeon(R) Platinum 8562Y+ CPU for all experiments. Dataset splits are 80\% for training and 20\% for testing. \cref{tab:appx_training_sampling} shows training and sampling time for all datasets across window of sizes 24, 64 and 128. We train the time series diffusion model following the Diffusion-TS settings~\cite{yuan2024diffusion}, and generate 10,000 watermarked synthetic samples using \algo for each sampling run. 

\input{src/tables/appx_training_sampling}

\clearpage
\section{Additional experimental results}
\label{appx:addexp}


\subsection{Quality performance}

\cref{tab:appx_quality_diff_len} shows the quality of synthetic time series generated in 64 and 128 window sizes. \algo remains stable across all datasets and even comparable to the quality of non-watermarked samples.

\input{src/tables/appx_quality_diff_len}

\clearpage
\subsection{Post-editing attacks}

\cref{tab:appx_attack_24}--\ref{tab:appx_attack_128} show the detectability results of several post-editing attacks on 24 and 128 window sizes, respectively.

\input{src/tables/appx_attack_24}
\input{src/tables/appx_attack_128}

\clearpage
\subsection{Reconstruction attack}

We implement the reconstruction attack using the original diffusion model. Specifically, we first applied the $q$-sampling process up to half of the total diffusion steps (i.e., midpoint timestep), and then performed reverse sampling starting from this midpoint. The results obtained from this approach are presented in \cref{tab:appx_recon_attack}. We observe that although the Z-score decreases, our watermark remains detectable.

\input{src/tables/appx_recon_attack}

\subsection{BDIA-DDIM on other baselines}

The baselines in our main experiments were not evaluated with BDIA-DDIM. This is because BDIA-DDIM tends to degrade data quality compared to standard DDIM, representing a trade-off for achieving lower inversion error. For many baselines, such as Tree-Ring and Gaussian Shading, the generated quality is already poor. Applying BDIA-DDIM in these cases might improve detectability but would further deteriorate quality, making the results less meaningful. And for TabWak, we include results with BDIA-DDIM applied to both TabWak and TabWak$^\top$ in \cref{tab:appx_bdia_ddim}. While BDIA-DDIM improves detectability for both variants, it comes at the cost of further quality degradation compared to the original results in \cref{tab:main_exp}. In contrast, \algo maintains stable performance across both quality metrics and Z-score, highlighting its robustness.

\input{src/tables/appx_bdia_ddim}

\clearpage
\subsection{Watermarked dataset on downstream tasks}

We implement time series forecasting and imputation as the downstream tasks, i.e, taking either the real, synthetic, or watermarked synthetic data to build a diffusion model that can predict the future or missing values of time series. We compared the mean squared error (MSE) between the predicted and actual values and summarized the results in \cref{tab:appx_forecasting_64} and \cref{tab:appx_imputation_64}. The results indicate that training on watermarked synthetic data has a minimal impact on forecasting and imputation performance compared to training on non-watermarked synthetic data.

\input{src/tables/appx_forecasting_64}
\input{src/tables/appx_imputation_64}

\clearpage
\subsection{Ablation study}

To assess the effectiveness of the \algo, we compare its full version with three distinct variants outlined in \cref{tab:appx_ablation_methods}. \cref{tab:appx_ablation} presents the quality of the watermarked time series data for sequences of length 24, and the detectability of the watermarks across lengths of 24, 64 and 128. \algo demonstrates a comparable quality performance and high detectability.

\input{src/tables/appx_ablation_methods}
\input{src/tables/appx_ablation}

\clearpage
\subsection{Challenging time series datasets}

We evaluate \algo on 2 additional and more challenging real-world datasets, (i) the ILI dataset, which records influenza-like illness cases in the United States, and (ii) the Weather dataset, which is sparse and noisy. Both datasets are standard benchmarks and are used in TimesNet~\cite{wu2023timesnet}. Results run on these datasets are shown in \cref{tab:appx_challenging_dataset}, where \algo achieves robust watermark detectability while preserving the quality of the synthetic data.

\input{src/tables/appx_challenging_dataset}

\subsection{TPR@0.1\%FPR performance}
\label{appx:tprs}

In \cref{fig:appx_tprs}, we present the TPR@0.1\%FPR metric against the number of samples across five datasets under 24, 64 and 128 window sizes. In most cases, \algo consistently outperforms other baselines, such as Gaussian Shading and TabWak$^\top$, by achieving significantly higher TPR values. Notably, \algo reaches a perfect 1.0 TPR@0.1\%FPR in the majority of scenarios, with 7 cases requiring only a single sample and 4 cases needing just 2 samples, demonstrating its strong detectability with minimal data requirements.

\input{src/tables/appx_tprs}

\clearpage
\subsection{TPR@0.1\%FPR on mixed dataset}

We construct a mixed dataset containing equal proportions ($1/3$ each) of real data, synthetic data without watermarks, and synthetic data with watermarks, totaling 100 trials with a window length of 64. We evaluate TPR@0.1\%FPR with 1, 10, and 20 samples per record, as shown in \cref{tab:appx_tpr_mixed_dataset}. For comparison, we select the methods with the best detectability: GS and TabWak$^\top$. The results demonstrate that \algo achieves 99 to 100\% true positive rates in this mixed data setting when the number of samples is 20. But GS completely fails to detect watermarks in the MuJoCo dataset with 0\% TPR across all sample sizes, while TabWak$^\top$ fails on both MuJoCo with 0\% TPR and Energy datasets with 0 to 1\% TPR.

\input{src/tables/appx_tpr_mixed_dataset}

\clearpage
\subsection{Preservation of key signal characteristics in watermarked time series data}

To assess the preservation of key signal characteristics, we add 4 additional metrics from TSGBench~\cite{ang2023tsgbench}: Marginal Distribution Difference (MDD), AutoCorrelation Difference (ACD), Skewness Difference (SD), and Kurtosis Difference (KD). These measures are designed to capture inter-series correlations and temporal dependencies, thereby evaluating how well the generated time series preserves the original characteristics. For all these metrics, the lower the score, the better. As shown in \cref{tab:appx_characteristics_data}, \algo consistently achieves scores very close to the non-watermarked (W/O) baseline across all datasets, indicating minimal distortion introduced by the watermark.

\input{src/tables/appx_characteristics_data}

\clearpage
\subsection{Hyperparameter evaluation}


\subsubsection{Intervals}

Interval, also referred to as $H$, is one of the key hyperparameters in our approach. Based on our experiments in \cref{tab:appx_intervals_24} (24-length), \cref{tab:appx_intervals_64} (64-length), and \cref{tab:appx_intervals_128} (128-length), we found that setting $H = 2$ yields the best results across most datasets. For instance, consider the Stocks dataset, which consists of 6 features and 24 time steps. When $H = 8$, the number of bit templates that can be generated is $2^{(3 \times 6)}$, whereas for $H = 2$, the number of bit templates increases significantly to $2^{(12 \times 6)}$. A lower $H$ value allows for the generation of a greater number of bit combinations, leading to a more diverse seed distribution. However, as shown in \cref{tab:appx_intervals_128}, for datasets such as fMRI, tuning $H$ can enhance detectability while preserving the quality of the synthetic data. This could be attributed to the inherently noisy nature of the fMRI dataset, where adjusting $H$ helps balance detectability and data fidelity.

\input{src/tables/appx_intervals_24}
\input{src/tables/appx_intervals_64}
\input{src/tables/appx_intervals_128}

\subsubsection{Bits}

We perform an empirical validation of the expected bit accuracy using simulations. Specifically, we set the synthetic time series data size as 24 time steps (window length) and 10 features, and evaluate the watermarking and detection pipeline using \algo. In this experiment, we intentionally omit both the forward diffusion and reverse (inversion) diffusion processes to focus solely on the effect of noise during reconstruction. Instead, we simulate the reconstruction noise directly by adding noise to the clean initial noise.

We generate initial samples using our watermarking method and simulate reconstruction error by adding noise with feature-specific means sampled from $\mathcal{N}(0, 5)$ and a shared variance $\sigma$. This results in a noise distribution of $\mathcal{N}(\mu_f, \sigma)$ per feature, where $\mu_f \sim \mathcal{N}(0, 5)$. We then apply watermark detection to the perturbed samples and compute the average bit accuracy.

\begin{figure}[ht]
    \centering
    \includegraphics[width=0.75\linewidth]{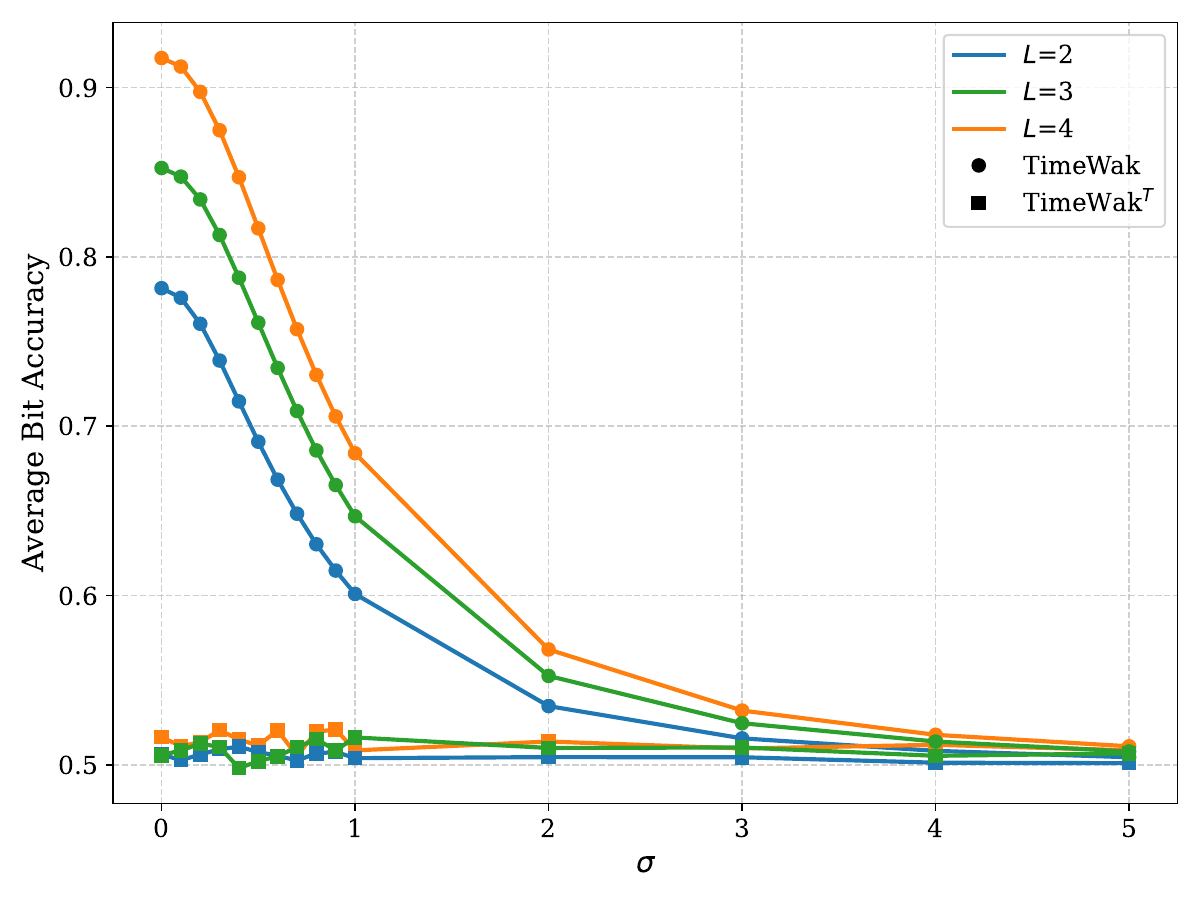}
    \caption{Average bit accuracy of different bit-length $L$.}
    \label{fig:avg_bit_acc}
\end{figure}

We run the simulation using 100,000 samples, grouped into trials of 2,000 samples each. This process is repeated across 50 independent rounds to compute the average bit accuracy. \cref{fig:avg_bit_acc} shows the results we get. We observe that a larger $L$ leads to higher bit accuracy, indicating better detectability. In addition, we evaluate a ``transposed'' version of \algo, denoted as \texttt{TimeWak}$^\top$, where the chained hash is applied along the feature dimension instead of the time axis. We find that the bit accuracy of this variant remains close to 0.5 and is significantly lower than that of the original \algo. This simulation further validates the importance of applying the watermark along the time axis, rather than across features, to ensure reliable detection.

Additionally, we present the values of bit-length $L$ used across different experiments in \cref{tab:appx_bits_24}--\ref{tab:appx_bits_128}. For bit-lengths greater than 2, we applied the valid bit mechanism from~\cite{zhu2025tabwak}. In general, a larger $L$ tends to improve watermark detectability. This is because, during bit accuracy calculation, a larger $L$ places more emphasis on the tail bits, which are less likely to be affected by reconstruction errors or noise. However, increasing $L$ also leads to lower sample quality, as it involves modifying more of the initial noise, making it deviate further from a standard Gaussian distribution. Our results show this trade-off holds across most scenarios, with the exception of the Stocks dataset.

\input{src/tables/appx_bits_24}
\input{src/tables/appx_bits_64}
\input{src/tables/appx_bits_128}

\clearpage
\subsection{Watermark detection overhead}

To evaluate the practical feasibility of real-time deployment, we measure the watermark detection overhead for \algo using a single NVIDIA L40S GPU and Intel(R) Xeon(R) Platinum 8562Y+ CPU. \cref{tab:appx_overhead} presents the computational overhead for watermark detection across different datasets and configurations. The results demonstrate that detection overhead remains consistently low, ranging from approximately 1.5 to 7.5 seconds depending on the dataset complexity and batch size. We consider this overhead acceptable for streaming scenarios, particularly given the security benefits provided by the watermarking system. Adapting the hashing mechanism to handle variable-length sequences without padding represents a promising direction for future research that could further enhance the method's applicability to diverse real-world scenarios.

\input{src/tables/appx_overhead}

\clearpage

\section{Synthetic samples}

\begin{figure}[ht]
    \centering
    \includegraphics[width=\linewidth]{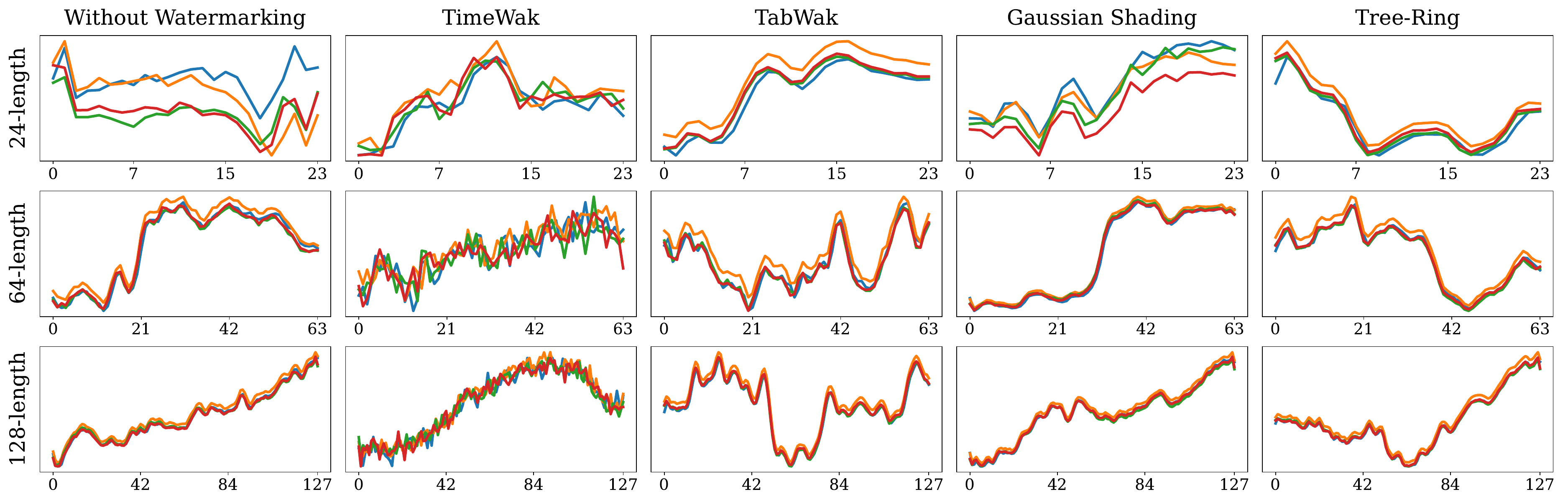}
    \caption{
    Non-watermarked (leftmost column) and watermarked (remaining columns) time series generated by \algo, TabWak, Gaussian Shading, and Tree-Ring watermarking for the Stocks dataset.}
    \label{fig:samples_stocks}
\end{figure}

\begin{figure}[ht]
    \centering
    \includegraphics[width=\linewidth]{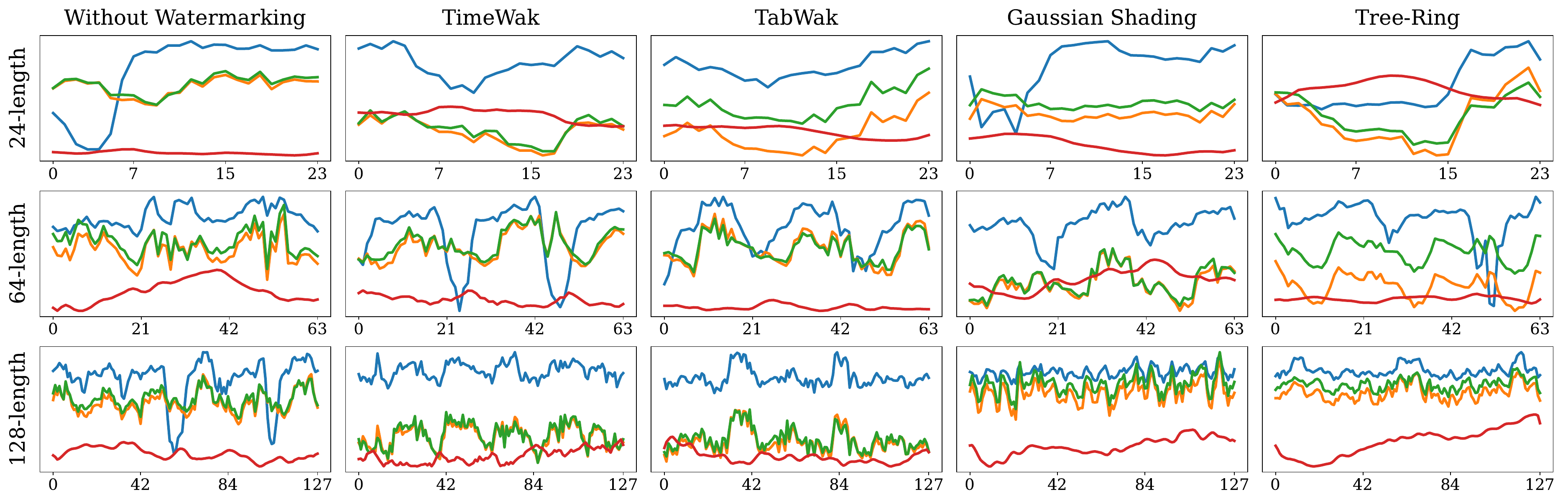}
    \caption{Non-watermarked (leftmost column) and watermarked (remaining columns) time series generated by \algo, TabWak, Gaussian Shading, and Tree-Ring watermarking for the ETTh dataset.}
    \label{fig:samples_etth}
\end{figure}

\begin{figure}[ht]
    \centering
    \includegraphics[width=\linewidth]{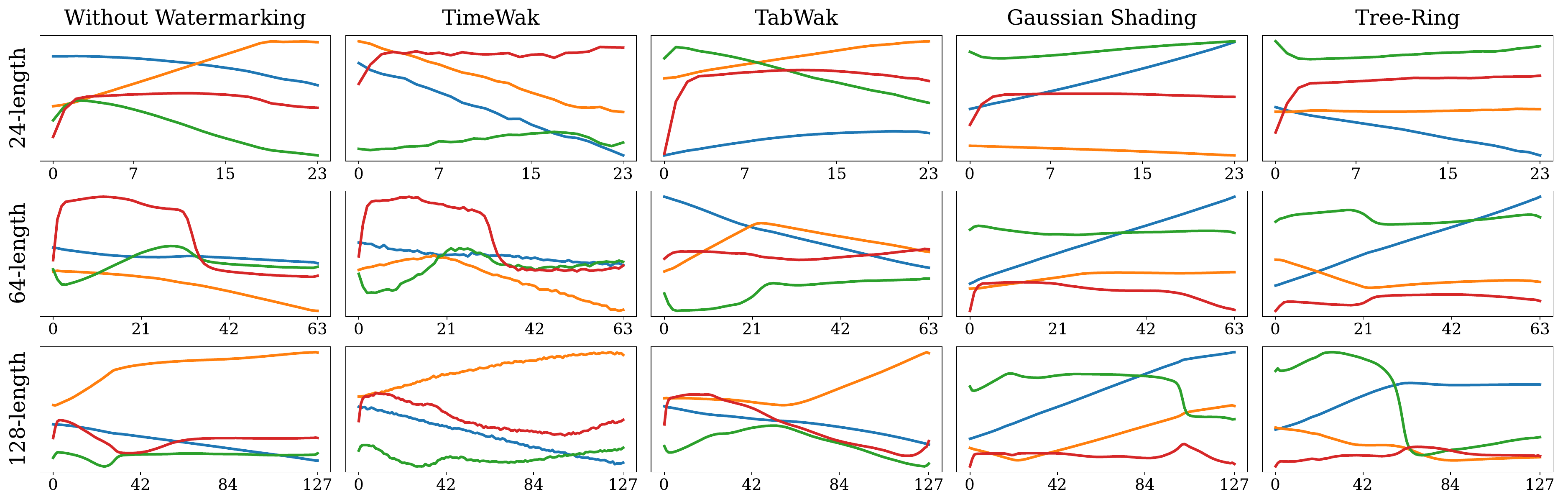}
    \caption{Non-watermarked (leftmost column) and watermarked (remaining columns) time series generated by \algo, TabWak, Gaussian Shading, and Tree-Ring watermarking for the MuJoCo dataset.}
    \label{fig:samples_mujoco}
\end{figure}

\begin{figure}[ht]
    \centering
    \includegraphics[width=\linewidth]{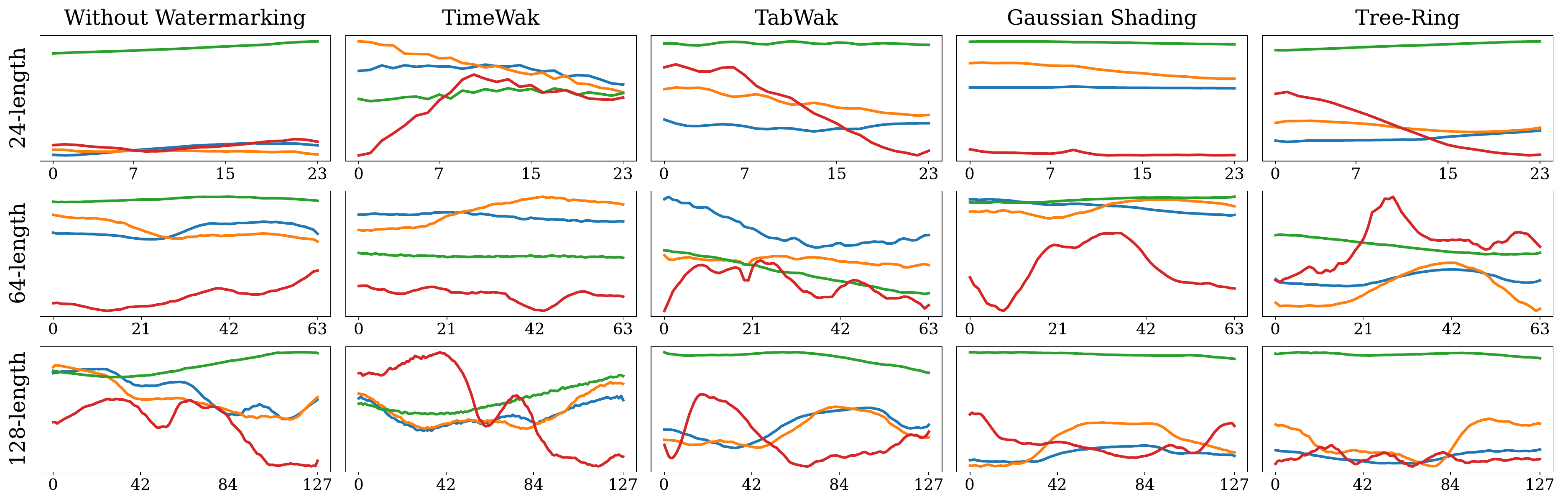}
    \caption{Non-watermarked (leftmost column) and watermarked (remaining columns) time series generated by \algo, TabWak, Gaussian Shading, and Tree-Ring watermarking for the Energy dataset.}
    \label{fig:samples_energy}
\end{figure}

\begin{figure}[ht]
    \centering
    \includegraphics[width=\linewidth]{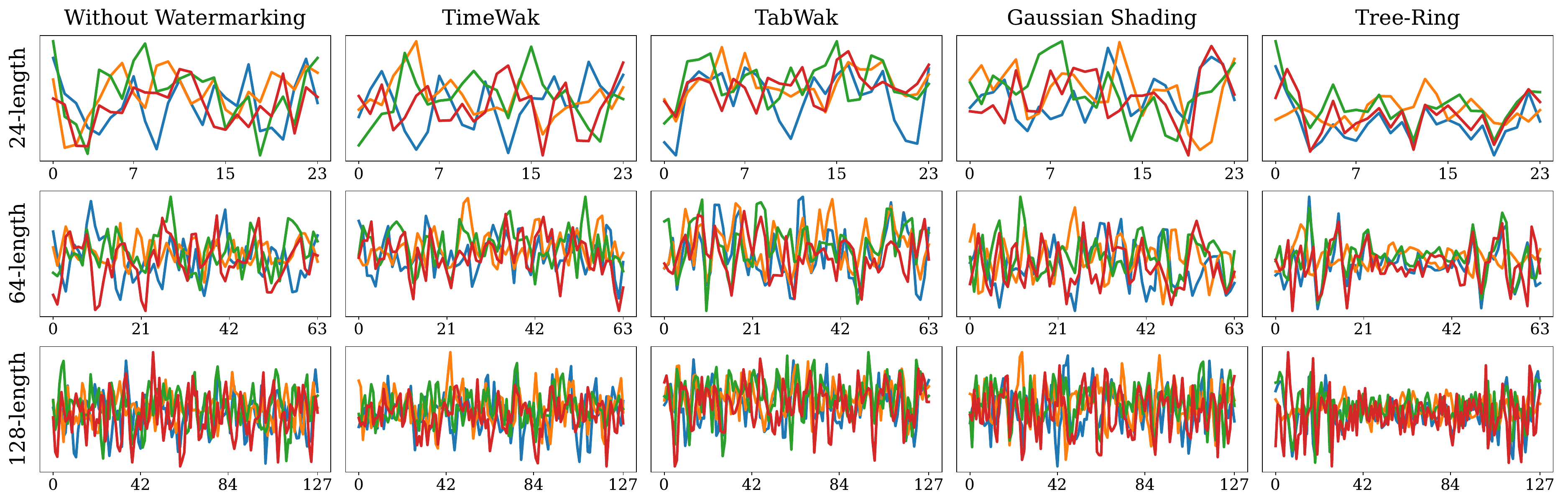}
    \caption{Non-watermarked (leftmost column) and watermarked (remaining columns) time series generated by \algo, TabWak, Gaussian Shading, and Tree-Ring watermarking for the fMRI dataset.}
    \label{fig:samples_fmri}
\end{figure}

Figures \ref{fig:samples_stocks}--\ref{fig:samples_fmri} show synthetic time series generated unconditionally by Diffusion-TS with/without watermark embedding. Each figure corresponds to one of the following datasets: Stocks, ETTh, MuJoCo, Energy, and fMRI. Within each figure, the columns represent the following algorithms: no watermark, \algo, TabWak, Gaussian Shading, and Tree-Ring watermarks. Up to 4 features are randomly selected from each dataset.

%% file: src/appx_nomenclature.tex
\renewcommand{\nomname}{}

\section{Nomenclature}
\label{appx:notation}
\printnomenclature[1.0in]
\nomenclature{$W$}{Time window length}
\nomenclature{$L$}{Bit length}
\nomenclature{$F$}{Number of features}
\nomenclature{$T$}{Total diffusion steps}
\nomenclature{$\mathbf{x}_0$}{Generated sequence of length $W$ and $F$ features}
\nomenclature{$\mathbf{x}_T$}{Noised sequence at timestep $T$}
\nomenclature{$\mathbf{x}_t^\textit{approx}$}{Approximate diffusion state at step $t$}
\nomenclature{$\mathcal{N}(0,I)$}{Standard normal distribution}
\nomenclature{$\alpha_t$}{Noise schedule parameter}
\nomenclature{$\beta_t$}{Variance schedule for noise addition}
\nomenclature{$\hat{\boldsymbol{\epsilon}}_{\boldsymbol{\theta}}$}{Noise estimator function}
\nomenclature{$\mathbf{s}^{w,f}$}{Watermark seed at timestep $w$ and feature $f$}
\nomenclature{$\mathbf{s}^{w,[:]}$}{Watermark seed across all features at timestep $w$}
\nomenclature{$\mathbf{s}^{[:],f}$}{Watermark seed across all timesteps at feature $f$}
\nomenclature{$\mathcal{H}(\kappa, w, \mathbf{s}^{w-1,[:]})$}{Temporal chained hashing function}
\nomenclature{$\pi_{\kappa}$}{Feature permutation function}
\nomenclature{$\Phi^{-1}$}{Inverse CDF of standard normal distribution}
\nomenclature{$\Phi$}{CDF of standard normal distribution}
\nomenclature{$\sigma_t$}{Standard deviation at timestep $t$}
\nomenclature{$H$}{Interval length for watermark partitioning}
\nomenclature{$n$}{Number of intervals in time window}
\nomenclature{$\kappa$}{Cryptographic key for watermarking}
\nomenclature{$\mathcal{U}(\{0,L-1\})^{F}$}{Vector in $\mathbb{R}^F$ with i.i.d. discrete uniform components over $\{0, \ldots, L-1\}$}
\nomenclature{$\mathcal{U}(0,1)$}{Continuous uniform distribution between 0 and 1}
\nomenclature{$\mathbf{x}_{t}$}{Diffusion state at step $t$}
\nomenclature{$\gamma$}{Scaling factor in BDIA inversion}
\nomenclature{$a_t, b_t$}{BDIA parameters for inversion process}
\nomenclature{$\Delta_{t}$}{Lipschitz constant for noise estimator sensitivity at time step $t$}
\nomenclature{$\epsilon$}{Perturbation error in inversion}
\nomenclature{$\mathbb{I}[\cdot]$}{Indicator function}
\nomenclature{$\mathcal{W}^*$}{Set of valid timesteps for comparison}

%% file: src/appx_diffusion.tex
\section{Diffusion and diffusion inversion}

\subsection{Time series diffusion model}

This section covers the necessary background knowledge for time series diffusion models.

\begin{figure}[ht!]
    \centering
    \includegraphics[width=0.9\linewidth]{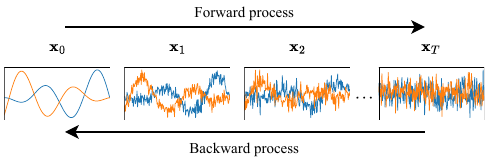}
    \caption{Forward and backward diffusion process. $\mathbf{x}_0$ denotes the initial signal window and $\mathbf{x}_T$ corresponds to the fully diffused version of the signal obtained after $T$ forward diffusion steps.}
    \label{fig:fwdbckwd2}
    \vspace{-10pt}
\end{figure}

Denoising Diffusion Probabilistic Models (DDPMs) are a powerful generative models, especially for time series synthesis~\cite{DDPM, alcaraz2022, kollovieh_2023, yuan2024diffusion}. As shown in \cref{fig:fwdbckwd2}, they work by iteratively forward noising data and then learning to invert this process through during the backward step~\cite{DDPM}. For a time series window, one step of forward noise is given by:
\begin{equation}
\label{eq:noising}
    \mathbf{x}_{t} = \sqrt{1 - \beta_t} \mathbf{x}_{t-1} + \sqrt{\beta_t} \epsilon_t.
\end{equation}
In this formulation, \( \mathbf{x}_{t} \) represents a multivariate time series signal after undergoing \( t \) steps of noise addition. The term \( \beta_t \) denotes the noise variance at step \( t \). The noise component \( \epsilon_t \) is sampled from a normal distribution, \( \mathcal{N}(0, I) \). Using this notation, \( \mathbf{x}_0 \) and \( \mathbf{x}_T \) define a noise-free sequence and a fully noised sequence, respectively, for an arbitrary window slice, where \( T \) is the total number of noise steps. Without iteratively applying Equation \eqref{eq:noising}, an intermediate noising step can be efficiently computed using a \emph{reparameterization trick}~\cite{DDPM}:  
\begin{equation}
    \mathbf{x}_{t} = \sqrt{\bar{\alpha}_t} \mathbf{x}_{0} + \sqrt{1 - \bar{\alpha}_t} \epsilon,
    \label{eq:ffwdnoising2}
\end{equation}
where \( \bar{\alpha}_t \) is the cumulative product of noise reduction factors up to step \( t \), expressed as \( \bar{\alpha}_t = \prod_{s=1}^t \alpha_s = \prod_{s=1}^t (1 - \beta_s) \), with \( \beta_s \) being the noise variance at timestep \( s \) and \( \epsilon \sim \mathcal{N}(0, I) \).  

The denoising process reverses the noising steps to reconstruct \( \mathbf{x}_0 \) from \( \mathbf{x}_{t} \), ideally restoring the original signal. This is achieved by training a neural network, \( \boldsymbol{\epsilon}_{\boldsymbol{\theta}} \), to estimate the noise component at each step. A single reverse step is given by:
\begin{equation}
    \tilde{\mathbf{x}}_{t-1} = \frac{1}{\sqrt{\alpha_t}} \left(\tilde{\mathbf{x}}_{t} - \frac{\beta_t}{\sqrt{1 - \bar{\alpha}_t}} \boldsymbol{\epsilon}_{\boldsymbol{\theta}}(\tilde{\mathbf{x}}_{t}, t) \right) + \sigma_t z.
\end{equation}
Here, \( \tilde{\mathbf{x}}_t \) and \( \tilde{\mathbf{x}}_{t-1} \) denote the signal estimates at time steps \( t \) and \( t-1 \), respectively, where \( \tilde{\mathbf{x}}_j = \mathbf{x}_j \) at the final noising step. The term \( \sigma_t \) is typically a function of \( \beta \) to introduce stochasticity in the sampling process, and \( z \sim \mathcal{N}(0, I) \)~\cite{DDPM}. The objective is to iteratively refine the denoised sample so that \( \tilde{x}_0 \approx x_0 \).

\subsection{Diffusion models for time series}

Diffusion models show great promise in time series tasks, excelling in both forecasting~\cite{rasul_forecasting, alcaraz2022, li_forecasting} and generation~\cite{lim2023_generation, tabsyn}.
Among these, Denoising Diffusion Probabilistic Models (DDPMs)~\cite{DDPM} are a leading framework, which progressively denoise samples to reconstruct data from noise~\cite{diffusion_survey} and rely on stochastic noise addition during sampling~\cite{DDPM}.
On the other hand, Denoising Diffusion Implicit Models (DDIMs) use a deterministic sampling process that removes noise, enabling faster sampling and fewer steps to generate high-quality samples with greater predictability~\cite{DDIM_ICLR21}.
However, this reduces the model's ability to explore a wide range of outputs, leading to lower diversity and reduced robustness (i.e., consistency in generating diverse and reliable samples)~\cite{DDPM, DDIM_ICLR21}.

Diffusion-TS, which serves as the backbone model of \algo presented in this paper, employs a DDPM combined with seasonal-trend decomposition to better capture underlying structures and dependencies of multivariate time series~\cite{yuan2024diffusion}. It also introduces a Fourier-based loss to optimize reconstruction, improving accuracy by better matching the frequency components~\cite{yuan2024diffusion}.
Its innovation lies in the integration of seasonal-trend decomposition with DDPMs, and the use of the Fourier-based loss to enhance the model’s ability to capture complex temporal patterns.

Some diffusion models that synthesize time series data include ScoreGrad~\cite{yan_scoregrad_2021}, $\text{SSSD}^{S4}$~\cite{alcaraz2022}, TSDiff~\cite{kollovieh_2023}, but some also incorporate transformer-based elements like TimeGrad~\cite{rasul_forecasting}, CSDI~\cite{tashiro_csdi_2021}, and TDSTF~\cite{chang_tdstf_2024}.
Diffusion-TS~\cite{yuan2024diffusion} effectively addresses weaknesses in these models. Unlike ScoreGrad and TimeGrad which use autoregressive models~\cite{yan_scoregrad_2021, rasul_forecasting}, it avoids error accumulation and slow inference over long horizons by using DDPM, a stable diffusion-based framework. It outperforms $\text{SSSD}^{S4}$ and TSDiff~\cite{alcaraz2022, kollovieh_2023} by replacing resource-intensive S4 layers with an efficient latent layer that simplifies handling multivariate data.
Diffusion-TS handles incomplete datasets better than CSDI~\cite{tashiro_csdi_2021} by avoiding the need for explicit pairing of observed and missing data during training, while being more adaptable and efficient on diverse datasets compared to TDSTF~\cite{chang_tdstf_2024}, which struggles with real-time forecasting.


\subsection{DDIM and DDIM inversion}


The Denoising Diffusion Implicit Model (DDIM)~\citep{DDIM_ICLR21} offers deterministic diffusion and sampling, extending the traditional Markovian diffusion process to a broader class of non-Markovian processes. Given a initial noise vector $\mathbf{x}_T$ and a neural network $\boldsymbol{\epsilon}_{\boldsymbol{\theta}}$ that predicts the noise $\boldsymbol{\epsilon}_{\boldsymbol{\theta}}(\mathbf{x}_t, t)$ at each timestep $t$, the DDIM sampling step to generate sample $\mathbf{x}_{t-1}$ from $\mathbf{x}_t$, is defined as:
\begin{equation}
\mathbf{x}_{t-1}=\sqrt{\alpha_{t-1}}\left(\frac{\mathbf{x}_t-\sqrt{1-\alpha_t} \boldsymbol{\epsilon}_{\boldsymbol{\theta}}\left(\mathbf{x}_t, t\right)}{\sqrt{\alpha_t}}\right)+\sqrt{1-\alpha_{t-1}-\sigma_t^2} \cdot \boldsymbol{\epsilon}_{\boldsymbol{\theta}}\left(\mathbf{x}_t, t\right)+\sigma_t \epsilon_t,
\end{equation}
where $\alpha_1, \dots, \alpha_T$ are computed from a predefined variance schedule, $\epsilon_t \sim \mathcal{N}(0, I)$ denotes standard Gaussian noise independent of $\mathbf{x}_t$, and the $\sigma_t$ values can be varied to yield different generative processes. Setting $\sigma_t$ to 0 for all $t$ makes the sampling process deterministic:
\begin{equation}
\mathbf{x}_{t-1}=\sqrt{\frac{\alpha_{t-1}}{\alpha_t}} \mathbf{x}_t+\left(\sqrt{1-\alpha_{t-1}}-\sqrt{\frac{\alpha_{t-1}}{\alpha_t}-\alpha_{t-1}}\right) \boldsymbol{\epsilon}_{\boldsymbol{\theta}}\left(\mathbf{x}_t, t\right).
\end{equation}
This sampling process ensures the same latent matrix $\mathbf{x}_0$ is consistently generated by a given noise matrix $\mathbf{x}_T$.

Having large $T$ values (being limited with small steps) allows to cross the timesteps in the backward direction toward increasing noise levels, which gives out a deterministic diffusion process from $\mathbf{x}_0$ to $\mathbf{x}_T$; this is also known as DDIM inversion:
\begin{equation}
\mathbf{x}_{t+1}\approx	\sqrt{\frac{\alpha_{t+1}}{\alpha_t}} \mathbf{x}_t+\left(\sqrt{1-\alpha_{t+1}}-\sqrt{\frac{\alpha_{t+1}}{\alpha_t}-\alpha_{t+1}}\right) \boldsymbol{\epsilon}_{\boldsymbol{\theta}}\left(\mathbf{x}_t, t\right).
\end{equation}

%% file: src/appx_proof.tex
\section{Proof}
\label{app:proof}
\mytheorem*

\textbf{Proof:} 

For \( t = 1 \), we have:
\begin{equation}
    \boldsymbol{\delta}_1 = \mathbf{x}_1^{\textit{approx}} - \mathbf{x}_1^{\textit{orig}} = \boldsymbol{\epsilon}.
\end{equation}
For \( t = 2 \), using the recurrence,
\begin{equation}
    \mathbf{x}_2 = \frac{\mathbf{x}_0}{\gamma} - \frac{a_1 \mathbf{x}_1 + b_1 \hat{\boldsymbol{\epsilon}}_{\boldsymbol{\theta}}(\mathbf{x}_1,1)}{\gamma} + \left(\frac{\mathbf{x}_1}{a_2} - \frac{b_2}{a_2} \hat{\boldsymbol{\epsilon}}_{\boldsymbol{\theta}}(\mathbf{x}_1,2) \right).
\end{equation}
Subtracting the approximated and original cases and using \cref{assump}, we obtain:
\begin{equation}
    \boldsymbol{\delta}_2 = -\frac{a_1 \boldsymbol{\delta}_1}{\gamma} + \frac{\boldsymbol{\delta}_1}{a_2} - \frac{b_1}{\gamma} \Delta_{1} \boldsymbol{\delta}_1 - \frac{b_2}{a_2} \Delta_{1}\boldsymbol{\delta}_1.
\end{equation}
Thus, defining
\begin{equation}
    C_2 = \left| \frac{1}{a_2} - \frac{a_1}{\gamma} \right| + \frac{b_1}{\gamma} \Delta_{1} + \frac{b_2}{a_2} \Delta_{1},
\end{equation}
we bound
\begin{equation}
    \|\boldsymbol{\delta}_2\| \leq C_2 \|\boldsymbol{\delta}_1\| = C_2 \|\boldsymbol{\epsilon}\|.
\end{equation}

Suppose for some \( t \geq 2 \), there exists a constant \( C_t \) such that
\begin{equation}
    \|\boldsymbol{\delta}_t\| \leq C_t \|\boldsymbol{\epsilon}\|.
\end{equation}
For \( t+1 \), using the recurrence:
\begin{equation}
    \mathbf{x}_{t+1} = \frac{\mathbf{x}_{t-1}}{\gamma} - \frac{a_t \mathbf{x}_t + b_t \hat{\boldsymbol{\epsilon}}_{\boldsymbol{\theta}}(\mathbf{x}_t,t)}{\gamma} + \left(\frac{\mathbf{x}_t}{a_{t+1}} - \frac{b_{t+1}}{a_{t+1}} \hat{\boldsymbol{\epsilon}}_{\boldsymbol{\theta}}(\mathbf{x}_t,t+1) \right).
\end{equation}
Taking differences, we obtain
\begin{equation}
    \boldsymbol{\delta}_{t+1} = \frac{\boldsymbol{\delta}_{t-1}}{\gamma} - \frac{a_t \boldsymbol{\delta}_t}{\gamma} - \frac{b_t}{\gamma} \Delta_{t} \boldsymbol{\delta}_t + \frac{\boldsymbol{\delta}_t}{a_{t+1}} - \frac{b_{t+1}}{a_{t+1}} \Delta_{t} \boldsymbol{\delta}_t.
\end{equation}
Bounding the terms, we define:
\begin{equation}
    C_{t+1} = C_t \left( \left| \frac{1}{\gamma} - \frac{a_t}{\gamma} + \frac{1}{a_{t+1}} \right| + \frac{b_t}{\gamma} \Delta_{t} + \frac{b_{t+1}}{a_{t+1}} \Delta_{t} \right).
\end{equation}
Thus, we conclude:
\begin{equation}
    \|\boldsymbol{\delta}_{t+1}\| \leq C_{t+1} \|\boldsymbol{\epsilon}\|.
\end{equation}

\smallskip
\noindent
By induction, we obtain:
\begin{equation}
    C_T = \prod_{t=1}^{T-1} \left( \left| \frac{1}{\gamma} - \frac{a_{i}}{\gamma} + \frac{1}{a_{t+1}} \right| + \frac{b_{t}}{\gamma} \Delta_t + \frac{b_{t+1}}{a_{t+1}} \Delta_{t} \right).
\end{equation}

Thus, the perturbation remains bounded for all \( T \), i.e.
\begin{equation}
    \|\delta_T\| \leq |\epsilon| \prod_{t=1}^{T-1} \left( \left| \frac{1}{\gamma} - \frac{a_t}{\gamma} + \frac{1}{a_{t+1}} \right| + \frac{b_t}{\gamma} \Delta_{t} + \frac{b_{t+1}}{a_{t+1}} \Delta_{t} \right),
\end{equation}

completing the proof. \hfill \(\square\)

\begin{figure}[ht!]
    \centering
    \includegraphics[width=0.75\linewidth]{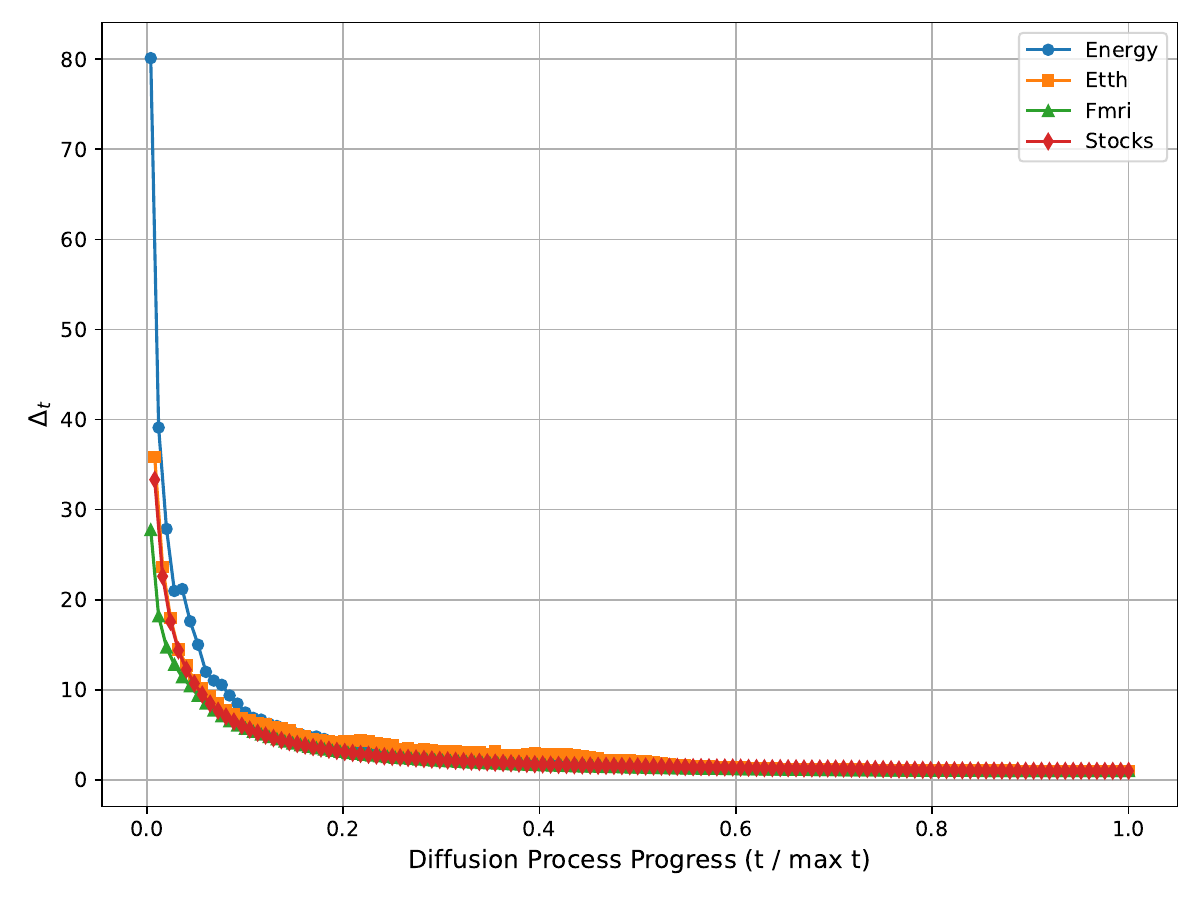}
    \caption{\( \Delta_t \) for different datasets.}
    \label{fig:delta_t_plot}
\end{figure}

To further validate \cref{assump}, we empirically computed \( \Delta_t \) across four different datasets, using 10,000 samples from each. Specifically, \( \Delta_t \) is calculated as the maximum ratio for different \( t \) using the \( L_1 \) norm of the data samples, as illustrated in \cref{fig:delta_t_plot}.

We also generate 10,000 samples with 64-length sequences across five datasets, computing the final output $\mathbf{x}_0$ and the one step prior $\mathbf{x}_1$. Results in \cref{tab:l1_x0_x1} show consistently small \( L_1 \) norm between $\mathbf{x}_1$ and $\mathbf{x}_0$, with an average of $5.1 \times 10^{-3}$ to $7.0 \times 10^{-3}$ and maximum of $<0.23$, validating our $\epsilon$-exact approximation. These empirically small errors confirm \cref{thm:original}'s theoretical bounds and demonstrate reliable watermark detection despite the approximation.

\begin{table}[ht!]
\centering
\caption{$L_1$ norm between $\mathbf{x}_1$ and $\mathbf{x}_0$ for 64-length sequences over 10,000 samples.}
\label{tab:l1_x0_x1}
\vskip 0.15in
\begin{tabular}{lcc}
\toprule
Dataset & Avg. $L_1$ $\left(\times 10^{-3}\right)$ & Max. $L_1$ \\
\midrule
Stocks & 7.031 & 0.082 \\
ETTh & 6.776 & 0.104 \\
MuJoCo & 5.146 & 0.081 \\
Energy & 5.687 & 0.230 \\
fMRI & 5.945 & 0.070 \\
\bottomrule
\end{tabular}
\end{table}

%% file: src/tables/appx_datasets.tex
\begin{table}[H]
\vskip -0.15in
\centering

\caption{Details of datasets used in experiments.}
\label{tab:appx_dataset}%
\vskip 0.15in

\begin{tabular}{l c c l}

\toprule

Dataset & Number of Rows & Number of Features & Source \\

\midrule

Stocks & 3,773 & 6 & \href{https://finance.yahoo.com/quote/GOOG/history?p=GOOG}{https://finance.yahoo.com/quote/GOOG} \\
ETTh & 17,420 & 7 & \href{https://github.com/zhouhaoyi/ETDataset}{https://github.com/zhouhaoyi/ETDataset} \\
MuJoCo & 10,000 & 14 & \href{https://github.com/google-deepmind/dm_control}{https://github.com/deepmind/dm$\_$control} \\
Energy & 19,711 & 28 & \href{https://archive.ics.uci.edu/ml/datasets/Appliances+energy+prediction}{https://archive.ics.uci.edu/ml/datasets} \\
fMRI & 10,000 & 50 & \href{https://www.fmrib.ox.ac.uk/datasets/netsim/}{https://www.fmrib.ox.ac.uk/datasets} \\

\bottomrule

\end{tabular}%
\end{table}%

%% file: src/appx_baselines.tex

\paragraph{Tree-Ring} We adapt the latent-representation-based Tree-Ring watermark for multivariate time series in the data space. Initially, the Tree-Ring watermark was proposed for images with square dimensions as it places the circular ring watermark pattern centrally. However, multivariate time series often have rectangular dimensions with varying numbers of features and timesteps. To accommodate this structure, we apply a flexible ring pattern with a predefined radius indicating the outermost circle of the watermark. Finally, we embed and detect the watermark in the Fourier domain.

\paragraph{Gaussian Shading} Similar to Tree-Ring, we implement Gaussian Shading from the image domain to multivariate time series by embedding the watermark directly in the data space. To maintain coherence and efficiency, we use a single control seed across all samples to avoid the need for additional indices for each sample.



\paragraph{Heads Tails Watermark} The HTW is a post-watermarking technique designed for univariate time series. To extend its applicability, we adapt it for multivariate time series by iterating through each variate in the generated synthetic sample. During evaluation, the watermarked time series is reversed, and processing each series independently. In short, we treat each variate as a single series.

\paragraph{TabWak} TabWak was originally designed for the tabular data domain, where it operates effectively in the latent space. We extend its application to multivariate time series in the data space. However, tabular data primarily captures feature dependencies, while time series data inherently relies on both feature and temporal dependencies. Therefore, we implement another version called TabWak$^\top$ by transposing the watermarking direction onto the temporal axis. Similarly, we evaluate the watermark detectability on a sample by sample basis.

%% file: src/tables/appx_training_sampling.tex
\begin{table}[H]
\vskip -0.15in
\centering

\caption{Details of training and sampling time.}
\label{tab:appx_training_sampling}%
\vskip 0.15in

\begin{tabular}{l c c c}

\toprule

Dataset & Window Size & Training Time ($\sim$ min.) & Sampling Time ($\sim$ min.) \\

\midrule

\multirow{3}{*}{Stocks} & 24 & 6.1 & 0.4 \\
& 64 & 6.2 & 2.1 \\
& 128 & 8.0 & 4.6 \\

\cmidrule{1-4}

\multirow{3}{*}{ETTh} & 24 & 11.6 & 0.4 \\
& 64 & 12.7 & 2.4 \\
& 128 & 13.7 & 5.2 \\

\cmidrule{1-4}

\multirow{3}{*}{MuJoCo} & 24 & 9.7 & 0.7 \\
& 64 & 9.9 & 4.9 \\
& 128 & 10.6 & 12.6 \\

\cmidrule{1-4}

\multirow{3}{*}{Energy} & 24 & 23.5 & 1.4 \\
& 64 & 24.2 & 10.5 \\
& 128 & 24.3 & 23.0 \\

\cmidrule{1-4}

\multirow{3}{*}{fMRI} & 24 & 16.8 & 1.8 \\
& 64 & 17.5 & 13.5 \\
& 128 & 18.1 & 28.5 \\

\bottomrule

\end{tabular}%
\end{table}%

%% file: src/tables/appx_quality_diff_len.tex
\begin{table}[ht!]

\centering
\caption{
Results of synthetic time series quality. No watermarking (`W/O') is included.}
\label{tab:appx_quality_diff_len}

\vskip 0.15in

\resizebox{\linewidth}{!}{
\begin{tabular}{l l c c c c c c c c}

\toprule

& & \multicolumn{4}{c}{64-length $\downarrow$} & \multicolumn{4}{c}{128-length $\downarrow$} \\

\cmidrule(r){3-6}
\cmidrule(l){7-10}

Dataset & Method & Context-FID & Correlational & Discriminative & Predictive & Context-FID & Correlational & Discriminative & Predictive \\

\midrule

\multirow{7}{*}{Stocks} & W/O & \val{0.444}{0.114} & \val{0.030}{0.023} & \val{0.104}{0.014} & \val{0.037}{0.000} & \val{0.536}{0.135} & \val{0.020}{0.015} & \val{0.148}{0.019} & \val{0.037}{0.000} \\
& TR & \val{1.812}{0.210} & \val{0.102}{0.012} & \val{0.206}{0.038} & \val{0.038}{0.001} & \val{3.555}{0.868} & \val{0.130}{0.025} & \val{0.276}{0.030} & \val{0.041}{0.004} \\
& GS & \val{1.643}{0.126} & \val{0.011}{0.005} & \val{0.252}{0.072} & \val{0.042}{0.001} & \val{2.647}{0.262} & \val{0.025}{0.015} & \val{0.186}{0.071} & \val{0.040}{0.000} \\
& HTW & \val{0.426}{0.073} & \val{0.026}{0.016} & \val{0.105}{0.032} & \val{0.037}{0.000} & \val{0.622}{0.092} & \val{0.017}{0.013} & \val{0.152}{0.086} & \val{0.037}{0.000} \\
& TabWak & \val{0.242}{0.045} & \val{0.011}{0.008} & \val{0.141}{0.021} & \val{0.037}{0.000} & \val{0.394}{0.034} & \val{0.009}{0.009} & \val{0.150}{0.031} & \val{0.037}{0.000} \\
& TabWak$^\top$ & \val{0.284}{0.091} & \val{0.005}{0.005} & \val{0.099}{0.025} & \val{0.037}{0.000} & \val{0.367}{0.035} & \val{0.010}{0.007} & \val{0.129}{0.054} & \val{0.037}{0.000} \\
& \algo & \val{0.387}{0.054} & \val{0.017}{0.017} & \val{0.092}{0.041} & \val{0.037}{0.000} & \val{0.316}{0.044} & \val{0.021}{0.024} & \val{0.140}{0.029} & \val{0.037}{0.000} \\

\cmidrule{1-10}

\multirow{7}{*}{ETTh} & W/O & \val{0.384}{0.034} & \val{0.070}{0.011} & \val{0.106}{0.009} & \val{0.115}{0.006} & \val{1.086}{0.070} & \val{0.098}{0.021} & \val{0.166}{0.013} & \val{0.111}{0.008} \\
& TR & \val{2.224}{0.209} & \val{0.217}{0.011} & \val{0.284}{0.032} & \val{0.131}{0.004} & \val{2.450}{0.128} & \val{0.270}{0.017} & \val{0.284}{0.063} & \val{0.138}{0.005} \\
& GS & \val{3.398}{0.384} & \val{0.248}{0.025} & \val{0.356}{0.029} & \val{0.154}{0.001} & \val{4.998}{0.603} & \val{0.233}{0.027} & \val{0.381}{0.042} & \val{0.162}{0.010} \\
& HTW & \val{0.372}{0.027} & \val{0.076}{0.006} & \val{0.118}{0.024} & \val{0.116}{0.006} & \val{1.119}{0.065} & \val{0.091}{0.014} & \val{0.165}{0.010} & \val{0.120}{0.006} \\
& TabWak & \val{0.597}{0.044} & \val{0.346}{0.027} & \val{0.176}{0.020} & \val{0.122}{0.007} & \val{1.931}{0.254} & \val{0.466}{0.058} & \val{0.248}{0.017} & \val{0.128}{0.005} \\
& TabWak$^\top$ & \val{0.503}{0.045} & \val{0.095}{0.043} & \val{0.119}{0.012} & \val{0.120}{0.010} & \val{1.477}{0.075} & \val{0.129}{0.037} & \val{0.143}{0.014} & \val{0.112}{0.011} \\
& \algo & \val{0.297}{0.038} & \val{0.133}{0.040} & \val{0.097}{0.015} & \val{0.115}{0.003} & \val{1.090}{0.100} & \val{0.135}{0.057} & \val{0.174}{0.007} & \val{0.110}{0.009} \\

\cmidrule{1-10}

\multirow{7}{*}{MuJoCo} & W/O & \val{0.103}{0.016} & \val{0.341}{0.025} & \val{0.023}{0.015} & \val{0.007}{0.000} & \val{0.179}{0.011} & \val{0.290}{0.025} & \val{0.055}{0.027} & \val{0.005}{0.001} \\
& TR & \val{2.627}{0.230} & \val{0.948}{0.064} & \val{0.270}{0.128} & \val{0.016}{0.005} & \val{2.348}{0.236} & \val{0.865}{0.025} & \val{0.358}{0.008} & \val{0.008}{0.001} \\
& GS & \val{7.162}{1.301} & \val{1.034}{0.087} & \val{0.447}{0.011} & \val{0.011}{0.002} & \val{4.797}{1.290} & \val{1.335}{0.041} & \val{0.454}{0.023} & \val{0.007}{0.001} \\
& HTW & \val{0.316}{0.034} & \val{0.334}{0.044} & \val{0.248}{0.079} & \val{0.011}{0.001} & \val{0.431}{0.051} & \val{0.309}{0.064} & \val{0.255}{0.174} & \val{0.009}{0.001} \\
& TabWak & \val{0.372}{0.064} & \val{0.671}{0.083} & \val{0.137}{0.027} & \val{0.007}{0.001} & \val{0.369}{0.073} & \val{0.589}{0.059} & \val{0.175}{0.042} & \val{0.006}{0.002} \\
& TabWak$^\top$ & \val{0.238}{0.019} & \val{0.339}{0.059} & \val{0.087}{0.025} & \val{0.006}{0.001} & \val{0.275}{0.033} & \val{0.333}{0.066} & \val{0.084}{0.028} & \val{0.006}{0.001} \\
& \algo & \val{0.108}{0.014} & \val{0.413}{0.062} & \val{0.038}{0.021} & \val{0.007}{0.001} & \val{0.155}{0.016} & \val{0.316}{0.022} & \val{0.046}{0.030} & \val{0.005}{0.001} \\

\cmidrule{1-10}

\multirow{7}{*}{Energy} & W/O & \val{0.112}{0.016} & \val{1.032}{0.289} & \val{0.124}{0.008} & \val{0.251}{0.001} & \val{0.120}{0.011} & \val{0.798}{0.213} & \val{0.202}{0.073} & \val{0.249}{0.000} \\
& TR & \val{0.902}{0.093} & \val{3.503}{0.589} & \val{0.427}{0.072} & \val{0.307}{0.005} & \val{1.195}{0.111} & \val{2.303}{0.569} & \val{0.498}{0.001} & \val{0.287}{0.003} \\
& GS & \val{2.205}{0.192} & \val{3.277}{0.129} & \val{0.479}{0.010} & \val{0.310}{0.005} & \val{3.680}{0.444} & \val{4.233}{0.287} & \val{0.474}{0.039} & \val{0.280}{0.006} \\
& HTW & \val{0.133}{0.015} & \val{1.045}{0.357} & \val{0.135}{0.013} & \val{0.251}{0.001} & \val{0.133}{0.017} & \val{0.822}{0.310} & \val{0.103}{0.043} & \val{0.249}{0.001} \\
& TabWak & \val{0.168}{0.021} & \val{1.811}{0.530} & \val{0.136}{0.013} & \val{0.251}{0.000} & \val{0.201}{0.020} & \val{2.001}{0.440} & \val{0.156}{0.077} & \val{0.250}{0.001} \\
& TabWak$^\top$ & \val{0.237}{0.029} & \val{1.321}{0.252} & \val{0.138}{0.015} & \val{0.252}{0.000} & \val{0.274}{0.019} & \val{1.211}{0.196} & \val{0.136}{0.023} & \val{0.249}{0.001} \\
& \algo & \val{0.143}{0.019} & \val{1.662}{0.298} & \val{0.145}{0.019} & \val{0.251}{0.000} & \val{0.148}{0.027} & \val{1.687}{0.328} & \val{0.140}{0.057} & \val{0.249}{0.000} \\

\cmidrule{1-10}

\multirow{7}{*}{fMRI} & W/O & \val{0.435}{0.033} & \val{1.899}{0.075} & \val{0.268}{0.150} & \val{0.100}{0.000} & \val{0.859}{0.058} & \val{1.823}{0.064} & \val{0.209}{0.263} & \val{0.100}{0.000} \\
& TR & \val{3.358}{0.402} & \val{13.699}{0.132} & \val{0.411}{0.158} & \val{0.141}{0.001} & \val{5.391}{0.919} & \val{13.000}{0.084} & \val{0.434}{0.072} & \val{0.149}{0.002} \\
& GS & \val{0.756}{0.098} & \val{8.378}{0.028} & \val{0.500}{0.001} & \val{0.104}{0.001} & \val{1.046}{0.052} & \val{5.941}{0.048} & \val{0.499}{0.001} & \val{0.106}{0.001} \\
& HTW & \val{0.413}{0.017} & \val{1.822}{0.111} & \val{0.338}{0.032} & \val{0.100}{0.000} & \val{0.811}{0.051} & \val{1.757}{0.053} & \val{0.063}{0.064} & \val{0.100}{0.000} \\
& TabWak & \val{0.655}{0.137} & \val{6.532}{0.158} & \val{0.242}{0.302} & \val{0.108}{0.001} & \val{1.114}{0.135} & \val{5.967}{0.033} & \val{0.153}{0.218} & \val{0.111}{0.001} \\
& TabWak$^\top$ & \val{0.554}{0.049} & \val{1.955}{0.058} & \val{0.331}{0.039} & \val{0.100}{0.000} & \val{0.919}{0.024} & \val{1.807}{0.024} & \val{0.265}{0.201} & \val{0.100}{0.000} \\
& \algo & \val{0.441}{0.035} & \val{1.786}{0.043} & \val{0.314}{0.041} & \val{0.100}{0.000} & \val{0.855}{0.072} & \val{1.704}{0.060} & \val{0.298}{0.227} & \val{0.100}{0.000} \\

\bottomrule

\end{tabular}}
\end{table}

%% file: src/tables/appx_attack_24.tex
\begin{table}[ht!]

\centering
\caption{Results of robustness against post-editing attacks. Average Z-score on 24-length sequences, including un-attacked scores from \cref{tab:main_exp}. Best results are in \textbf{bold}, and second-best are \underline{underlined}.}
\label{tab:appx_attack_24}
\vskip 0.15in

\resizebox{\linewidth}{!}{
\begin{tabular}{l l c c c c c c c}

\toprule



& & \multirow[b]{2}{*}{\makecell{Without \\ Attack}}  & \multicolumn{2}{c}{Offset $\uparrow$} & \multicolumn{2}{c}{Random Crop $\uparrow$} & \multicolumn{2}{c}{Min-Max Insertion $\uparrow$} \\

\cmidrule(r){4-5}
\cmidrule(rl){6-7}
\cmidrule(l){8-9}

Dataset & Method & & 5\% & 30\% & 5\% & 30\% & 5\% & 30\% \\

\midrule

\multirow{6}{*}{Stocks} & TR & \val{0.43}{0.04} & \val{0.41}{0.04} & \val{0.26}{0.05} & \underline{\val{61.01}{1.18}} & \textbf{\val{76.69}{1.17}} & \val{0.18}{0.05} & \val{0.81}{0.08} \\
& GS & \underline{\val{86.07}{0.74}} & \underline{\val{85.74}{0.71}} & \underline{\val{84.38}{0.67}} & \val{-34.10}{0.69} & \val{3.08}{0.58} & \underline{\val{82.31}{0.75}} & \textbf{\val{57.07}{0.72}} \\
& HTW & \val{4.45}{0.62} & \val{4.45}{0.62} & \val{4.45}{0.62} & \val{-0.30}{0.39} & \val{-0.40}{0.30} & \val{3.96}{0.72} & \val{1.80}{0.71} \\
& TabWak & \val{-67.22}{1.17} & \val{-66.62}{1.23} & \val{-75.89}{1.10} & \val{-137.99}{1.23} & \val{-72.71}{1.76} & \val{-70.11}{1.14} & \val{-89.22}{0.64} \\
& TabWak$^\top$ & \val{55.39}{0.86} & \val{55.59}{0.74} & \val{54.76}{0.78} & \val{9.32}{0.94} & \val{7.43}{0.73} & \val{45.69}{0.73} & \val{21.19}{0.56} \\
& \algo & \textbf{\val{182.10}{0.73}} & \textbf{\val{181.98}{0.84}} & \textbf{\val{180.68}{0.77}} & \textbf{\val{62.29}{1.01}} & \underline{\val{9.33}{1.04}} & \textbf{\val{155.72}{0.91}} & \underline{\val{56.86}{1.06}} \\

\cmidrule{1-9}

\multirow{6}{*}{ETTh} & TR & \val{7.84}{0.12} & \val{7.94}{0.12} & \val{8.19}{0.13} & \val{26.95}{0.32} & \textbf{\val{34.36}{0.26}} & \val{14.11}{0.28} & \underline{\val{28.72}{0.50}} \\
& GS & \val{101.07}{1.17} & \val{99.45}{1.09} & \underline{\val{105.70}{1.08}} & \textbf{\val{98.99}{1.21}} & \val{2.10}{1.37} & \underline{\val{98.78}{1.13}} & \textbf{\val{73.67}{1.30}} \\
& HTW & \val{3.43}{0.83} & \val{3.43}{0.83} & \val{3.43}{0.83} & \val{0.38}{0.77} & \val{-1.08}{0.47} & \val{3.09}{0.82} & \val{1.54}{0.66} \\
& TabWak & \val{-14.95}{1.08} & \val{-7.92}{0.97} & \val{36.49}{1.08} & \val{-11.76}{1.31} & \underline{\val{8.65}{1.38}} & \val{-7.95}{0.93} & \val{21.70}{0.82} \\
& TabWak$^\top$ & \underline{\val{109.35}{0.91}} & \underline{\val{110.22}{0.85}} & \val{103.34}{0.82} & \underline{\val{50.64}{0.96}} & \val{1.06}{1.02} & \val{80.98}{0.87} & \val{21.83}{0.95} \\
& \algo & \textbf{\val{134.83}{0.95}} & \textbf{\val{130.50}{0.99}} & \textbf{\val{118.65}{1.03}} & \val{35.30}{1.14} & \val{5.99}{1.04} & \textbf{\val{101.53}{0.96}} & \val{20.77}{0.95} \\

\cmidrule{1-9}

\multirow{6}{*}{MuJoCo} & TR & \val{1.38}{0.03} & \val{1.34}{0.03} & \val{1.22}{0.03} & \val{5.31}{0.05} & \val{5.65}{0.06} & \val{2.42}{0.05} & \val{5.53}{0.09} \\
& GS & \val{21.13}{0.77} & \val{23.95}{0.71} & \val{27.15}{0.76} & \val{23.06}{0.86} & \val{-10.38}{0.72} & \val{22.89}{0.83} & \val{22.89}{1.06} \\
& HTW & \val{2.89}{0.54} & \val{2.89}{0.54} & \val{2.89}{0.54} & \val{0.85}{0.48} & \val{-0.47}{0.40} & \val{2.65}{0.53} & \val{1.51}{0.43} \\
& TabWak & \underline{\val{31.30}{1.07}} & \underline{\val{31.59}{0.96}} & \underline{\val{39.05}{1.05}} & \underline{\val{38.80}{1.11}} & \textbf{\val{25.04}{1.13}} & \underline{\val{31.60}{0.92}} & \underline{\val{28.70}{0.59}} \\
& TabWak$^\top$ & \val{-4.85}{0.87} & \val{-6.27}{0.90} & \val{0.74}{0.78} & \val{-15.92}{0.88} & \val{-56.15}{1.50} & \val{-10.95}{1.01} & \val{-43.08}{1.17} \\
& \algo & \textbf{\val{85.69}{1.08}} & \textbf{\val{77.78}{1.26}} & \textbf{\val{49.48}{1.12}} & \textbf{\val{48.03}{1.17}} & \underline{\val{11.46}{1.18}} & \textbf{\val{74.94}{1.10}} & \textbf{\val{38.87}{1.14}} \\

\cmidrule{1-9}

\multirow{6}{*}{Energy} & TR & \val{9.51}{0.09} & \val{9.30}{0.10} & \val{8.77}{0.08} & \textbf{\val{110.47}{0.23}} & \textbf{\val{129.09}{0.19}} & \val{17.65}{0.15} & \val{35.69}{0.18} \\
& GS & \underline{\val{51.22}{0.88}} & \underline{\val{50.27}{0.96}} & \underline{\val{47.31}{0.97}} & \val{11.65}{0.72} & \underline{\val{38.70}{1.34}} & \underline{\val{48.80}{0.97}} & \underline{\val{39.80}{0.94}} \\
& HTW & \val{3.06}{0.36} & \val{3.06}{0.36} & \val{3.06}{0.36} & \val{0.56}{0.35} & \val{-0.13}{0.23} & \val{2.81}{0.36} & \val{1.61}{0.30} \\
& TabWak & \val{3.26}{0.89} & \val{0.52}{1.03} & \val{-2.91}{1.08} & \val{3.27}{0.63} & \val{3.54}{0.53} & \val{2.20}{0.87} & \val{-2.08}{0.49} \\
& TabWak$^\top$ & \val{40.82}{0.81} & \val{38.47}{0.72} & \val{40.20}{0.82} & \underline{\val{33.83}{1.08}} & \val{-7.47}{1.50} & \val{31.89}{0.89} & \val{5.37}{0.93} \\
& \algo & \textbf{\val{231.28}{1.45}} & \textbf{\val{228.22}{1.71}} & \textbf{\val{185.48}{1.52}} & \val{-9.20}{0.96} & \val{-1.54}{1.02} & \textbf{\val{189.15}{1.44}} & \textbf{\val{56.39}{1.16}} \\

\cmidrule{1-9}

\multirow{6}{*}{fMRI} & TR & \val{6.49}{0.05} & \val{6.43}{0.05} & \val{6.12}{0.05} & \val{5.42}{0.05} & \val{4.40}{0.05} & \val{5.54}{0.05} & \val{1.37}{0.07} \\
& GS & \underline{\val{420.02}{1.44}} & \underline{\val{416.83}{1.43}} & \underline{\val{401.86}{1.72}} & \underline{\val{360.27}{1.40}} & \underline{\val{171.38}{1.16}} & \underline{\val{386.10}{1.54}} & \textbf{\val{245.44}{1.28}} \\
& HTW & \val{4.32}{0.22} & \val{4.32}{0.22} & \val{4.32}{0.22} & \val{2.92}{0.27} & \val{-0.39}{0.28} & \val{3.86}{0.26} & \val{1.78}{0.25} \\
& TabWak & \val{84.02}{1.04} & \val{83.55}{1.08} & \val{78.94}{1.16} & \val{78.63}{1.24} & \val{27.05}{1.30} & \val{76.49}{1.17} & \val{50.55}{1.33} \\
& TabWak$^\top$ & \textbf{\val{464.67}{0.50}} & \textbf{\val{464.04}{0.48}} & \textbf{\val{458.81}{0.57}} & \textbf{\val{400.47}{0.72}} & \textbf{\val{202.57}{1.04}} & \textbf{\val{403.49}{0.66}} & \underline{\val{168.64}{0.75}} \\
& \algo & \val{379.51}{0.82} & \val{378.95}{0.85} & \val{374.99}{0.78} & \val{277.64}{0.78} & \val{77.74}{1.05} & \val{327.13}{0.86} & \val{133.68}{0.85} \\

\bottomrule

\end{tabular}}
\end{table}

%% file: src/tables/appx_attack_128.tex
\begin{table}[ht!]

\centering
\caption{Results of robustness against post-editing attacks. Average Z-score on 128-length sequences, including un-attacked scores from \cref{tab:main_exp}. Best results are in \textbf{bold}, and second-best are \underline{underlined}.}
\label{tab:appx_attack_128}
\vskip 0.15in

\resizebox{\linewidth}{!}{
\begin{tabular}{l l c c c c c c c}

\toprule



& & \multirow[b]{2}{*}{\makecell{Without \\ Attack}}  & \multicolumn{2}{c}{Offset $\uparrow$} & \multicolumn{2}{c}{Random Crop $\uparrow$} & \multicolumn{2}{c}{Min-Max Insertion $\uparrow$} \\

\cmidrule(r){4-5}
\cmidrule(rl){6-7}
\cmidrule(l){8-9}

Dataset & Method & & 5\% & 30\% & 5\% & 30\% & 5\% & 30\% \\

\midrule

\multirow{6}{*}{Stocks} & TR & \val{0.08}{0.10} & \val{0.05}{0.10} & \val{0.02}{0.12} & \textbf{\val{93.62}{0.99}} & \textbf{\val{119.45}{1.16}} & \val{6.87}{0.22} & \val{23.98}{0.55} \\
& GS & \underline{\val{172.23}{1.08}} & \underline{\val{167.39}{1.10}} & \underline{\val{145.89}{1.83}} & \val{13.22}{1.43} & \val{-2.23}{1.03} & \underline{\val{144.05}{0.98}} & \underline{\val{64.79}{1.01}} \\
& HTW & \val{10.39}{1.33} & \val{10.39}{1.33} & \val{10.39}{1.33} & \val{-1.30}{0.44} & \val{-1.42}{0.33} & \val{9.05}{1.67} & \val{3.76}{1.63} \\
& TabWak & \val{-10.49}{1.28} & \val{49.98}{3.31} & \val{36.78}{3.15} & \val{69.44}{3.32} & \underline{\val{70.32}{2.93}} & \val{9.26}{1.41} & \val{33.81}{0.81} \\
& TabWak$^\top$ & \val{129.39}{0.90} & \val{131.70}{0.97} & \val{129.83}{1.08} & \val{7.70}{1.04} & \val{-4.86}{1.20} & \val{80.70}{1.03} & \val{15.70}{0.57} \\
& \algo & \textbf{\val{550.05}{1.18}} & \textbf{\val{535.16}{0.96}} & \textbf{\val{525.09}{1.20}} & \underline{\val{83.81}{2.15}} & \val{15.42}{1.02} & \textbf{\val{385.68}{2.48}} & \textbf{\val{85.38}{1.37}} \\

\cmidrule{1-9}

\multirow{6}{*}{ETTh} & TR & \val{6.18}{0.16} & \val{6.14}{0.16} & \val{6.08}{0.14} & \val{24.37}{0.28} & \underline{\val{36.22}{0.21}} & \val{23.02}{0.25} & \underline{\val{55.97}{0.40}} \\
& GS & \underline{\val{327.47}{4.44}} & \underline{\val{322.35}{5.15}} & \textbf{\val{296.89}{5.10}} & \textbf{\val{189.84}{2.98}} & \val{8.49}{1.16} & \textbf{\val{296.61}{4.10}} & \textbf{\val{172.81}{2.05}} \\
& HTW & \val{6.84}{2.22} & \val{6.84}{2.22} & \val{6.84}{2.22} & \val{0.79}{1.90} & \val{-2.94}{1.15} & \val{6.00}{2.16} & \val{2.59}{1.54} \\
& TabWak & \val{-20.57}{0.96} & \val{-18.36}{1.21} & \val{-3.56}{1.91} & \val{37.32}{2.48} & \textbf{\val{78.54}{2.07}} & \val{-7.75}{0.79} & \val{22.65}{0.45} \\
& TabWak$^\top$ & \val{235.03}{1.48} & \val{227.57}{1.33} & \val{194.84}{1.65} & \val{101.64}{1.73} & \val{23.82}{1.45} & \val{155.78}{1.24} & \val{23.70}{0.91} \\
& \algo & \textbf{\val{340.36}{2.06}} & \textbf{\val{333.66}{1.93}} & \underline{\val{276.99}{2.04}} & \underline{\val{107.52}{1.90}} & \val{7.13}{1.14} & \underline{\val{244.68}{1.72}} & \val{44.75}{1.21} \\

\cmidrule{1-9}

\multirow{6}{*}{MuJoCo} & TR & \val{1.31}{0.04} & \val{1.33}{0.04} & \val{1.50}{0.04} & \val{6.93}{0.07} & \val{6.08}{0.05} & \val{10.52}{0.12} & \val{25.61}{0.20} \\
& GS & \underline{\val{39.63}{0.71}} & \underline{\val{38.07}{0.67}} & \underline{\val{32.10}{0.81}} & \underline{\val{37.22}{0.89}} & \val{10.61}{0.85} & \underline{\val{38.46}{0.89}} & \underline{\val{31.58}{1.16}} \\
& HTW & \val{4.20}{1.37} & \val{4.20}{1.37} & \val{4.20}{1.37} & \val{1.73}{1.18} & \val{-2.13}{0.99} & \val{3.78}{1.29} & \val{2.02}{0.90} \\
& TabWak & \val{-3.00}{1.04} & \val{-5.68}{0.92} & \val{-3.91}{1.16} & \val{-8.02}{1.07} & \underline{\val{16.29}{0.61}} & \val{-0.31}{0.87} & \val{2.90}{0.55} \\
& TabWak$^\top$ & \val{3.91}{0.88} & \val{5.10}{1.00} & \val{9.07}{0.97} & \val{3.36}{0.90} & \textbf{\val{24.65}{1.15}} & \val{25.56}{1.00} & \textbf{\val{41.50}{1.00}} \\
& \algo & \textbf{\val{123.36}{1.43}} & \textbf{\val{139.07}{1.49}} & \textbf{\val{111.27}{1.15}} & \textbf{\val{93.19}{1.46}} & \val{12.44}{1.01} & \textbf{\val{71.22}{1.21}} & \val{7.26}{0.95} \\

\cmidrule{1-9}

\multirow{6}{*}{Energy} & TR & \val{22.58}{0.19} & \val{21.92}{0.18} & \val{20.13}{0.17} & \textbf{\val{167.48}{0.64}} & \textbf{\val{207.75}{0.39}} & \underline{\val{68.40}{0.29}} & \textbf{\val{148.79}{0.50}} \\
& GS & \underline{\val{45.42}{1.05}} & \underline{\val{43.41}{0.96}} & \underline{\val{30.53}{0.93}} & \underline{\val{71.60}{0.94}} & \underline{\val{67.39}{1.11}} & \val{44.20}{1.16} & \underline{\val{44.13}{1.18}} \\
& HTW & \val{5.42}{1.00} & \val{5.42}{1.00} & \val{5.42}{1.00} & \val{0.24}{0.62} & \val{-1.11}{0.43} & \val{4.86}{0.95} & \val{2.47}{0.67} \\
& TabWak & \val{0.57}{0.87} & \val{-21.06}{0.96} & \val{-28.01}{1.32} & \val{-25.69}{0.82} & \val{-10.44}{0.76} & \val{-0.07}{0.63} & \val{-7.15}{0.36} \\
& TabWak$^\top$ & \val{26.00}{1.12} & \val{33.08}{0.98} & \val{19.25}{0.95} & \val{9.30}{1.14} & \val{10.84}{1.92} & \val{24.21}{0.95} & \val{33.38}{0.69} \\
& \algo & \textbf{\val{245.37}{2.88}} & \textbf{\val{307.31}{1.98}} & \textbf{\val{183.59}{1.74}} & \val{9.57}{0.97} & \val{2.48}{0.84} & \textbf{\val{171.81}{1.77}} & \val{21.44}{1.10} \\

\cmidrule{1-9}

\multirow{6}{*}{fMRI} & TR & \val{9.94}{0.04} & \val{9.93}{0.04} & \val{9.88}{0.04} & \val{10.47}{0.05} & \val{8.96}{0.04} & \val{16.62}{0.06} & \val{39.21}{0.13} \\
& GS & \underline{\val{701.90}{0.72}} & \val{701.47}{0.65} & \val{699.01}{0.63} & \val{621.08}{1.08} & \underline{\val{256.13}{1.63}} & \underline{\val{667.14}{2.43}} & \textbf{\val{528.66}{2.82}} \\
& HTW & \val{9.43}{0.61} & \val{9.43}{0.61} & \val{9.43}{0.61} & \val{6.75}{0.67} & \val{-1.09}{0.65} & \val{8.24}{0.66} & \val{3.45}{0.57} \\
& TabWak & \val{47.29}{0.83} & \val{43.52}{0.79} & \val{24.95}{0.86} & \val{-57.42}{2.04} & \val{71.03}{3.25} & \val{78.43}{4.67} & \val{0.17}{8.00} \\
& TabWak$^\top$ & \textbf{\val{1031.96}{0.79}} & \textbf{\val{1031.53}{0.78}} & \textbf{\val{1030.14}{0.72}} & \textbf{\val{889.62}{0.84}} & \textbf{\val{383.15}{1.37}} & \textbf{\val{801.16}{0.83}} & \underline{\val{229.37}{1.15}} \\
& \algo & \val{526.81}{13.12} & \underline{\val{834.78}{1.10}} & \underline{\val{834.62}{1.24}} & \underline{\val{632.77}{1.07}} & \val{160.62}{1.02} & \val{651.80}{1.13} & \val{216.50}{0.94} \\

\bottomrule

\end{tabular}}
\end{table}

%% file: src/tables/appx_recon_attack.tex
\begin{table}[ht!]

\centering
\caption{Results of synthetic time series quality and watermark detectability. Comparing \algo and \algo$_{recon}$ under reconstruction attack. Quality metrics and Z-score are for 24-length sequences.}
\label{tab:appx_recon_attack}
\vskip 0.15in

\resizebox{\linewidth}{!}{
\begin{tabular}{ l l c c c c c }

\toprule

Dataset & Method & Context-FID $\downarrow$ & Correlational $\downarrow$ & Discriminative $\downarrow$ & Predictive $\downarrow$ & Z-score $\uparrow$ \\

\midrule

\multirow{2}{*}{Stocks} & \algo & \val{0.277}{0.019} & \val{0.020}{0.018} & \val{0.120}{0.039} & \val{0.038}{0.000} & \val{182.10}{0.73} \\
& \algo$_{recon}$ & \val{4.570}{0.502} & \val{0.031}{0.028} & \val{0.393}{0.073} & \val{0.148}{0.017} & \val{179.41}{0.81} \\

\cmidrule{1-7}

\multirow{2}{*}{ETTh} & \algo & \val{0.237}{0.017} & \val{0.212}{0.043} & \val{0.102}{0.014} & \val{0.122}{0.002} & \val{134.83}{0.95} \\
& \algo$_{recon}$ & \val{1.743}{0.225} & \val{0.138}{0.012} & \val{0.290}{0.009} & \val{0.158}{0.002} & \val{82.11}{2.58} \\

\cmidrule{1-7}

\multirow{2}{*}{MuJoCo} & \algo & \val{0.089}{0.017} & \val{0.532}{0.137} & \val{0.044}{0.021} & \val{0.008}{0.001} & \val{85.69}{1.08} \\
& \algo$_{recon}$ & \val{0.925}{0.047} & \val{0.622}{0.063} & \val{0.261}{0.010} & \val{0.008}{0.002} & \val{81.97}{1.33} \\

\cmidrule{1-7}

\multirow{2}{*}{Energy} & \algo & \val{0.121}{0.016} & \val{1.977}{0.750} & \val{0.142}{0.008} & \val{0.254}{0.000} & \val{231.28}{1.45} \\
& \algo$_{recon}$ & \val{5.158}{0.568} & \val{6.678}{0.087} & \val{0.444}{0.006} & \val{0.263}{0.002} & \val{39.99}{2.03} \\

\cmidrule{1-7}

\multirow{2}{*}{fMRI} & \algo & \val{0.199}{0.010} & \val{2.006}{0.053} & \val{0.122}{0.033} & \val{0.100}{0.000} & \val{379.51}{0.82} \\
& \algo$_{recon}$ & \val{0.595}{0.036} & \val{2.374}{0.105} & \val{0.431}{0.017} & \val{0.102}{0.000} & \val{457.90}{0.81} \\

\bottomrule

\end{tabular}
}
\end{table}

%% file: src/tables/appx_bdia_ddim.tex
\begin{table}[ht!]

\centering
\caption{Results of synthetic time series quality and watermark detectability. All results are applied with BDIA-DDIM. Quality metrics are for 24-length sequences.}
\label{tab:appx_bdia_ddim}
\vskip 0.15in

\resizebox{\linewidth}{!}{
\begin{tabular}{l l c c c c c c c}

\toprule

& & \multicolumn{4}{c}{Quality Metric $\downarrow$} & \multicolumn{3}{c}{Z-score $\uparrow$} \\

\cmidrule(r){3-6}
\cmidrule(l){7-9}

Dataset & Method & Context-FID & Correlational & Discriminative & Predictive & 24-length  & 64-length & 128-length \\

\midrule

\multirow{3}{*}{Stocks} & TabWak & \val{0.267}{0.042} & \val{0.017}{0.014} & \val{0.122}{0.029} & \val{0.039}{0.000} & \val{43.10}{0.75} & \val{92.56}{0.37} & \val{89.18}{0.39} \\
& TabWak$^\top$ & \val{0.273}{0.103} & \val{0.011}{0.008} & \val{0.115}{0.042} & \val{0.037}{0.000} & \val{117.12}{0.15} & \val{170.41}{0.14} & \val{267.27}{0.18} \\
& \algo & \val{0.277}{0.019} & \val{0.020}{0.018} & \val{0.120}{0.039} & \val{0.038}{0.000} & \val{182.10}{0.73} & \val{395.34}{1.24} & \val{550.05}{1.18} \\

\cmidrule{1-9}

\multirow{3}{*}{ETTh} & TabWak & \val{0.431}{0.033} & \val{0.436}{0.020} & \val{0.133}{0.029} & \val{0.134}{0.002} & \val{1.73}{0.76} & \val{16.26}{0.98} & \val{31.59}{0.87} \\
& TabWak$^\top$ & \val{0.454}{0.049} & \val{0.132}{0.018} & \val{0.104}{0.024} & \val{0.119}{0.006} & \val{149.16}{0.62} & \val{220.72}{0.84} & \val{315.34}{1.16} \\
& \algo & \val{0.237}{0.017} & \val{0.212}{0.043} & \val{0.102}{0.014} & \val{0.122}{0.002} & \val{134.83}{0.95} & \val{236.08}{1.63} & \val{340.36}{2.06} \\

\cmidrule{1-9}

\multirow{3}{*}{MuJoCo} & TabWak & \val{0.489}{0.036} & \val{0.958}{0.067} & \val{0.204}{0.056} & \val{0.010}{0.004} & \val{13.97}{1.07} & \val{20.52}{0.85} & \val{4.23}{0.93} \\
& TabWak$^\top$ & \val{0.270}{0.024} & \val{0.378}{0.033} & \val{0.128}{0.015} & \val{0.008}{0.002} & \val{68.32}{1.14} & \val{93.55}{1.06} & \val{228.10}{1.67} \\
& \algo & \val{0.089}{0.017} & \val{0.532}{0.137} & \val{0.044}{0.021} & \val{0.008}{0.001} & \val{85.69}{1.08} & \val{56.45}{1.26} & \val{123.36}{1.43} \\

\cmidrule{1-9}

\multirow{3}{*}{Energy} & TabWak & \val{0.189}{0.022} & \val{2.915}{0.410} & \val{0.166}{0.012} & \val{0.255}{0.000} & \val{3.10}{0.63} & \val{-9.67}{0.72} & \val{-18.64}{0.71} \\
& TabWak$^\top$ & \val{0.199}{0.007} & \val{1.648}{0.188} & \val{0.137}{0.019} & \val{0.265}{0.004} & \val{246.86}{1.30} & \val{258.45}{1.48} & \val{302.78}{1.98} \\
& \algo & \val{0.121}{0.016} & \val{1.977}{0.750} & \val{0.142}{0.008} & \val{0.254}{0.000} & \val{231.28}{1.45} & \val{267.53}{2.60} & \val{245.37}{2.88} \\

\cmidrule{1-9}

\multirow{3}{*}{fMRI} & TabWak & \val{0.319}{0.019} & \val{6.772}{0.129} & \val{0.484}{0.007} & \val{0.110}{0.001} & \val{57.99}{1.06} & \val{268.67}{0.88} & \val{-94.33}{0.99} \\
& TabWak$^\top$ & \val{0.317}{0.028} & \val{2.185}{0.148} & \val{0.218}{0.031} & \val{0.100}{0.000} & \val{471.86}{0.45} & \val{739.39}{0.60} & \val{1014.93}{0.74} \\
& \algo & \val{0.199}{0.010} & \val{2.006}{0.053} & \val{0.122}{0.033} & \val{0.100}{0.000} & \val{379.51}{0.82} & \val{595.68}{1.03} & \val{526.81}{13.12} \\

\bottomrule

\end{tabular}}
\end{table}

%% file: src/tables/appx_forecasting_64.tex
\begin{table}[ht!]

\centering
\caption{Results of 64-length time series forecasting that trains on real and synthetic data (Synth$_{\mathrm{W/O}}$ and Synth$_{\mathrm{\algo}}$) and tests on real data with a 24 timesteps forecast horizon. MSE values $\times 10^{-3}$.}
\label{tab:appx_forecasting_64}
\vskip 0.15in

\begin{tabular}{ l l c }

\toprule

Dataset & Training Data & MSE $\downarrow$ \\

\midrule

\multirow{3}{*}{Stocks} & Real & 2.119 \\
& Synth$_{\mathrm{W/O}}$ & 2.022 \\
& Synth$_{\mathrm{\algo}}$ & 2.014 \\

\cmidrule{1-3}

\multirow{3}{*}{ETTh} & Real & 6.678 \\
& Synth$_{\mathrm{W/O}}$ & 8.541 \\
& Synth$_{\mathrm{\algo}}$ & 8.655 \\

\cmidrule{1-3}

\multirow{3}{*}{MuJoCo} & Real & 1.312 \\
& Synth$_{\mathrm{W/O}}$ & 1.615 \\
& Synth$_{\mathrm{\algo}}$ & 1.741 \\

\cmidrule{1-3}

\multirow{3}{*}{Energy} & Real & 12.715 \\
& Synth$_{\mathrm{W/O}}$ & 13.480 \\
& Synth$_{\mathrm{\algo}}$ & 13.717 \\

\cmidrule{1-3}

\multirow{3}{*}{fMRI} & Real & 36.423 \\
& Synth$_{\mathrm{W/O}}$ & 67.796 \\
& Synth$_{\mathrm{\algo}}$ & 67.944 \\

\bottomrule

\end{tabular}
\end{table}

%% file: src/tables/appx_imputation_64.tex
\begin{table}[ht!]

\centering
\caption{Results of 64-length time series imputation that trains on real and synthetic data (Synth$_{\mathrm{W/O}}$ and Synth$_{\mathrm{\algo}}$) and tests on real data with 70\% missing ratio. MSE values $\times 10^{-3}$.}
\label{tab:appx_imputation_64}
\vskip 0.15in

\begin{tabular}{ l l c }

\toprule

Dataset & Training Data & MSE $\downarrow$ \\

\midrule

\multirow{3}{*}{Stocks} & Real & 1.020 \\
& Synth$_{\mathrm{W/O}}$ & 0.855 \\
& Synth$_{\mathrm{\algo}}$ & 0.858 \\

\cmidrule{1-3}

\multirow{3}{*}{ETTh} & Real & 1.526 \\
& Synth$_{\mathrm{W/O}}$ & 1.842 \\
& Synth$_{\mathrm{\algo}}$ & 1.963 \\

\cmidrule{1-3}

\multirow{3}{*}{MuJoCo} & Real & 0.101 \\
& Synth$_{\mathrm{W/O}}$ & 0.364 \\
& Synth$_{\mathrm{\algo}}$ & 0.387 \\

\cmidrule{1-3}

\multirow{3}{*}{Energy} & Real & 7.926 \\
& Synth$_{\mathrm{W/O}}$ & 8.258 \\
& Synth$_{\mathrm{\algo}}$ & 8.279 \\

\cmidrule{1-3}

\multirow{3}{*}{fMRI} & Real & 27.439 \\
& Synth$_{\mathrm{W/O}}$ & 45.412 \\
& Synth$_{\mathrm{\algo}}$ & 47.349 \\

\bottomrule

\end{tabular}
\end{table}

%% file: src/tables/appx_ablation_methods.tex
\begin{table}[ht!]

\centering
\caption{List of methods, including \algo, to be compared.}
\label{tab:appx_ablation_methods}%
\vskip 0.15in

\begin{tabular}{l c c}

\toprule

Method & Watermarking Direction & Sampling \\

\midrule

SpatDDIM & Spatial & DDIM \\
SpatBDIA & Spatial & BDIA-DDIM \\
TempDDIM & Temporal & DDIM \\
\algo & Temporal & BDIA-DDIM \\

\bottomrule

\end{tabular}%
\end{table}%

%% file: src/tables/appx_ablation.tex
\begin{table}[ht!]

\centering
\caption{Results of synthetic time series quality and watermark detectability. Quality metrics are for 24-length sequences. All methods are originally found in \cref{tab:appx_ablation_methods}. Best results are in \textbf{bold}, and second-best are \underline{underlined}.}
\label{tab:appx_ablation}
\vskip 0.15in

\resizebox{\linewidth}{!}{
\begin{tabular}{l l c c c c c c c}

\toprule

& & \multicolumn{4}{c}{Quality Metric $\downarrow$} & \multicolumn{3}{c}{Z-score $\uparrow$} \\

\cmidrule(r){3-6}
\cmidrule(l){7-9}

Dataset & Method & Context-FID & Correlational & Discriminative & Predictive & 24-length  & 64-length & 128-length \\

\midrule

\multirow{4}{*}{Stocks} & SpatDDIM & \underline{\val{0.233}{0.025}} & \textbf{\val{0.012}{0.005}} & \val{0.127}{0.019} & \textbf{\val{0.037}{0.000}} & \val{11.06}{1.07} & \val{0.18}{0.77} & \val{-1.03}{0.96} \\
& SpatBDIA & \textbf{\val{0.199}{0.024}} & \val{0.024}{0.028} & \underline{\val{0.124}{0.023}} & \textbf{\val{0.037}{0.000}} & \underline{\val{126.97}{1.43}} & \underline{\val{156.84}{0.80}} & \underline{\val{170.13}{1.59}} \\
& TempDDIM & \val{0.277}{0.054} & \underline{\val{0.013}{0.005}} & \val{0.124}{0.043} & \underline{\val{0.038}{0.000}} & \val{25.14}{0.89} & \val{43.24}{0.87} & \val{58.82}{0.93} \\
& \algo & \val{0.277}{0.019} & \val{0.020}{0.018} & \textbf{\val{0.120}{0.039}} & \underline{\val{0.038}{0.000}} & \textbf{\val{182.10}{0.73}} & \textbf{\val{395.34}{1.24}} & \textbf{\val{550.05}{1.18}} \\

\cmidrule{1-9}

\multirow{4}{*}{ETTh} & SpatDDIM & \val{0.249}{0.020} & \textbf{\val{0.145}{0.026}} & \textbf{\val{0.094}{0.011}} & \val{0.124}{0.002} & \val{10.80}{1.06} & \val{11.13}{0.81} & \val{7.29}{0.95} \\
& SpatBDIA & \underline{\val{0.246}{0.015}} & \val{0.150}{0.034} & \val{0.098}{0.011} & \val{0.123}{0.007} & \val{28.11}{1.10} & \val{40.67}{1.21} & \val{47.73}{1.04} \\
& TempDDIM & \val{0.249}{0.013} & \underline{\val{0.150}{0.020}} & \underline{\val{0.097}{0.010}} & \textbf{\val{0.121}{0.005}} & \underline{\val{63.78}{0.97}} & \underline{\val{102.06}{1.16}} & \underline{\val{152.63}{1.29}} \\
& \algo & \textbf{\val{0.237}{0.017}} & \val{0.212}{0.043} & \val{0.102}{0.014} & \underline{\val{0.122}{0.002}} & \textbf{\val{134.83}{0.95}} & \textbf{\val{236.08}{1.63}} & \textbf{\val{340.36}{2.06}} \\

\cmidrule{1-9}

\multirow{4}{*}{MuJoCo} & SpatDDIM & \val{0.091}{0.008} & \underline{\val{0.476}{0.049}} & \val{0.062}{0.027} & \val{0.008}{0.002} & \val{-4.62}{0.91} & \val{15.42}{0.96} & \val{7.12}{1.22} \\
& SpatBDIA & \underline{\val{0.090}{0.016}} & \val{0.491}{0.078} & \underline{\val{0.051}{0.023}} & \val{0.008}{0.002} & \val{11.73}{0.94} & \underline{\val{18.02}{0.95}} & \underline{\val{25.42}{0.90}} \\
& TempDDIM & \val{0.098}{0.010} & \textbf{\val{0.450}{0.047}} & \val{0.059}{0.027} & \textbf{\val{0.007}{0.001}} & \underline{\val{21.85}{0.84}} & \val{-0.88}{0.99} & \val{-2.58}{1.03} \\
& \algo & \textbf{\val{0.089}{0.017}} & \val{0.532}{0.137} & \textbf{\val{0.044}{0.021}} & \underline{\val{0.008}{0.001}} & \textbf{\val{85.69}{1.08}} & \textbf{\val{56.45}{1.26}} & \textbf{\val{123.36}{1.43}} \\

\cmidrule{1-9}

\multirow{4}{*}{Energy} & SpatDDIM & \val{0.135}{0.021} & \underline{\val{1.814}{0.373}} & \underline{\val{0.142}{0.013}} & \textbf{\val{0.253}{0.000}} & \val{1.55}{0.90} & \val{6.75}{1.00} & \val{-0.70}{0.98} \\
& SpatBDIA & \val{0.142}{0.027} & \val{2.104}{0.254} & \val{0.149}{0.025} & \textbf{\val{0.253}{0.000}} & \underline{\val{52.86}{0.90}} & \underline{\val{63.56}{1.26}} & \underline{\val{81.46}{1.11}} \\
& TempDDIM & \textbf{\val{0.110}{0.019}} & \textbf{\val{1.724}{0.270}} & \val{0.142}{0.023} & \underline{\val{0.254}{0.000}} & \val{1.72}{0.92} & \val{3.64}{0.90} & \val{2.24}{0.93} \\
& \algo & \underline{\val{0.121}{0.016}} & \val{1.977}{0.750} & \textbf{\val{0.142}{0.008}} & \underline{\val{0.254}{0.000}} & \textbf{\val{231.28}{1.45}} & \textbf{\val{267.53}{2.60}} & \textbf{\val{245.37}{2.88}} \\

\cmidrule{1-9}

\multirow{4}{*}{fMRI} & SpatDDIM & \val{0.198}{0.023} & \val{2.014}{0.046} & \val{0.139}{0.030} & \val{0.101}{0.001} & \val{90.69}{0.94} & \val{61.12}{0.87} & \val{81.31}{0.70} \\
& SpatBDIA & \textbf{\val{0.188}{0.004}} & \textbf{\val{1.974}{0.074}} & \underline{\val{0.124}{0.035}} & \underline{\val{0.101}{0.000}} & \val{76.48}{0.89} & \val{93.74}{0.82} & \val{75.93}{0.66} \\
& TempDDIM & \underline{\val{0.193}{0.018}} & \val{2.097}{0.086} & \val{0.143}{0.020} & \underline{\val{0.101}{0.000}} & \textbf{\val{381.15}{0.81}} & \textbf{\val{617.91}{0.98}} & \textbf{\val{828.89}{1.01}} \\
& \algo & \val{0.199}{0.010} & \underline{\val{2.006}{0.053}} & \textbf{\val{0.122}{0.033}} & \textbf{\val{0.100}{0.000}} & \underline{\val{379.51}{0.82}} & \underline{\val{595.68}{1.03}} & \underline{\val{526.81}{13.12}} \\

\bottomrule

\end{tabular}}
\end{table}

%% file: src/tables/appx_challenging_dataset.tex
\begin{table}[ht!]

\centering
\caption{Results of synthetic time series quality and watermark detectability for 64-length sequences. No watermarking (`W/O') is included. Best results are in \textbf{bold}, and second-best are \underline{underlined}.}
\label{tab:appx_challenging_dataset}
\vskip 0.15in

\resizebox{\linewidth}{!}{
\begin{tabular}{ l l c c c c c }

\toprule

Dataset & Method & Context-FID $\downarrow$ & Correlational $\downarrow$ & Discriminative $\downarrow$ & Predictive $\downarrow$ & Z-score $\uparrow$ \\

\midrule

\multirow{7}{*}{Illness} & W/O & \val{0.411}{0.040} & \val{0.073}{0.061} & \val{0.147}{0.102} & \val{0.028}{0.002} & - \\
& TR & \val{1.530}{0.177} & \val{0.149}{0.056} & \val{0.286}{0.096} & \val{0.035}{0.003} & \val{5.09}{0.06} \\
& GS & \val{0.734}{0.030} & \val{0.159}{0.035} & \val{0.397}{0.054} & \val{0.030}{0.001} & \underline{\val{78.18}{0.84}} \\
& HTW & \val{0.439}{0.040} & \textbf{\val{0.069}{0.041}} & \val{0.217}{0.060} & \val{0.032}{0.002} & \val{7.37}{0.75} \\
& TabWak & \textbf{\val{0.239}{0.030}} & \underline{\val{0.070}{0.036}} & \val{0.131}{0.132} & \textbf{\val{0.027}{0.001}} & \val{-2.06}{0.88} \\
& TabWak$^\top$ & \val{0.295}{0.045} & \val{0.071}{0.035} & \underline{\val{0.114}{0.041}} & \val{0.028}{0.002} & \val{21.26}{1.13} \\
& \algo & \underline{\val{0.240}{0.009}} & \val{0.076}{0.050} & \textbf{\val{0.111}{0.074}} & \underline{\val{0.028}{0.001}} & \textbf{\val{151.03}{1.60}} \\

\cmidrule{1-7}

\multirow{7}{*}{Weather} & W/O & \val{0.647}{0.079} & \val{1.429}{0.089} & \val{0.175}{0.011} & \val{0.002}{0.000} & - \\
& TR & \val{3.381}{0.577} & \val{2.518}{0.039} & \val{0.388}{0.020} & \textbf{\val{0.002}{0.000}} & \val{0.40}{0.02} \\
& GS & \val{4.495}{0.804} & \val{2.194}{0.140} & \val{0.446}{0.008} & \textbf{\val{0.002}{0.000}} & \val{15.36}{0.92} \\
& HTW & \underline{\val{0.712}{0.062}} & \val{1.463}{0.098} & \val{0.190}{0.013} & \textbf{\val{0.002}{0.000}} & \val{4.82}{0.83} \\
& TabWak & \val{0.751}{0.088} & \val{1.571}{0.168} & \val{0.200}{0.007} & \textbf{\val{0.002}{0.000}} & \val{-1.23}{1.00} \\
& TabWak$^\top$ & \textbf{\val{0.588}{0.081}} & \underline{\val{1.369}{0.089}} & \textbf{\val{0.178}{0.012}} & \textbf{\val{0.002}{0.000}} & \underline{\val{39.58}{0.89}} \\
& \algo & \val{0.717}{0.071} & \textbf{\val{0.951}{0.088}} & \underline{\val{0.184}{0.007}} & \textbf{\val{0.002}{0.000}} & \textbf{\val{205.53}{1.86}} \\

\bottomrule

\end{tabular}
}
\end{table}

%% file: src/tables/appx_tprs.tex
\newcommand{\plot}[1]{\raisebox{-0.5\height}{\includegraphics[width=0.195\linewidth]{#1}}}

\begin{figure}[ht!]
\begin{center}
\setlength{\tabcolsep}{0pt}
\renewcommand{\arraystretch}{0}

\begin{tabular}{ m{1em} c c c c c }

& \small Stocks & \small ETTh & \small MuJoCo & \small Energy & \small fMRI \\

\rotatebox{90}{\small 24-length} & \plot{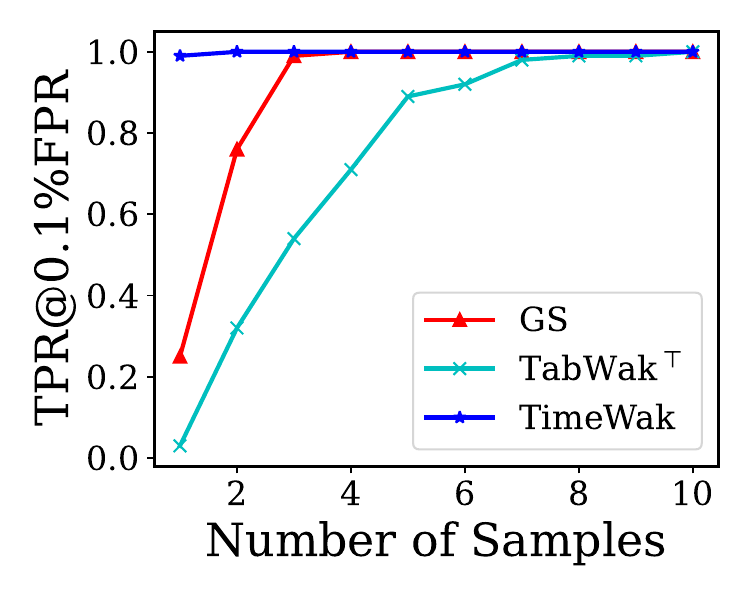} & \plot{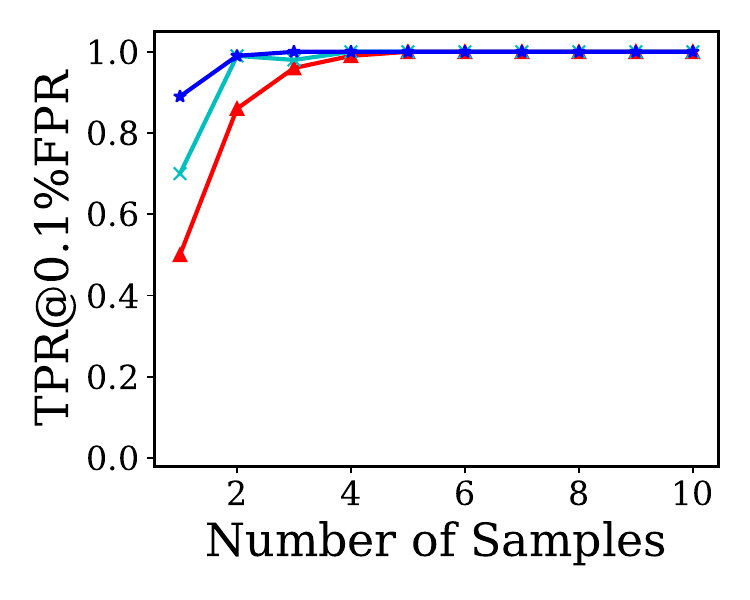} & \plot{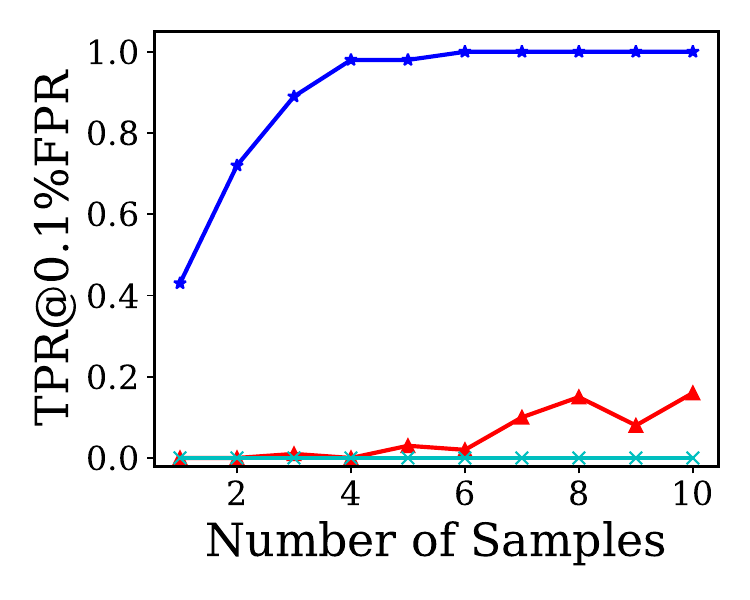} & \plot{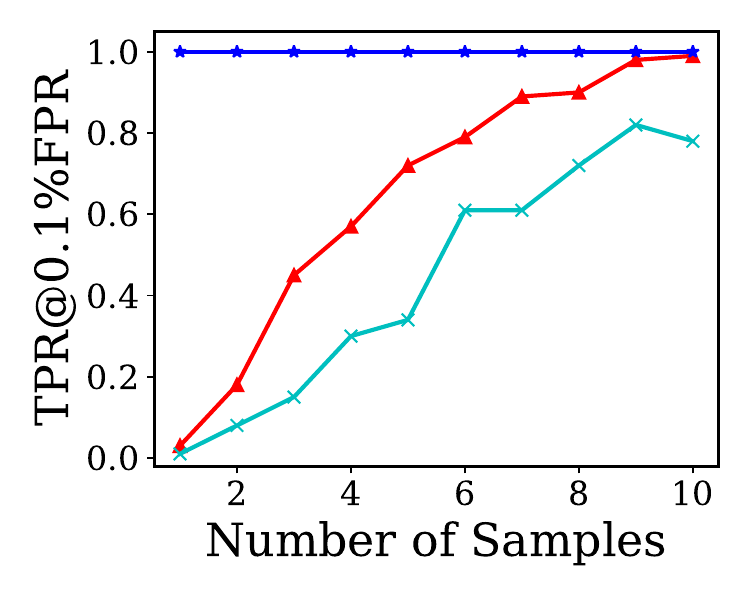} & \plot{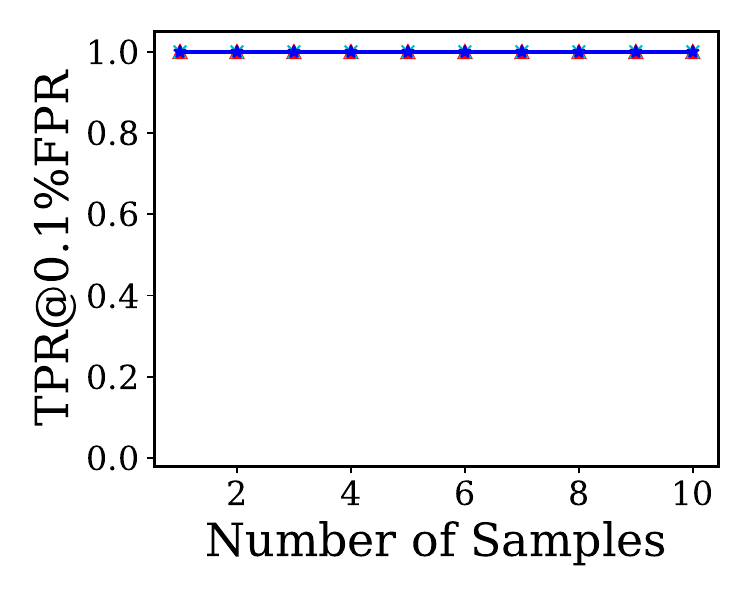} \\

\rotatebox{90}{\small 64-length} & \plot{src/figures/tprs/tpr-stocks-64.pdf} & \plot{src/figures/tprs/tpr-etth-64.pdf} & \plot{src/figures/tprs/tpr-mujoco-64.pdf} & \plot{src/figures/tprs/tpr-energy-64.pdf} & \plot{src/figures/tprs/tpr-fmri-64.pdf} \\

\rotatebox{90}{\small 128-length} & \plot{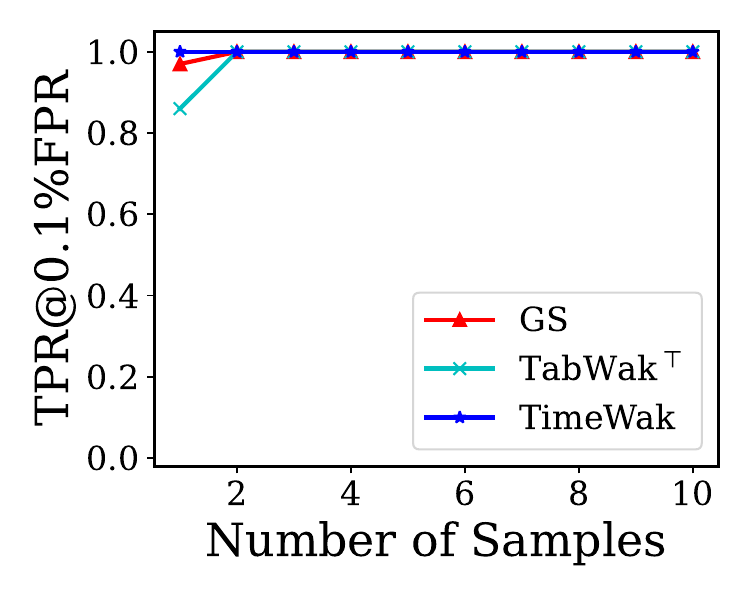} & \plot{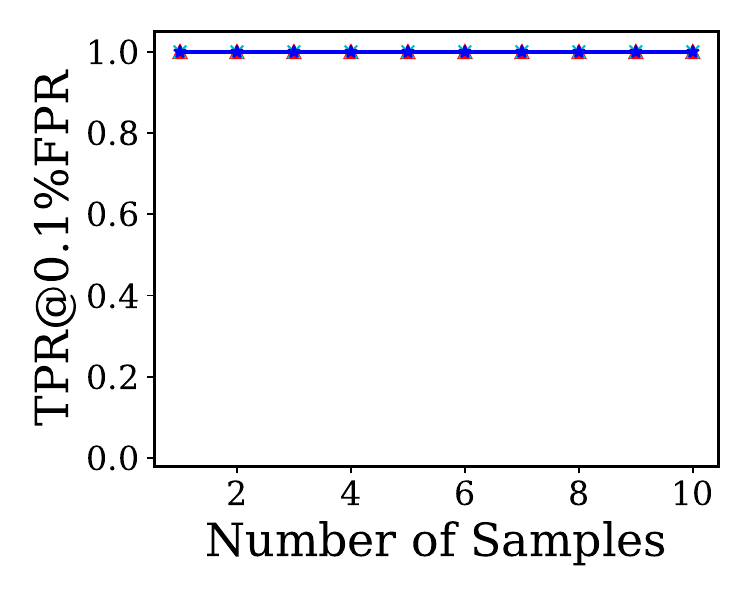} & \plot{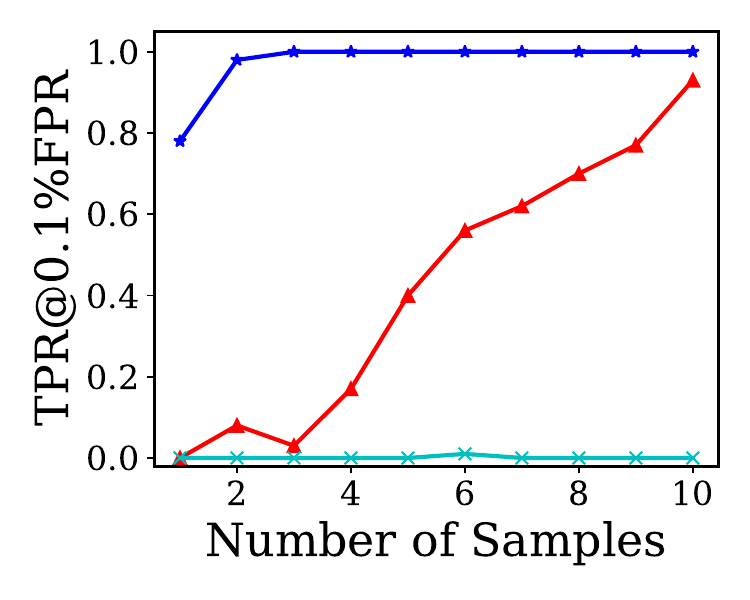} & \plot{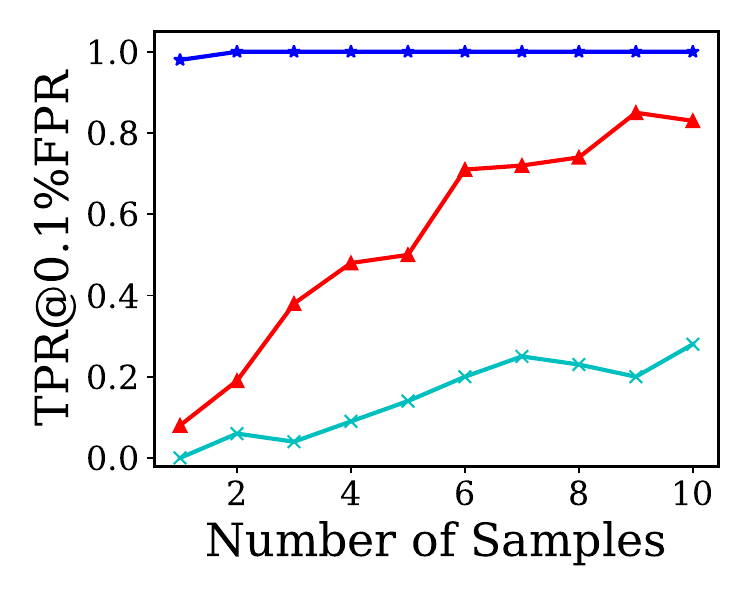} & \plot{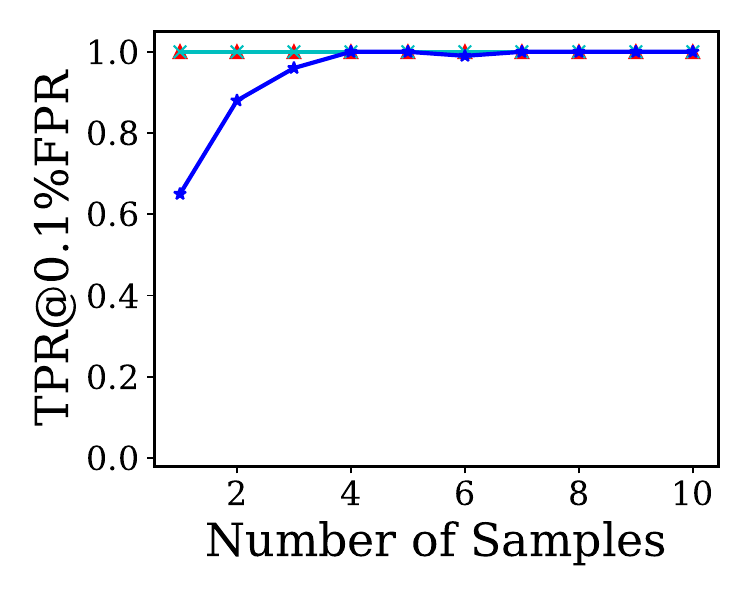} \\

\end{tabular}

\end{center}
\caption{TPR@0.1\%FPR against the number of samples across five datasets under different window sizes.}
\label{fig:appx_tprs}
\end{figure}

%% file: src/tables/appx_tpr_mixed_dataset.tex
\begin{table}[ht!]

\centering
\caption{Results of TPR@0.1\%FPR on mixed dataset when the number of samples is 1, 10, and 20.}
\label{tab:appx_tpr_mixed_dataset}
\vskip 0.15in

\begin{tabular}{llccc}

\toprule

Dataset & Method & 1 $\uparrow$ & 10 $\uparrow$ & 20 $\uparrow$ \\

\midrule

\multirow{3}{*}{Stocks} & GS & \textbf{0.33} & 0.87 & \textbf{0.99} \\
& TabWak$^\top$ & 0.13 & 0.47 & 0.68 \\
& \algo & 0.22 & \textbf{0.96} & \textbf{0.99} \\

\cmidrule{1-5}

\multirow{3}{*}{ETTh} & GS & 0.43 & \textbf{0.98} & \textbf{1.0} \\
& TabWak$^\top$ & \textbf{0.49} & 0.91 & 0.99 \\
& \algo & 0.38 & 0.92 & \textbf{1.0} \\

\cmidrule{1-5}

\multirow{3}{*}{MuJoCo} & GS & 0.0 & 0.0 & 0.0 \\
& TabWak$^\top$ & 0.0 & 0.0 & 0.0 \\
& \algo & \textbf{0.34} & \textbf{0.84} & \textbf{0.99} \\

\cmidrule{1-5}

\multirow{3}{*}{Energy} & GS & \textbf{0.42} & \textbf{1.0} & \textbf{1.0} \\
& TabWak$^\top$ & 0.01 & 0.0 & 0.0 \\
& \algo & 0.32 & 0.97 & \textbf{1.0} \\

\cmidrule{1-5}

\multirow{3}{*}{fMRI} & GS & \textbf{0.4} & 0.98 & \textbf{1.0} \\
& TabWak$^\top$ & 0.31 & \textbf{1.0} & \textbf{1.0} \\
& \algo & 0.26 & 0.97 & \textbf{1.0} \\

\bottomrule

\end{tabular}

\end{table}

%% file: src/tables/appx_characteristics_data.tex
\begin{table}[ht!]

\centering
\caption{Results of Marginal Distribution Difference (MDD), AutoCorrelation Difference (ACD), Skewness Difference (SD), and Kurtosis Difference (KD) for 64-length sequences.}
\label{tab:appx_characteristics_data}
\vskip 0.15in

\begin{tabular}{llcccc}

\toprule

Dataset & Method & MDD $\downarrow$ & ACD $\downarrow$ & SD $\downarrow$ & KD $\downarrow$ \\

\midrule

\multirow{7}{*}{Stocks} & W/O & 0.491 & 0.044 & 0.473 & 1.522 \\
& TR & 0.881 & 0.445 & 0.759 & 4.408 \\
& GS & 1.063 & 0.444 & 0.727 & 4.889 \\
& HTW & 0.491 & \textbf{0.044} & 0.473 & 1.522 \\
& TabWak & 0.470 & 0.078 & 0.180 & 0.544 \\
& TabWak$^\top$ & 0.478 & 0.098 & \textbf{0.060} & 0.852 \\
& \algo & \textbf{0.431} & 0.068 & 0.103 & \textbf{0.295} \\

\cmidrule{1-6}

\multirow{7}{*}{ETTh} & W/O & 0.176 & 0.421 & 0.255 & 1.359 \\
& TR & 0.546 & 1.128 & 0.633 & 3.474 \\
& GS & 0.712 & 1.581 & 0.723 & 2.725 \\
& HTW & 0.384 & \textbf{0.459} & 0.276 & 1.468 \\
& TabWak & 0.211 & 0.532 & 0.245 & 1.422 \\
& TabWak$^\top$ & \textbf{0.183} & 0.753 & \textbf{0.141} & \textbf{0.707} \\
& \algo & 0.184 & 0.511 & 0.197 & 1.109 \\

\cmidrule{1-6}

\multirow{7}{*}{MuJoCo} & W/O & 0.379 & 0.225 & 0.062 & 0.306 \\
& TR & 1.296 & 1.494 & 0.426 & 1.555 \\
& GS & 1.500 & 1.550 & 0.533 & 1.221 \\
& HTW & 0.941 & 0.434 & 0.076 & \textbf{0.287} \\
& TabWak & 0.420 & \textbf{0.262} & 0.087 & 0.311 \\
& TabWak$^\top$ & 0.521 & 0.532 & 0.084 & 0.351 \\
& \algo & \textbf{0.394} & 0.300 & \textbf{0.069} & 0.340 \\

\cmidrule{1-6}

\multirow{7}{*}{Energy} & W/O & 0.188 & 0.157 & 0.112 & 0.703 \\
& TR & 0.558 & 0.606 & 0.355 & 1.681 \\
& GS & 0.660 & 0.958 & 0.373 & 1.015 \\
& HTW & 0.363 & \textbf{0.148} & \textbf{0.108} & 0.701 \\
& TabWak & 0.235 & 0.270 & 0.126 & 0.618 \\
& TabWak$^\top$ & 0.263 & 0.336 & 0.112 & 0.682 \\
& \algo & \textbf{0.215} & 0.241 & 0.111 & \textbf{0.601} \\

\cmidrule{1-6}

\multirow{7}{*}{fMRI} & W/O & 0.099 & 0.153 & 0.041 & 0.128 \\
& TR & 0.447 & 1.356 & 0.066 & 0.428 \\
& GS & 0.818 & 1.435 & 0.096 & 0.174 \\
& HTW & 0.452 & 0.151 & 0.048 & 0.133 \\
& TabWak & 0.126 & 0.159 & 0.043 & 0.138 \\
& TabWak$^\top$ & 0.127 & 0.261 & \textbf{0.040} & \textbf{0.114} \\
& \algo & \textbf{0.120} & \textbf{0.148} & \textbf{0.040} & 0.132 \\

\bottomrule

\end{tabular}

\end{table}

%% file: src/tables/appx_intervals_24.tex
\begin{table}[ht!]

\centering
\caption{Results of synthetic time series quality and watermark detectability with different intervals on \algo. Quality metrics and Z-score are for 24-length sequences.}
\label{tab:appx_intervals_24}
\vskip 0.15in

\resizebox{\linewidth}{!}{
\begin{tabular}{l c c c c c c}

\toprule

Dataset & Interval & Context-FID $\downarrow$ & Correlational $\downarrow$ & Discriminative $\downarrow$ & Predictive $\downarrow$ & Z-score $\uparrow$ \\

\midrule

\multirow{3}{*}{Stocks} & 2 & \val{0.277}{0.019} & \val{0.020}{0.018} & \val{0.120}{0.039} & \val{0.038}{0.000} & \val{182.10}{0.73} \\
& 4 & \val{0.419}{0.098} & \val{0.006}{0.002} & \val{0.162}{0.022} & \val{0.039}{0.000} & \val{202.45}{1.11} \\
& 8 & \val{1.006}{0.085} & \val{0.023}{0.018} & \val{0.197}{0.015} & \val{0.041}{0.000} & \val{216.56}{1.43} \\

\cmidrule{1-7}

\multirow{3}{*}{ETTh} & 2 & \val{0.237}{0.017} & \val{0.212}{0.043} & \val{0.102}{0.014} & \val{0.122}{0.002} & \val{134.83}{0.95} \\
& 4 & \val{0.463}{0.094} & \val{0.435}{0.047} & \val{0.113}{0.018} & \val{0.130}{0.001} & \val{166.50}{1.27} \\
& 8 & \val{0.925}{0.140} & \val{0.576}{0.082} & \val{0.150}{0.014} & \val{0.135}{0.005} & \val{167.54}{1.36} \\

\cmidrule{1-7}

\multirow{3}{*}{MuJoCo} & 2 & \val{0.089}{0.017} & \val{0.532}{0.137} & \val{0.044}{0.021} & \val{0.008}{0.001} & \val{85.69}{1.08} \\
& 4 & \val{0.148}{0.031} & \val{0.739}{0.086} & \val{0.087}{0.020} & \val{0.007}{0.000} & \val{71.95}{1.36} \\
& 8 & \val{0.327}{0.059} & \val{1.179}{0.168} & \val{0.177}{0.019} & \val{0.009}{0.002} & \val{76.90}{1.29} \\

\cmidrule{1-7}

\multirow{3}{*}{Energy} & 2 & \val{0.121}{0.016} & \val{1.977}{0.750} & \val{0.142}{0.008} & \val{0.254}{0.000} & \val{231.28}{1.45} \\
& 4 & \val{0.186}{0.017} & \val{3.315}{0.395} & \val{0.159}{0.011} & \val{0.254}{0.000} & \val{266.77}{1.99} \\
& 8 & \val{0.363}{0.093} & \val{5.402}{0.499} & \val{0.184}{0.024} & \val{0.255}{0.000} & \val{279.40}{1.90} \\

\cmidrule{1-7}

\multirow{3}{*}{fMRI} & 2 & \val{0.199}{0.010} & \val{2.006}{0.053} & \val{0.122}{0.033} & \val{0.100}{0.000} & \val{379.51}{0.82} \\
& 4 & \val{0.191}{0.013} & \val{2.117}{0.124} & \val{0.125}{0.034} & \val{0.101}{0.000} & \val{464.70}{0.92} \\
& 8 & \val{0.204}{0.021} & \val{2.354}{0.110} & \val{0.171}{0.043} & \val{0.103}{0.000} & \val{506.95}{1.10} \\

\bottomrule

\end{tabular}
}
\end{table}

%% file: src/tables/appx_intervals_64.tex
\begin{table}[ht!]

\centering
\caption{Results of synthetic time series quality and watermark detectability with different intervals on \algo. Quality metrics and Z-score are for 64-length sequences.}
\label{tab:appx_intervals_64}
\vskip 0.15in

\resizebox{\linewidth}{!}{
\begin{tabular}{l c c c c c c}

\toprule

Dataset & Interval & Context-FID $\downarrow$ & Correlational $\downarrow$ & Discriminative $\downarrow$ & Predictive $\downarrow$ & Z-score $\uparrow$ \\

\midrule

\multirow{3}{*}{Stocks} & 2 & \val{0.387}{0.054} & \val{0.017}{0.017} & \val{0.092}{0.041} & \val{0.037}{0.000} & \val{395.34}{1.24} \\
& 4 & \val{0.466}{0.099} & \val{0.021}{0.006} & \val{0.077}{0.025} & \val{0.037}{0.000} & \val{397.68}{1.31} \\
& 8 & \val{1.053}{0.099} & \val{0.017}{0.013} & \val{0.155}{0.037} & \val{0.037}{0.000} & \val{406.56}{1.77} \\

\cmidrule{1-7}

\multirow{3}{*}{ETTh} & 2 & \val{0.297}{0.038} & \val{0.133}{0.040} & \val{0.097}{0.015} & \val{0.115}{0.003} & \val{236.08}{1.63} \\
& 4 & \val{0.514}{0.027} & \val{0.335}{0.040} & \val{0.098}{0.031} & \val{0.119}{0.008} & \val{272.70}{1.79} \\
& 8 & \val{0.911}{0.048} & \val{0.540}{0.030} & \val{0.147}{0.038} & \val{0.123}{0.008} & \val{268.14}{2.00} \\

\cmidrule{1-7}

\multirow{3}{*}{MuJoCo} & 2 & \val{0.108}{0.014} & \val{0.413}{0.062} & \val{0.038}{0.021} & \val{0.007}{0.001} & \val{56.45}{1.26} \\
& 4 & \val{0.205}{0.024} & \val{0.522}{0.024} & \val{0.088}{0.031} & \val{0.007}{0.002} & \val{58.75}{1.18} \\
& 8 & \val{0.338}{0.068} & \val{0.768}{0.077} & \val{0.159}{0.020} & \val{0.007}{0.001} & \val{68.29}{1.19} \\

\cmidrule{1-7}

\multirow{3}{*}{Energy} & 2 & \val{0.143}{0.019} & \val{1.662}{0.298} & \val{0.145}{0.019} & \val{0.251}{0.000} & \val{267.53}{2.60} \\
& 4 & \val{0.195}{0.017} & \val{2.760}{0.504} & \val{0.150}{0.011} & \val{0.252}{0.000} & \val{289.07}{2.77} \\
& 8 & \val{0.407}{0.087} & \val{4.285}{0.229} & \val{0.170}{0.025} & \val{0.252}{0.000} & \val{345.97}{3.48} \\

\cmidrule{1-7}

\multirow{3}{*}{fMRI} & 2 & \val{0.441}{0.035} & \val{1.786}{0.043} & \val{0.314}{0.041} & \val{0.100}{0.000} & \val{595.68}{1.03} \\
& 4 & \val{0.425}{0.027} & \val{1.834}{0.122} & \val{0.273}{0.076} & \val{0.100}{0.000} & \val{749.10}{1.08} \\
& 8 & \val{0.469}{0.027} & \val{1.823}{0.065} & \val{0.294}{0.141} & \val{0.100}{0.000} & \val{817.23}{1.20} \\

\bottomrule

\end{tabular}
}
\end{table}

%% file: src/tables/appx_intervals_128.tex
\begin{table}[ht!]

\centering
\caption{Results of synthetic time series quality and watermark detectability with different intervals on \algo. Quality metrics and Z-score are for 128-length sequences.}
\label{tab:appx_intervals_128}
\vskip 0.15in

\resizebox{\linewidth}{!}{
\begin{tabular}{l c c c c c c}

\toprule

Dataset & Interval & Context-FID $\downarrow$ & Correlational $\downarrow$ & Discriminative $\downarrow$ & Predictive $\downarrow$ & Z-score $\uparrow$ \\

\midrule

\multirow{3}{*}{Stocks} & 2 & \val{0.316}{0.044} & \val{0.021}{0.024} & \val{0.140}{0.029} & \val{0.037}{0.000} & \val{550.05}{1.18} \\
& 4 & \val{0.410}{0.104} & \val{0.019}{0.017} & \val{0.140}{0.067} & \val{0.037}{0.000} & \val{571.54}{1.22} \\
& 8 & \val{1.132}{0.551} & \val{0.035}{0.021} & \val{0.217}{0.019} & \val{0.038}{0.000} & \val{599.41}{1.39} \\

\cmidrule{1-7}

\multirow{3}{*}{ETTh} & 2 & \val{1.090}{0.100} & \val{0.135}{0.057} & \val{0.174}{0.007} & \val{0.110}{0.009} & \val{340.36}{2.06} \\
& 4 & \val{1.445}{0.119} & \val{0.333}{0.034} & \val{0.173}{0.026} & \val{0.114}{0.003} & \val{374.69}{2.39} \\
& 8 & \val{1.838}{0.099} & \val{0.497}{0.054} & \val{0.173}{0.021} & \val{0.116}{0.003} & \val{392.83}{2.39} \\

\cmidrule{1-7}

\multirow{3}{*}{MuJoCo} & 2 & \val{0.155}{0.016} & \val{0.316}{0.022} & \val{0.046}{0.030} & \val{0.005}{0.001} & \val{123.36}{1.43} \\
& 4 & \val{0.172}{0.038} & \val{0.410}{0.057} & \val{0.083}{0.011} & \val{0.006}{0.000} & \val{178.16}{1.60} \\
& 8 & \val{0.330}{0.077} & \val{0.685}{0.055} & \val{0.124}{0.026} & \val{0.006}{0.002} & \val{170.77}{1.93} \\

\cmidrule{1-7}

\multirow{3}{*}{Energy} & 2 & \val{0.148}{0.027} & \val{1.687}{0.328} & \val{0.140}{0.057} & \val{0.249}{0.000} & \val{245.37}{2.88} \\
& 4 & \val{0.261}{0.025} & \val{3.246}{0.345} & \val{0.114}{0.030} & \val{0.250}{0.001} & \val{354.38}{2.32} \\
& 8 & \val{0.506}{0.114} & \val{4.550}{0.745} & \val{0.147}{0.009} & \val{0.251}{0.000} & \val{395.20}{2.91} \\

\cmidrule{1-7}

\multirow{3}{*}{fMRI} & 2 & \val{0.855}{0.072} & \val{1.704}{0.060} & \val{0.298}{0.227} & \val{0.100}{0.000} & \val{526.81}{13.12} \\
& 4 & \val{0.884}{0.124} & \val{1.708}{0.050} & \val{0.348}{0.201} & \val{0.100}{0.000} & \val{1022.29}{1.75} \\
& 8 & \val{0.866}{0.075} & \val{1.698}{0.050} & \val{0.345}{0.195} & \val{0.100}{0.000} & \val{1097.78}{1.57} \\

\bottomrule

\end{tabular}
}
\end{table}

%% file: src/tables/appx_bits_24.tex
\begin{table}[ht!]

\centering
\caption{Results of synthetic time series quality and watermark detectability with different bits on \algo. Quality metrics and Z-score are for 24-length sequences.}
\label{tab:appx_bits_24}
\vskip 0.15in

\resizebox{\linewidth}{!}{
\begin{tabular}{l c c c c c c }

\toprule

Dataset & Bit & Context-FID $\downarrow$ & Correlational $\downarrow$ & Discriminative $\downarrow$ & Predictive $\downarrow$ & Z-score $\uparrow$ \\

\midrule

\multirow{3}{*}{Stocks} & 2 & \val{0.277}{0.019} & \val{0.020}{0.018} & \val{0.120}{0.039} & \val{0.038}{0.000} & \val{182.10}{0.73} \\
& 3 & \val{0.214}{0.039} & \val{0.024}{0.019} & \val{0.130}{0.033} & \val{0.038}{0.000} & \val{194.87}{0.56} \\
& 4 & \val{0.328}{0.110} & \val{0.023}{0.026} & \val{0.155}{0.027} & \val{0.038}{0.000} & \val{182.58}{0.33} \\

\cmidrule{1-7}

\multirow{3}{*}{ETTh} & 2 & \val{0.237}{0.017} & \val{0.212}{0.043} & \val{0.102}{0.014} & \val{0.122}{0.002} & \val{134.83}{0.95} \\
& 3 & \val{0.211}{0.020} & \val{0.206}{0.031} & \val{0.093}{0.003} & \val{0.124}{0.003} & \val{162.25}{0.89} \\
& 4 & \val{0.238}{0.025} & \val{0.225}{0.043} & \val{0.095}{0.016} & \val{0.124}{0.001} & \val{149.74}{0.66} \\

\cmidrule{1-7}

\multirow{3}{*}{MuJoCo} & 2 & \val{0.089}{0.017} & \val{0.532}{0.137} & \val{0.044}{0.021} & \val{0.008}{0.001} & \val{85.69}{1.08} \\
& 3 & \val{0.092}{0.022} & \val{0.520}{0.105} & \val{0.054}{0.014} & \val{0.008}{0.001} & \val{73.09}{1.29} \\
& 4 & \val{0.099}{0.019} & \val{0.524}{0.079} & \val{0.056}{0.013} & \val{0.007}{0.000} & \val{67.67}{1.23} \\

\cmidrule{1-7}

\multirow{3}{*}{Energy} & 2 & \val{0.121}{0.016} & \val{1.977}{0.750} & \val{0.142}{0.008} & \val{0.254}{0.000} & \val{231.28}{1.45} \\
& 3 & \val{0.121}{0.014} & \val{1.799}{0.395} & \val{0.156}{0.023} & \val{0.254}{0.001} & \val{268.24}{1.69} \\
& 4 & \val{0.143}{0.015} & \val{1.721}{0.347} & \val{0.155}{0.010} & \val{0.254}{0.000} & \val{269.83}{1.60} \\

\cmidrule{1-7}

\multirow{3}{*}{fMRI} & 2 & \val{0.199}{0.010} & \val{2.006}{0.053} & \val{0.122}{0.033} & \val{0.100}{0.000} & \val{379.51}{0.82} \\
& 3 & \val{0.195}{0.008} & \val{1.987}{0.076} & \val{0.113}{0.031} & \val{0.101}{0.000} & \val{456.02}{0.67} \\
& 4 & \val{0.183}{0.012} & \val{2.032}{0.030} & \val{0.111}{0.026} & \val{0.101}{0.000} & \val{440.88}{0.55} \\

\bottomrule

\end{tabular}
}
\end{table}

%% file: src/tables/appx_bits_64.tex
\begin{table}[ht!]

\centering
\caption{Results of synthetic time series quality and watermark detectability with different bits on \algo. Quality metrics and Z-score are for 64-length sequences.}
\label{tab:appx_bits_64}
\vskip 0.15in

\resizebox{\linewidth}{!}{
\begin{tabular}{l c c c c c c}

\toprule

Dataset & Bit & Context-FID $\downarrow$ & Correlational $\downarrow$ & Discriminative $\downarrow$ & Predictive $\downarrow$ & Z-score $\uparrow$ \\

\midrule

\multirow{3}{*}{Stocks} & 2 & \val{0.387}{0.054} & \val{0.017}{0.017} & \val{0.092}{0.041} & \val{0.037}{0.000} & \val{395.34}{1.24} \\
& 3 & \val{0.312}{0.046} & \val{0.014}{0.006} & \val{0.121}{0.010} & \val{0.037}{0.000} & \val{334.15}{0.49} \\
& 4 & \val{0.251}{0.053} & \val{0.014}{0.018} & \val{0.095}{0.022} & \val{0.037}{0.000} & \val{309.59}{0.33} \\

\cmidrule{1-7}

\multirow{3}{*}{ETTh} & 2 & \val{0.297}{0.038} & \val{0.133}{0.040} & \val{0.097}{0.015} & \val{0.115}{0.003} & \val{236.08}{1.63} \\
& 3 & \val{0.369}{0.043} & \val{0.182}{0.036} & \val{0.102}{0.013} & \val{0.117}{0.003} & \val{249.67}{1.41} \\
& 4 & \val{0.365}{0.031} & \val{0.185}{0.030} & \val{0.102}{0.010} & \val{0.113}{0.007} & \val{261.13}{1.20} \\

\cmidrule{1-7}

\multirow{3}{*}{MuJoCo} & 2 & \val{0.108}{0.014} & \val{0.413}{0.062} & \val{0.038}{0.021} & \val{0.007}{0.001} & \val{56.45}{1.26} \\
& 3 & \val{0.136}{0.012} & \val{0.423}{0.051} & \val{0.073}{0.018} & \val{0.007}{0.002} & \val{84.07}{1.48} \\
& 4 & \val{0.126}{0.017} & \val{0.381}{0.063} & \val{0.036}{0.030} & \val{0.007}{0.001} & \val{96.73}{1.45} \\

\cmidrule{1-7}

\multirow{3}{*}{Energy} & 2 & \val{0.143}{0.019} & \val{1.662}{0.298} & \val{0.145}{0.019} & \val{0.251}{0.000} & \val{267.53}{2.60} \\
& 3 & \val{0.182}{0.047} & \val{1.284}{0.400} & \val{0.165}{0.019} & \val{0.251}{0.000} & \val{323.04}{2.37} \\
& 4 & \val{0.143}{0.010} & \val{1.460}{0.354} & \val{0.152}{0.023} & \val{0.251}{0.000} & \val{322.38}{2.25} \\

\cmidrule{1-7}

\multirow{3}{*}{fMRI} & 2 & \val{0.441}{0.035} & \val{1.786}{0.043} & \val{0.314}{0.041} & \val{0.100}{0.000} & \val{595.68}{1.03} \\
& 3 & \val{0.423}{0.024} & \val{1.782}{0.082} & \val{0.216}{0.175} & \val{0.100}{0.000} & \val{724.69}{0.85} \\
& 4 & \val{0.440}{0.027} & \val{1.783}{0.033} & \val{0.256}{0.106} & \val{0.100}{0.000} & \val{712.67}{0.59} \\

\bottomrule

\end{tabular}
}
\end{table}

%% file: src/tables/appx_bits_128.tex
\begin{table}[ht!]

\centering
\caption{Results of synthetic time series quality and watermark detectability with different bits on \algo. Quality metrics and Z-score are for 128-length sequences.}
\label{tab:appx_bits_128}
\vskip 0.15in

\resizebox{\linewidth}{!}{
\begin{tabular}{l c c c c c c}

\toprule

Dataset & Bit & Context-FID $\downarrow$ & Correlational $\downarrow$ & Discriminative $\downarrow$ & Predictive $\downarrow$ & Z-score $\uparrow$ \\

\midrule

\multirow{3}{*}{Stocks} & 2 & \val{0.316}{0.044} & \val{0.021}{0.024} & \val{0.140}{0.029} & \val{0.037}{0.000} & \val{550.05}{1.18} \\
& 3 & \val{0.380}{0.059} & \val{0.019}{0.015} & \val{0.176}{0.046} & \val{0.037}{0.000} & \val{459.23}{0.45} \\
& 4 & \val{0.391}{0.087} & \val{0.017}{0.020} & \val{0.134}{0.060} & \val{0.037}{0.000} & \val{427.53}{0.23} \\

\cmidrule{1-7}

\multirow{3}{*}{ETTh} & 2 & \val{1.090}{0.100} & \val{0.135}{0.057} & \val{0.174}{0.007} & \val{0.110}{0.009} & \val{340.36}{2.06} \\
& 3 & \val{1.111}{0.137} & \val{0.151}{0.040} & \val{0.153}{0.010} & \val{0.118}{0.005} & \val{352.26}{1.63} \\
& 4 & \val{1.173}{0.131} & \val{0.233}{0.058} & \val{0.166}{0.013} & \val{0.113}{0.006} & \val{362.55}{1.39} \\

\cmidrule{1-7}

\multirow{3}{*}{MuJoCo} & 2 & \val{0.155}{0.016} & \val{0.316}{0.022} & \val{0.046}{0.030} & \val{0.005}{0.001} & \val{123.36}{1.43} \\
& 3 & \val{0.183}{0.029} & \val{0.317}{0.068} & \val{0.062}{0.011} & \val{0.005}{0.000} & \val{183.47}{1.55} \\
& 4 & \val{0.150}{0.013} & \val{0.349}{0.028} & \val{0.051}{0.031} & \val{0.006}{0.001} & \val{174.45}{1.55} \\

\cmidrule{1-7}

\multirow{3}{*}{Energy} & 2 & \val{0.148}{0.027} & \val{1.687}{0.328} & \val{0.140}{0.057} & \val{0.249}{0.000} & \val{245.37}{2.88} \\
& 3 & \val{0.230}{0.037} & \val{1.154}{0.446} & \val{0.166}{0.054} & \val{0.249}{0.001} & \val{380.35}{2.49} \\
& 4 & \val{0.167}{0.014} & \val{1.700}{0.524} & \val{0.168}{0.069} & \val{0.249}{0.001} & \val{389.26}{2.01} \\

\cmidrule{1-7}

\multirow{3}{*}{fMRI} & 2 & \val{0.855}{0.072} & \val{1.704}{0.060} & \val{0.298}{0.227} & \val{0.100}{0.000} & \val{526.81}{13.12} \\
& 3 & \val{0.819}{0.010} & \val{1.688}{0.049} & \val{0.374}{0.193} & \val{0.100}{0.000} & \val{986.17}{1.08} \\
& 4 & \val{0.828}{0.053} & \val{1.713}{0.020} & \val{0.336}{0.209} & \val{0.100}{0.000} & \val{967.53}{0.86} \\

\bottomrule

\end{tabular}
}
\end{table}

%% file: src/tables/appx_overhead.tex
\begin{table}[ht!]

\centering
\caption{Watermark detection overhead in seconds for \algo when the batch size is 1 and 100.}
\label{tab:appx_overhead}
\vskip 0.15in

\begin{tabular}{lccc}

\toprule

Dataset & Window Size & 1 & 100 \\

\midrule

\multirow{3}{*}{Stocks} & 24 & 1.58 & 2.09 \\
& 64 & 1.50 & 2.27 \\
& 128 & 1.55 & 2.59 \\

\cmidrule{1-4}

\multirow{3}{*}{ETTh} & 24 & 1.72 & 2.23 \\
& 64 & 1.64 & 2.41 \\
& 128 & 1.63 & 2.62 \\

\cmidrule{1-4}

\multirow{3}{*}{MuJoCo} & 24 & 2.96 & 3.84 \\
& 64 & 2.90 & 4.25 \\
& 128 & 2.90 & 4.49 \\

\cmidrule{1-4}

\multirow{3}{*}{Energy} & 24 & 3.89 & 5.22 \\
& 64 & 3.88 & 5.73 \\
& 128 & 3.89 & 6.58 \\

\cmidrule{1-4}

\multirow{3}{*}{fMRI} & 24 & 4.69 & 6.31 \\
& 64 & 4.75 & 7.00 \\
& 128 & 4.81 & 7.47 \\

\bottomrule

\end{tabular}

\end{table}